\documentclass[journal=jacsat,manuscript=article]{achemso}

\usepackage[version=3]{mhchem} % Formula subscripts using \ce{}
\usepackage{algorithm} % Algorithm package
\usepackage{algpseudocode} % For nice algorithm visualization
\usepackage{subcaption}
\usepackage{braket}
\SectionNumbersOn  % numbering sections

\usepackage{soul}
\usepackage{xcolor}
\usepackage{soul}
\usepackage{xr}
\makeatletter
\newcommand*{\addFileDependency}[1]{% argument=file name and extension
  \typeout{(#1)}
  \@addtofilelist{#1}
  \IfFileExists{#1}{}{\typeout{No file #1.}}
}
\makeatother

\newcommand*{\myexternaldocument}[1]{%
    \externaldocument{#1}%
    \addFileDependency{#1.tex}%
    \addFileDependency{#1.aux}%
}
\usepackage{xr-hyper}
\myexternaldocument{supp_info} % Supporting information file

\author{Nonia Vaquero-Sabater}
 \affiliation{Donostia International Physics Center (DIPC), 20018 Donostia, Euskadi, Spain}
 \alsoaffiliation{Polimero eta Material Aurreratuak: Fisika, Kimika eta Teknologia Saila, Kimika Fakultatea, Euskal Herriko Unibertsitatea (EHU), PK 1072, 20080 Donostia, Euskadi, Spain}

\author{Abel Carreras}
\affiliation{Donostia International Physics Center (DIPC), 20018 Donostia, Euskadi, Spain}
\email{abelcarreras83@gmail.com}

\author{David Casanova}
\affiliation{Donostia International Physics Center (DIPC), 20018 Donostia, Euskadi, Spain}
\alsoaffiliation{IKERBASQUE, Basque Foundation for Science, 48009 Bilbao, Euskadi, Spain}
\email{david.casanova@dipc.org}

\title[]{Pruned-ADAPT-VQE: compacting molecular ans\"atze by removing irrelevant operators}

\abbreviations{}
\begin{document}


\begin{abstract}
The adaptive derivative-assembled problem-tailored variational quantum eigensolver (ADAPT-VQE) is one of the most widely used algorithms for electronic structure calculations in quantum computers. It adaptively selects operators based on their gradient, constructing ans\"atze that continuously evolve to match the energy landscape, helping avoid local traps and barren plateaus. However, this flexibility in reoptimization can lead to the inclusion of redundant or inefficient operators that have almost zero parameter value, barely contributing to the ansatz. We identify three phenomena responsible for the appearance of these operators: poor operator selection, operator reordering, and fading operators. In this work, we propose an automated, cost-free refinement method that removes unnecessary operators from the ansatz without disrupting convergence. Our approach evaluates each operator after ADAPT-VQE optimization by using a function that considers both its parameter value and position in the ansatz, striking a balance between eliminating low-coefficient operators while preserving the natural reduction of coefficients as the ansatz grows. Additionally, a dynamic threshold based on the parameters of recent operators enables efficient convergence. We apply this method to several molecular systems and find that it reduces ansatz size and accelerates convergence, particularly in cases with flat energy landscapes. The refinement process incurs, at most, a small additional computational cost and consistently improves or maintains ADAPT-VQE performance.
\end{abstract}

\clearpage
\section{Introduction}
Quantum simulation has long been regarded as one of the most promising applications of quantum computing, with the potential to achieve a significant quantum advantage \cite{bharti2022, McArdle2020}.
One of the earliest quantum algorithms designed for applications in quantum chemistry problems is the quantum phase estimation algorithm (QPEA).\cite{Aspuru-Guzik2005}
However, QPEA requires deep quantum circuits and extensive use of controlled operations, making it impractical for near-term quantum devices.
As an alternative, the Variational Quantum Eigensolver (VQE)\cite{Peruzzo:2014, McClean_2016} was introduced, better suited for the Noisy Intermediate-Scale Quantum (NISQ) era \cite{Arute_Arya_Babbush_Bacon_Bardin_Barends_Biswas_Boixo_Brandao_Buell_etal._2019, Preskill2018quantumcomputingi}. 
Despite its advantages, VQE still faces several challenges, including susceptibility to local traps \cite{Anschuetz2022, Bittel2021} and barren plateaus \cite{McClean2018, Arrasmith_2022}, which can hinder optimization and convergence.

The effectiveness of VQE strongly depends on the choice of ansatz. 
An ideal ansatz should be expressive enough to capture the exact solution while remaining shallow enough to be implemented within the coherence time of current quantum hardware. 
Additionally, it should have a minimal number of parameters to ensure efficient optimization and avoid unnecessary complexity in the classical optimization process. 
In molecular simulations, chemically inspired ans\"atze, such as Unitary Coupled-Cluster (UCC) methods \cite{Romero2018-zr, Lee2019}, are commonly used. 
These ans\"atze encode fermionic excitations applied to an initial state, such as Hartree-Fock. 
However, the direct encoding of fermionic excitations leads to deep circuits with a large number of two-qubit gates, making them impractical for NISQ devices.\cite{Yordanov2020}

To address these limitations, adaptive derivative-assembled problem-tailored VQE (ADAPT-VQE)\cite{Grimsley2019} was introduced as an iterative and problem-tailored approach. Instead of defining a fixed ansatz, ADAPT-VQE constructs the wavefunction dynamically by adding one fermionic excitation per iteration, selecting only the most relevant operators required for convergence. 
This adaptive strategy significantly reduces circuit depth compared to conventional VQE while ensuring that the ansatz remains compact and efficient. 
The resulting wavefunction takes the form:
\begin{equation}
    |\Psi\rangle = \prod_{i=1}^N e^{\theta_i\hat{A}_i} |\psi_0\rangle
    \label{eq:adapt_wf}
\end{equation}
where $N$ is the number of selected excitation operators ($\{\hat{A}_i\}$).
While standard UCC-based VQE optimizes all parameters $\{\theta_i\}$ simultaneously, ADAPT-VQE builds the ansatz incrementally. At each step, a new parameter is introduced (initially set to zero), while the previously optimized parameters are recycled (their values taken from the previous iteration). 
Then, all parameters are re-optimized. 
This strategy has shown to be very efficient, allowing ADAPT-VQE to improve convergence and reduce the likelihood of getting trapped in local minima.\cite{Grimsley2023}
This iterative refinement provides a warm start for parameterized circuits, a strategy demonstrated to be key in enhancing the efficiency of variational algorithms.\cite{Puig2025}

Despite its advantages, ADAPT-VQE faces significant challenges on current quantum hardware. 
As system complexity increases, the corresponding circuit depth grows, eventually exceeding the practical limits of today's devices.\cite{carreras2025limitationsquantumhardwaremolecular}
Also, high noise levels from gate errors, readout inaccuracies, and crosstalk degrade the accuracy of the results\cite{Dalton_Long_Yordanov_Smith_Barnes_Mertig_Arvidsson-Shukur_2024b}.
Moreover, mitigation methods add computational overhead, further limiting ourselves to the simplest techniques \cite{Kim_Eddins_Anand_Wei_van_den_Berg_Rosenblatt_Nayfeh_Wu_Zaletel_Temme_etal._2023}. 
Thus, while ADAPT-VQE remains a powerful algorithm, its practical application to significant molecular systems is currently infeasible due to circuit depth, amount of measurements, noise, and hardware constraints, requiring significant advancements in quantum technology \cite{Fedorov_Peng_Govind_Alexeev_2022a}.
Further algorithmic improvements aimed at reducing circuit depth are essential to enable its implementation on near-term quantum devices \cite{Yordanov2021, Anastasiou:tetris-adapt:2024, Tang:adapt:2021, Vaquero-Sabater_Carreras_Orús_Mayhall_Casanova_2024}.

The ADAPT-VQE method employs a gradient-based criterion to select the next excitation operator at each iteration, adding to the ansatz the operator with the largest gradient. This strategy has been shown to mitigate optimization challenges associated with barren plateaus.\cite{Grimsley2023}
However, the gradient-based criterion is not infallible. 
The operator with the highest gradient does not always lead to the greatest energy reduction, and in some cases, its contribution to the energy can be negligible, with nearly vanishing parameter values (see Supporting Information, Section S7). 
Moreover, during the iterative process, some operators that initially had a significant coefficient may eventually become negligible, contributing little to the total energy while increasing the overall ansatz size. 

Since such situations cannot be anticipated in advance, the gradient-driven, one-operator-at-a-time ansatz construction remains the most effective strategy currently available in ADAPT-VQE. 
However, one may wonder if there is some practical scheme that can, at least partially, correct this situation.
In this work, we explore a post-selection strategy aimed at removing negligible operators in order to reduce the size of the ansatz while preserving most of its expressivity. 
This idea has been previously explored,\cite{Ramoa_thesis} where under-performing operators were removed or blocked during several ADAPT-VQE iterations, leading to mixed results.   
Here, we propose an alternative protocol based on removing operators according to two simple criteria: the position of the operator within the ansatz and the magnitude of its associated coefficient. This approach is specifically designed with current NISQ devices in mind, where ansatz size has a significant impact on computational accuracy.

This study has two main objectives: (i) to identify and understand scenarios where certain excitation operators appear to perform poorly, exhibiting nearly zero-valued parameters, and (ii) to develop an efficient protocol for eliminating these redundant operators, thereby constructing more compact ans\"atze that yield shorter quantum circuits without compromising energy accuracy.

%------------------------------------------------
%------------------------------------------------
\section{Spotting superfluous operators} \label{sec:spotting_operators}
To begin, we aim to characterize cases where certain excitation operators contribute negligibly to the ansatz energy, exhibiting  near-zero associated parameters.
To illustrate this issue, we investigate the performance of ADAPT-VQE in a stretched linear \ce{H4} system (interatomic distance of 3.0~\AA). 
Highly correlated systems like this are particularly challenging, often requiring long ans\"atze to achieve chemical accuracy.  
For our simulations, we employ the 3-21G basis set, consisting of 8 orbitals (Figure S1) mapped to 16 qubits. 
Although ADAPT-VQE simulations in the literature commonly employ the minimal STO-3G basis set, in this study we opt for the larger 3-21G basis.
This allows for more complex wavefunction construction and better recovery of electron correlation.
For comparison, representative results using the STO-3G basis can be found in the Supporting Information (Section S10).

The excitation operator pool consists of UCC operators restricted to occupied-to-virtual spin-singlet adapted single and double excitations.\cite{Grimsley2019} 
We write these operators as $\hat{A}_i^a$ and $\hat{A}_{ij}^{ab}$ for single and double excitations, respectively, where $i,j$ represent the occupied orbitals in the Hartree-Fock determinant and $a,b$ span all virtual orbitals. 
For simplicity, along the text, we use the Mulliken notation for the orbitals abbreviated as $ng$ and $nu$, where $n=1-4$ is the orbital's number, and $g$ denotes gerade and $u$ ungerade symmetry.
See Section~\ref{sec:operators_SI} for additional information.
Explicit expressions for the spin-adapted unitary fermionic operators in terms of creation and annihilation operators can be found in the Supporting Information (Section~\ref{sec:operators_SI}).
The fermionic spin-adapted operators are transformed to qubit operators  using the Jordan–Wigner mapping.\cite{JW:mapping:1928}
It should be emphasized that, under this mapping, a single fermionic operator generally corresponds to a linear combination of multiple qubit operators. 
In practice, these qubit operators are implemented in quantum circuits through a Trotterization scheme with a limited number of Trotter steps. 
However, this approximation can break the spin symmetry of the ansatz, potentially leading to solutions that are strictly not spin-pure.
For completion, we have also tested our protocol with other pools\cite{Yordanov2021} and basis sets. The results can be found in the Supporting Information (Sections S9 and S10).

Numerical optimization of the ansatz parameters is carried out using the Broyden–-Fletcher–-Goldfarb–-Shanno algorithm.\cite{BFGS} 
Simulations are conducted with an in-house Python implementation of ADAPT-VQE, openly available on GitHub,\cite{VQEmulti} utilizing the NumPy,\cite{numpy} SciPy,\cite{2020SciPy} and OpenFermion\cite{mcclean2019openfermion} packages. 
All simulations in this study were performed under idealized conditions, neglecting both sampling and hardware-induced noise.

We define three convergence criteria for our simulations:
\begin{enumerate}
 \item The optimized coefficient of the most recently added operator becomes zero.
 \item The energy improvement is non-positive.
 \item A zero-valued coefficient is added or the last operator is removed, which effectively returns the ansatz to its previous form, causing the same operator to be selected and subsequently discarded.
\end{enumerate}
The simulation is considered converged when any of these criteria are met for ADAPT-VQE, or when only criteria 1 and 3 are met for Pruned-ADAPT-VQE.

\begin{figure}[H]
    \centering
    \includegraphics[width=10cm]{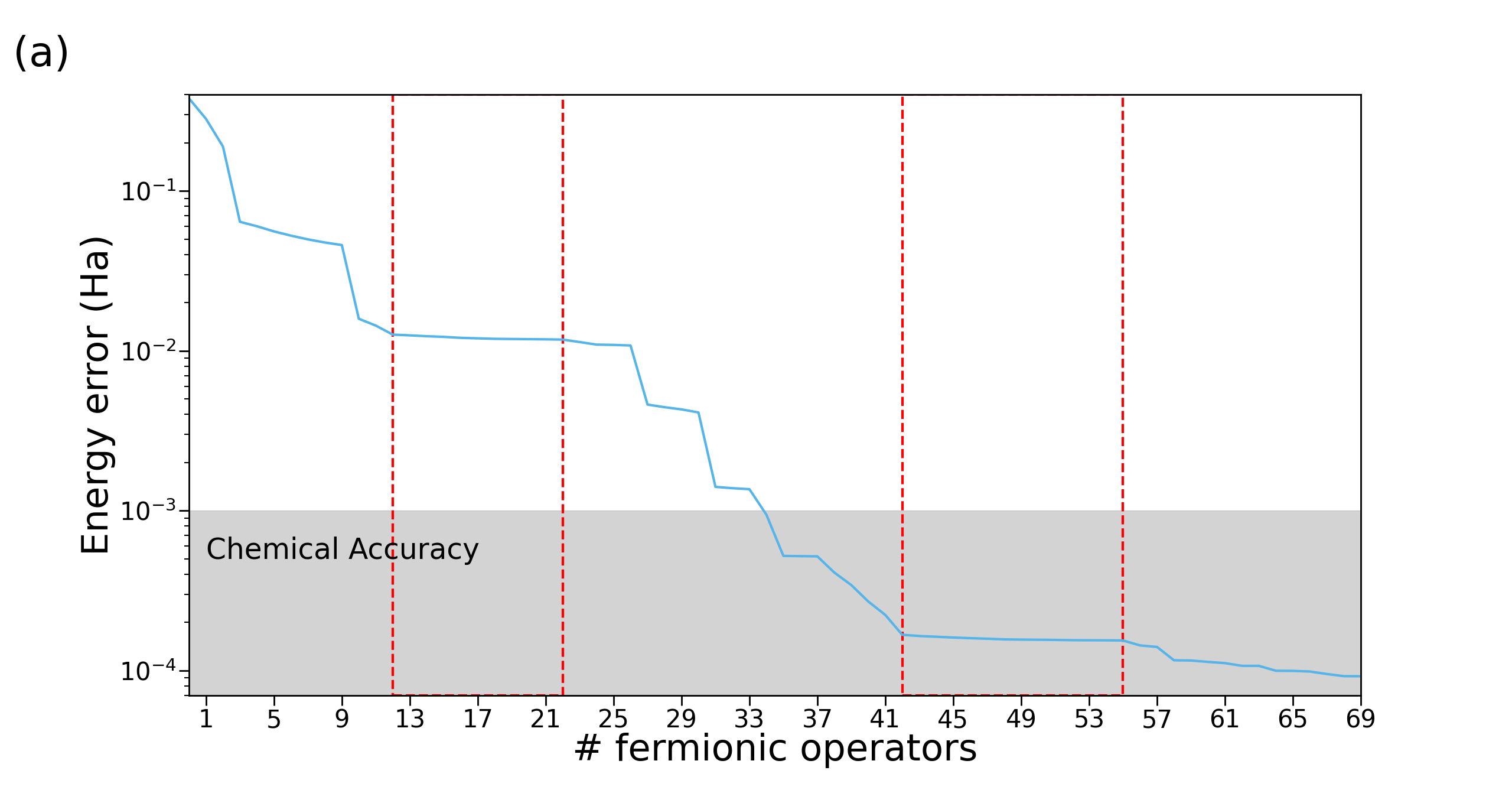}
    \includegraphics[width=10cm]{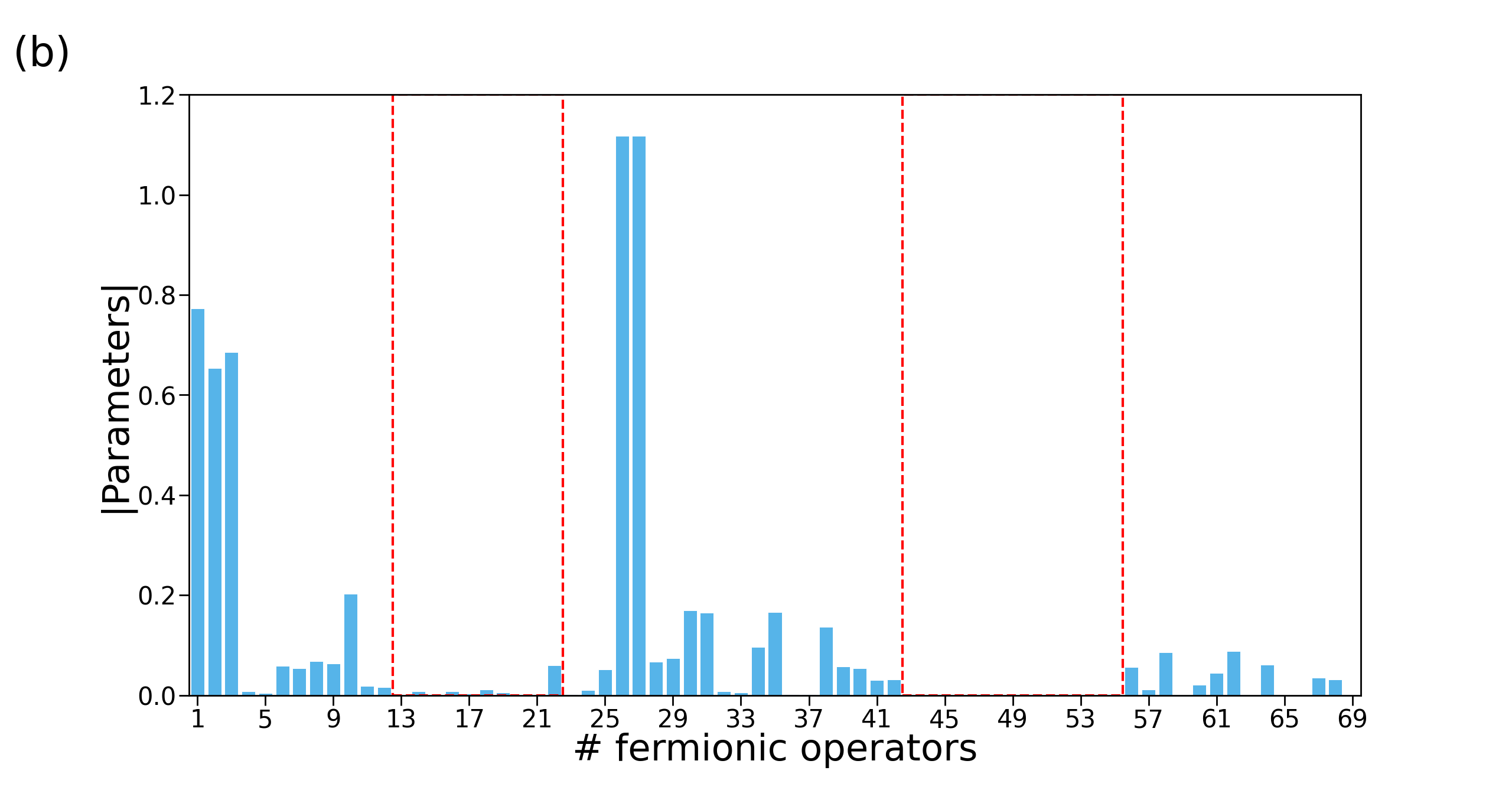}
    \includegraphics[width = 10cm]{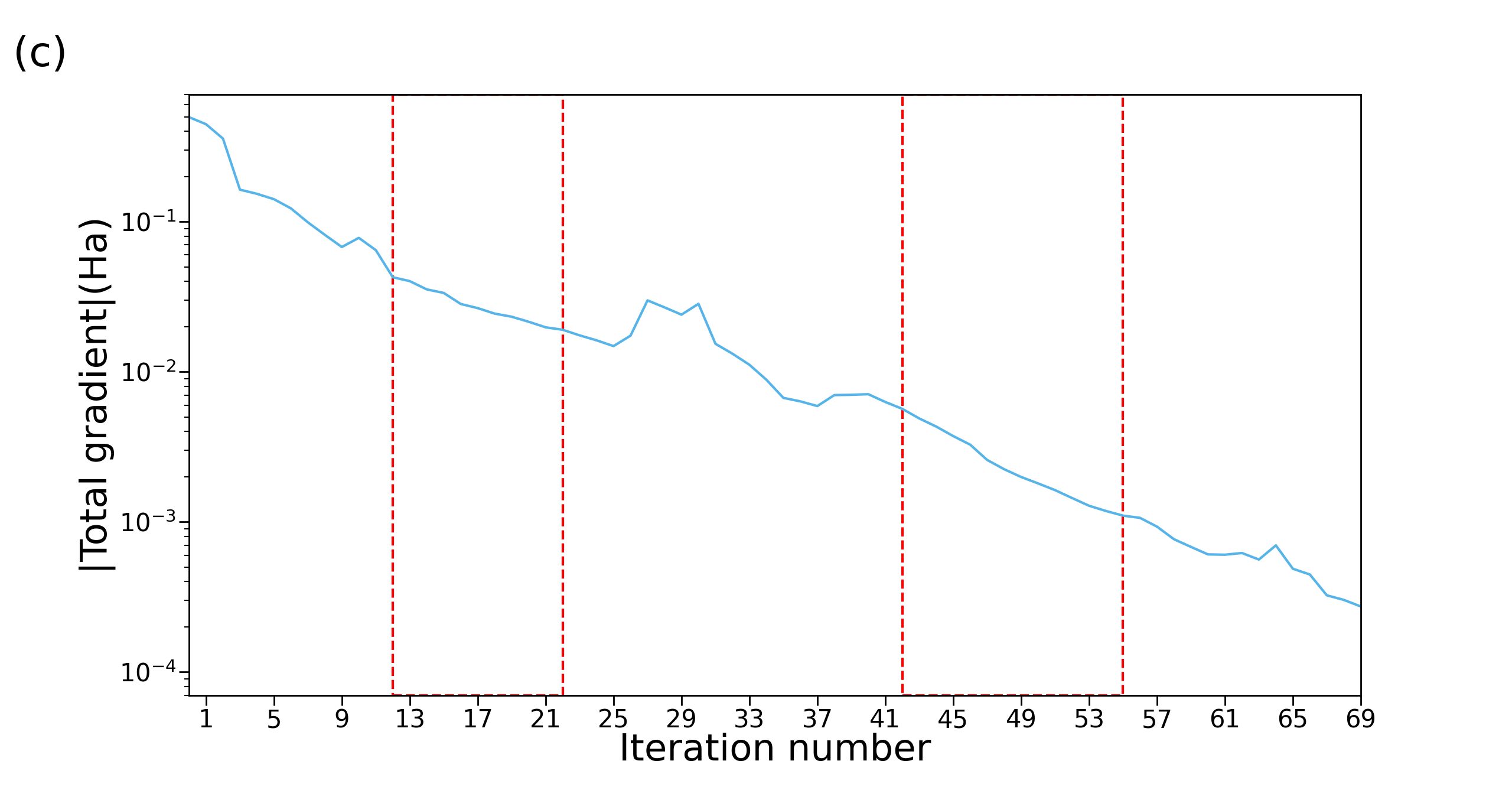}
    \caption{(a) Energy error (in a.u.) of the ADAPT-VQE with respect to FCI as a function of the number of operators, (b) distribution of absolute value of the ansatz parameters at $N=69$, and (c) the norm of the total gradient (sum of the gradients of all the operators in the pool) (in a.u.) obtained for linear \ce{H4} with interatomic distance of 3.0~\AA, with the 3-21G basis set. 
    Vertical dashed red lines delimit flat energy regions.}
    \label{fig:comparison_h4}
\end{figure}

Figure~\ref{fig:comparison_h4}a illustrates the evolution of the energy error in ADAPT-VQE relative to full configuration interaction (FCI) as a function of the number of operators in the ansatz. 
The distribution of the absolute value of the ansatz parameters at $N=69$ (Figure~\ref{fig:comparison_h4}b) reveals the presence of several operators with nearly zero-valued parameters.
Notably, contiguous operators with negligible parameter values correspond to flat regions in the energy profile, indicating that, despite being selected based on the gradient criterion, these operators do not meaningfully contribute to energy optimization. 
Consistently, in Section~\ref{sec:energy_diff_SI} we provide additional data showing the correlation between the magnitude of a parameter and the contribution of the associated operator to the energy.

Figure~\ref{fig:comparison_h4}c shows the gradient norm as a function of the iteration number, revealing the presence of gradient troughs\cite{Grimsley:adapt:2023} corresponding to the flat regions in the evolution of energy errors (Figure~\ref{fig:comparison_h4}a). 
As expected, the gradient norm increases concurrently with the parameter values, while the energy error decreases.

In Section~\ref{sec:energy_diff_SI} we present additional data that demonstrates the correlation  between the parameter value and the contribution of the associated operator to the energy.

A detailed analysis of the occurrence of operators with $\theta\approx0$ in the ADAPT-VQE ansatz for the linear \ce{H4} simulation reveals three underlying mechanisms: (i) poor or incorrect operator selection, (ii) operator reordering, and (iii) fading operators.
A similar analysis for the water molecule can be found in the Supporting Information (Section S6).

%------------------------------------------------
\subsection{Poor operator selection} \label{sec:poor_selection}
Typically, after adding a new operator and re-optimizing all the parameters, using the previously optimized values as initial guesses in the classical optimization (a technique we refer to as amplitude recycling), the ansatz gains expressivity, leading to a lower energy and reduced error relative to the exact solution. 
However, in some cases, the newly added operator $\hat{A}_N$ has little to no impact on the ansatz, resulting in a nearly zero parameter value $\theta_N\approx0$ from the moment it is introduced, and failing to significantly alter the energy.
We classify these instances as a consequence of poor (or incorrect) operator selection, since more optimal paths can be found (see Supporting Information, Section S7). 
The selection of these operators appear to be closely related to the presence of gradient troughs, which hinder the search for the optimal operator and result in regions with minimal error improvement, i.e., flat areas.

\begin{figure}[H]
    \centering
    \includegraphics[width=13cm]{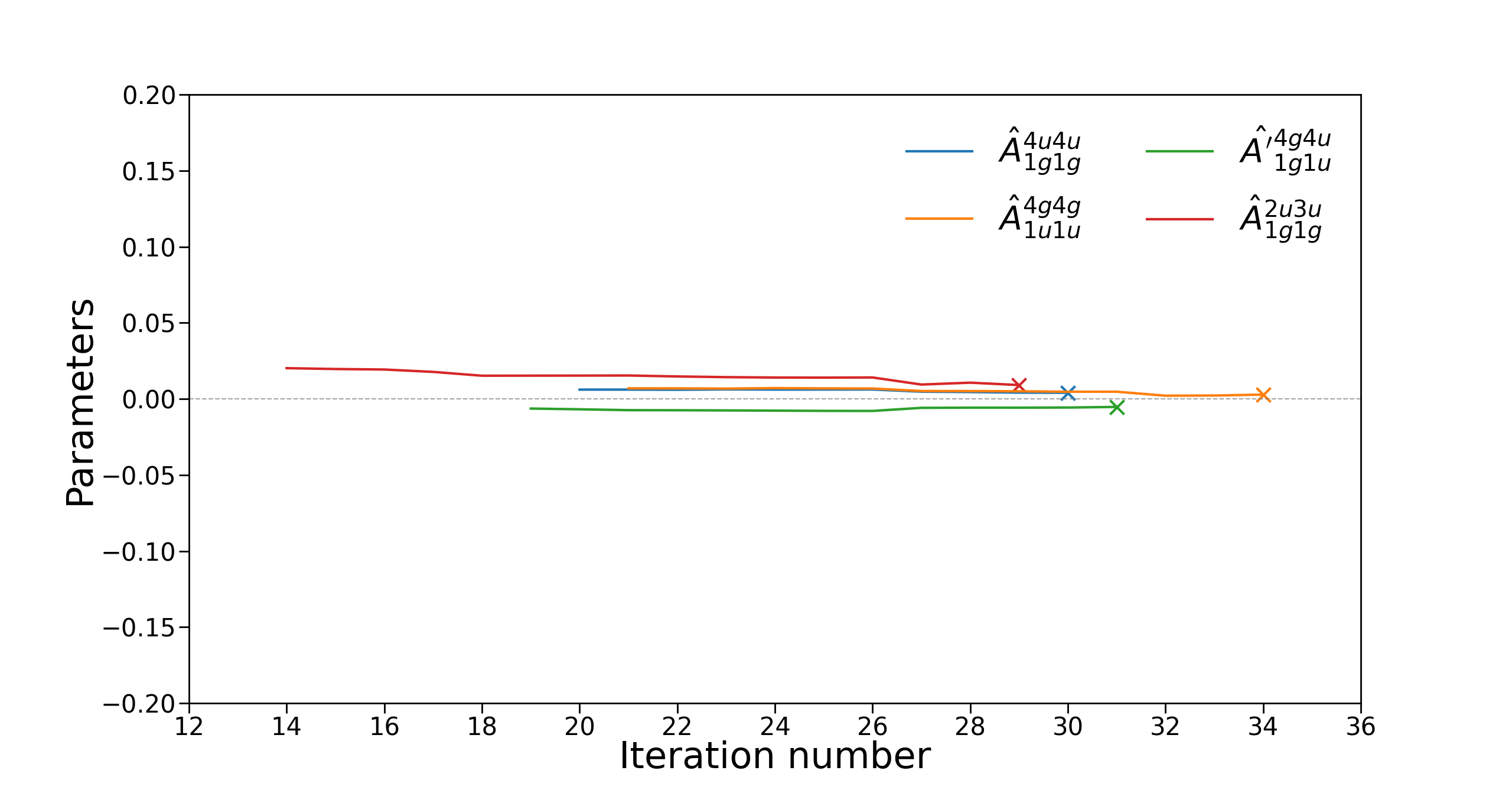}
    \caption{Parameter values of various operators from their introduction to iteration 34 for the simulation of linear \ce{H4} with interatomic distance of 3.0~\AA, with the 3-21G basis set.
    Cross markers indicate when the operator is removed by Pruned-ADAPT-VQE (see Section~\ref{sec:prune_performance}).}
    \label{fig:h4_linear_badops}
\end{figure}

As illustrated in Figure~\ref{fig:h4_linear_badops}, certain operators exhibit nearly zero parameter values immediately after being introduced into the ansatz, indicating that their selection did not meaningfully contribute to the optimization process. 

These operators exhibit consistently small parameter values from the moment they are introduced, remaining negligible throughout the entire simulation. 
This suggests that they were poorly or incorrectly selected and could potentially be removed from the ansatz without significant impact. To verify this, we compare the ADAPT-VQE energy at iteration 27 (just after the energy drop observed in Figure~\ref{fig:comparison_h4}) with the energy obtained after eliminating these operators, $\hat{A}_{1g1g}^{4u4u}$ and $\hat{A'}_{1g1u}^{4g4u}$, and reoptimizing the remaining parameter values. 
The resulting energy difference is minimal, approximately 0.046~mHa, reinforcing the need to develop strategies for identifying and eliminating inefficient operators.

%------------------------------------------------
\subsection{Operator reordering} \label{sec:op_reordering}
The adaptive nature of ADAPT-VQE allows the ansatz to dynamically adjust to the specific requirements of the problem being studied. 
As the wavefunction is iteratively constructed by adding excitation operators, the ansatz continuously evolves to improve accuracy. 
Notably, once an operator is added to the ansatz, it remains in the pool and is not removed.\cite{Grimsley2019} 
This can lead to the inclusion of multiple instances of the same operator.  
While such duplicity can sometimes be beneficial, we have identified cases where adding a duplicate operator causes a sudden drop in the parameter value of the previously included instance to near-zero value, no longer effectively contributing to the final wavefunction.
This behavior serves as an effective reordering mechanism, highlighting ADAPT-VQE’s ability to dynamically optimize the sequence of operators, an important feature given that excitation operators from UCC generally do not commute, i.e., different orderings of operators are not equivalent.\cite{Grimsley2020}

\begin{figure}[H]
    \centering
    \includegraphics[width=13cm]{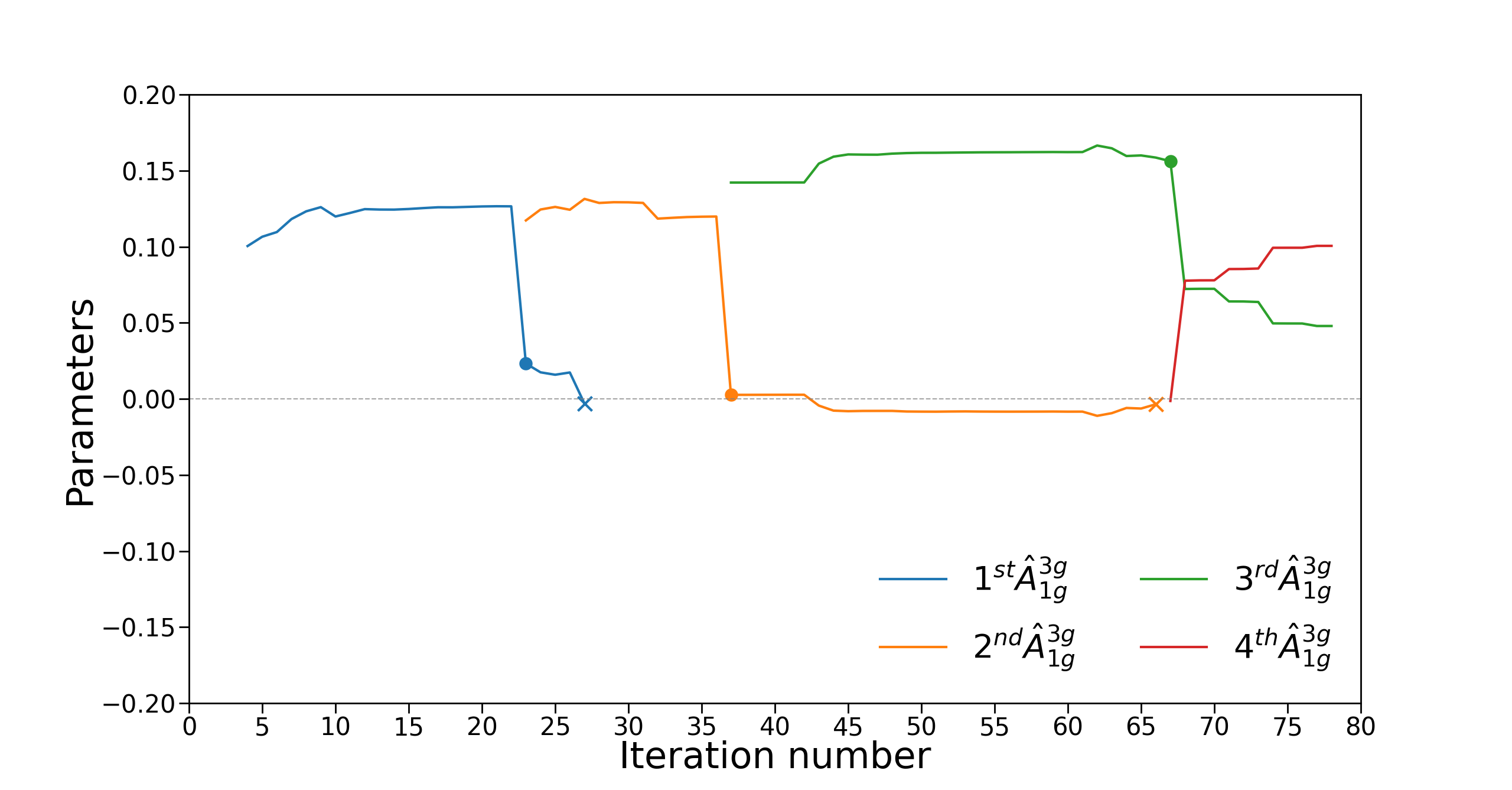}
    \caption{Parameter values of various operators from their introduction to iteration 80 in the simulation of linear \ce{H4} with interatomic distance of 3.0~\AA~and the 3-21G basis set. Full circles indicate introduction of another instance of the operator. 
    Cross markers indicate when the operator is removed by Pruned-ADAPT-VQE (see Section~\ref{sec:prune_performance}).
    }
    \label{fig:h4_linear_reorderedops_del}
\end{figure}

Figure~\ref{fig:h4_linear_reorderedops_del} illustrates this reordering effect, showing the progression of operator $\hat{A}_{1g}^{3g}$, which has been added four different times. In iteration 23, the second $\hat{A}_{1g}^{3g}$ is added, causing a sudden drop in the first operator parameter value.
This situation is repeated in iteration 37, when a third operator is added, causing the second $\hat{A}_{1g}^{3g}$ coefficient to drop to almost zero. This shows how operator $\hat{A}_{1g}^{3g}$ is relocated as the ansatz progresses.
This suggests that the ansatz dynamically adjusts the sequence of operators, effectively reorganizing them to better optimize the wavefunction construction.

In other cases, however, we observe that adding an additional instance of the same operator does not cause the parameter of the previous instance to drop to zero. Instead, the two instances share similar parameter values, such that their sum remains close to the value of the original parameter. 
This behavior suggests a form of quasi-commutation between that segment of the ansatz and the repeated operator, where placing the operator in either position results in a similar energy. Consequently, the optimizer converges to an intermediate configuration in which both instances retain a significant portion of the parameter.
This phenomenon is observed in Figure~\ref{fig:h4_linear_reorderedops_del}, where the operator $\hat{A}_{1g}^{3g}$ is added for a fourth time.

%------------------------------------------------
\subsection{Fading operators} \label{sec:fading_op}
\begin{figure}[H]
    \centering
    \includegraphics[width=12cm]{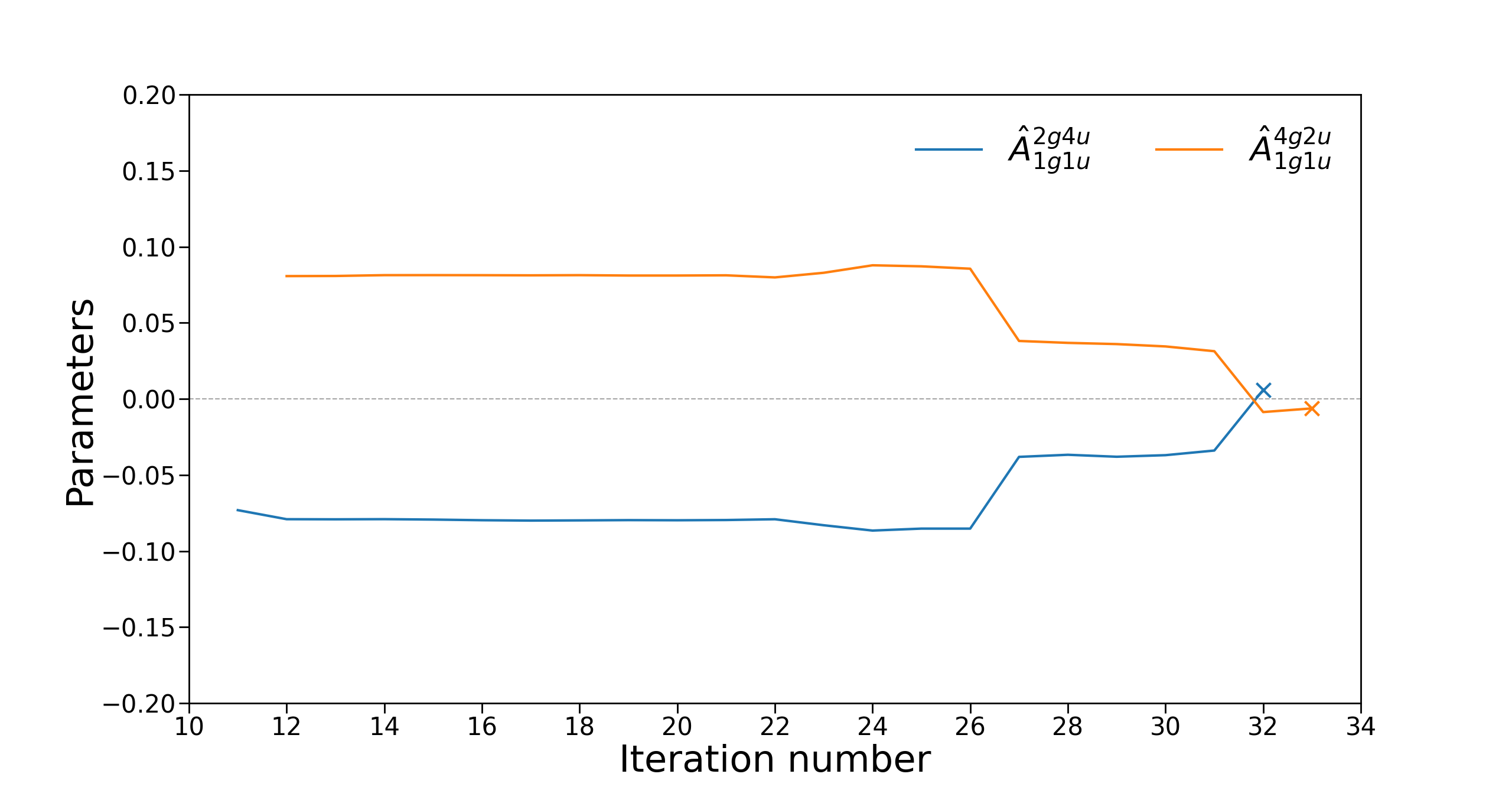}
    \caption{Parameter values of various operators from their introduction to iteration 34 in the simulation of linear \ce{H4} with interatomic distance of 3.0~\AA, with the 3-21G basis set. Cross markers indicate when the operator is removed by Pruned-ADAPT-VQE (see Section~\ref{sec:prune_performance}).
    }
    \label{fig:h4_linear_lostops}
\end{figure}

As the ansatz grows, some operators that initially play a significant role, having sizable parameter values, gradually become irrelevant. 
In other words, certain operators that were once crucial for describing the wavefunction eventually contribute little to the final solution. 
Predicting when such recalibrations will occur is generally challenging.  
One possible explanation is that the ansatz may initially converge to a local minimum where a given excitation operator is essential. 
However, as the ansatz expands and explores a larger Hilbert space, the algorithm may transition to a lower-energy minimum where the previously important operator is no longer needed, leading to a near-zero parameter value.
Additionally, this fading effect could be linked to the ``burrowing into the energy landscape" process,\cite{Grimsley2023} where qualitative changes in the wavefunction structure might render certain operators obsolete.  
Figure~\ref{fig:h4_linear_lostops} exemplifies this phenomenon, showing how operators $\hat{A}_{1g1u}^{2g4u}$ and $\hat{A}_{1g1u}^{4g2u}$, introduced in iterations 10 and 11 of ADAPT-VQE, initially carry significant weight but become negligible after iteration 30.

%------------------------------------------------
\subsection{Cooperative operator action} \label{sec:cooperative_op}
We observe that, in some cases, newly added operators initially have negligible parameter values but later become significant contributors to the ansatz. 
While their inclusion may initially appear to be a poor selection, the addition of subsequent operators can trigger a substantial increase in their parameter values.  
This suggests that certain operators, despite their seemingly insignificant impact at first, play a crucial role in unlocking specific regions of the Hilbert space. 
Their effectiveness emerges only when combined with other operators, highlighting the necessity of collective and cooperative action in the construction of an optimal ansatz.

\begin{figure}[H]
    \centering
    \includegraphics[width=12cm]{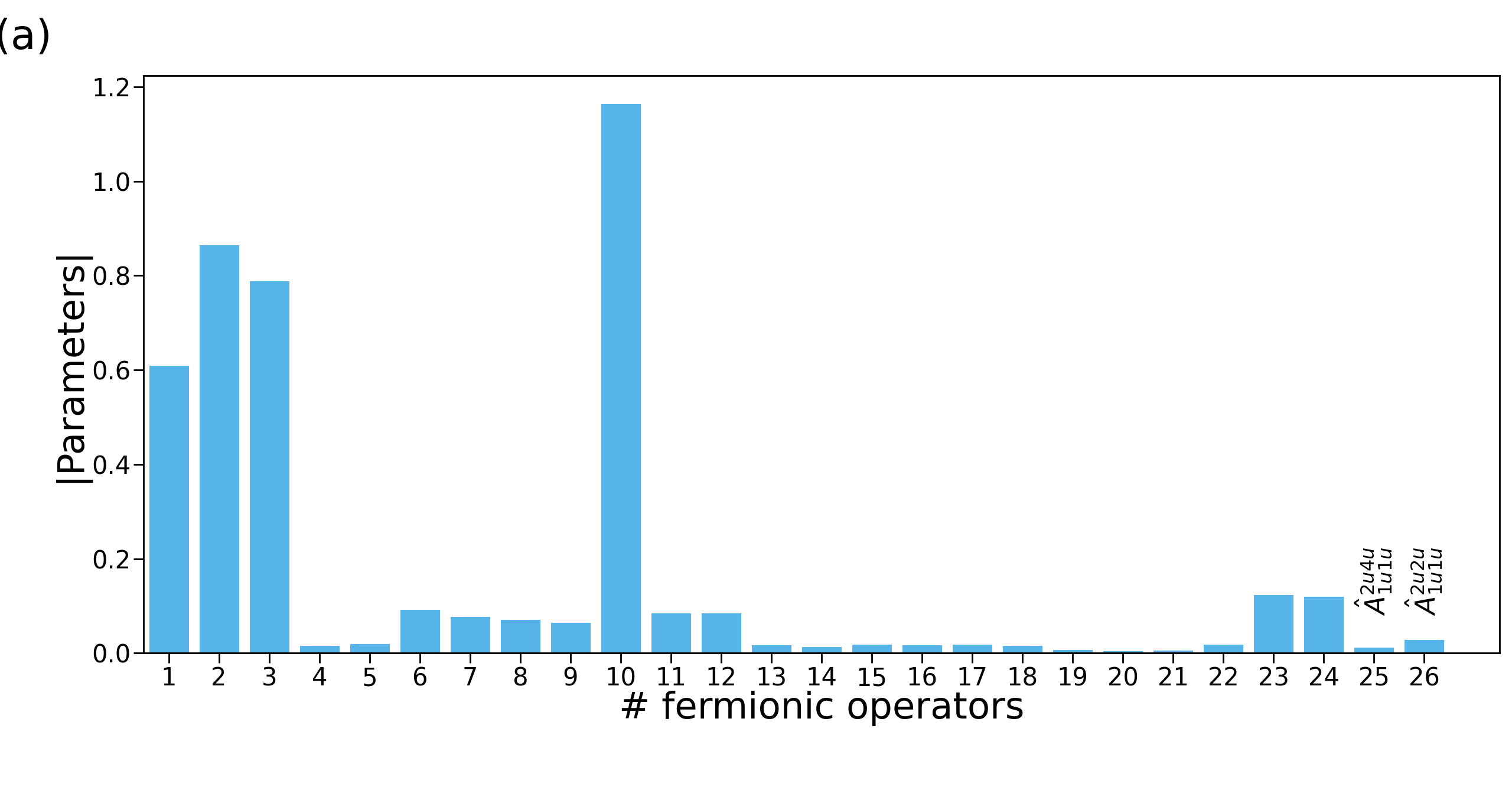}
    \includegraphics[width=12cm]{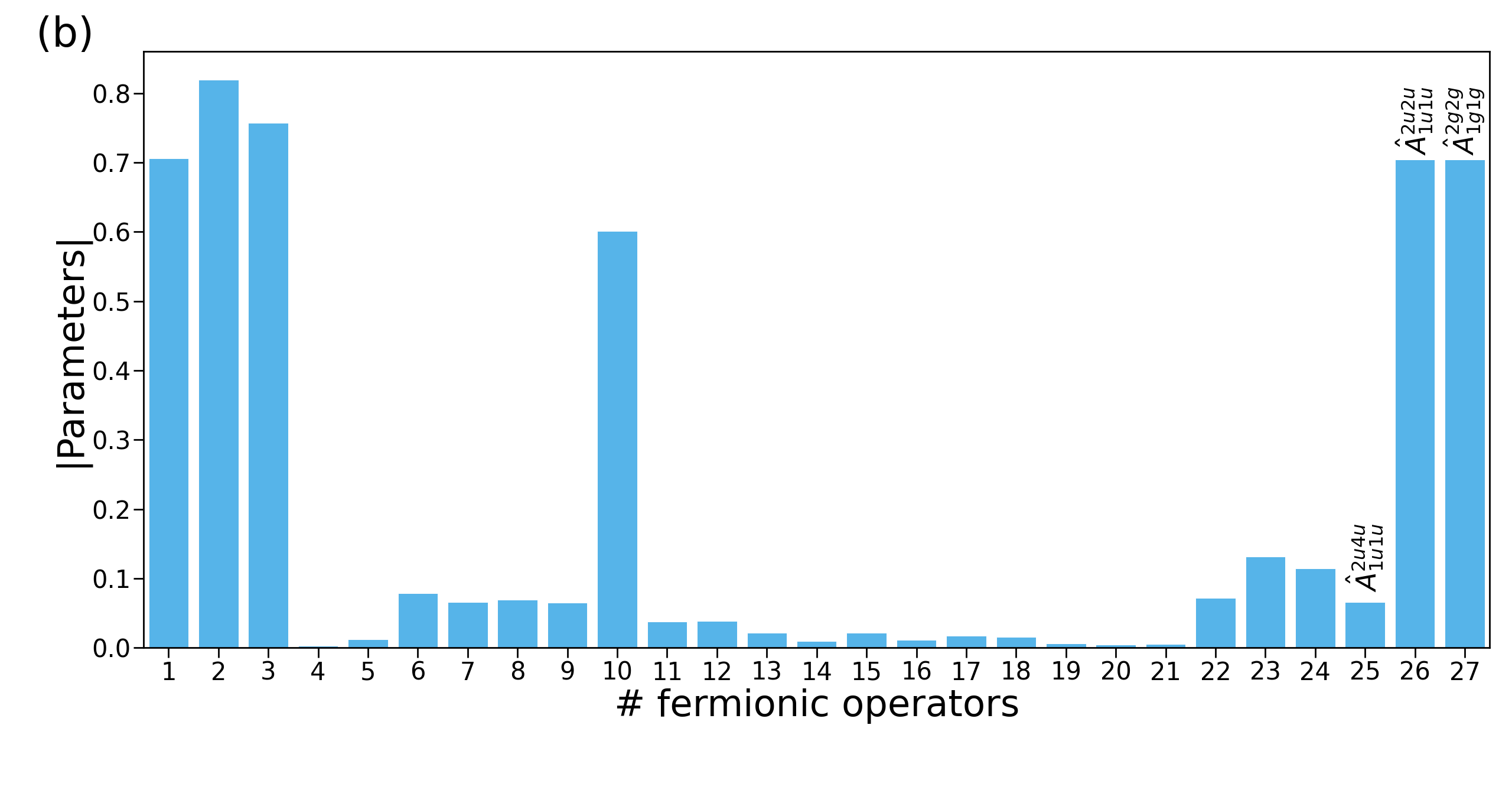}
    \caption{Absolute parameter value at iteration 26 (a) and 27 (b) of ADAPT-VQE for the simulation of linear \ce{H4} with interatomic distance of 3.0~\AA, with the 3-21G basis set.}
    \label{fig:h4_linear_energies_first27}
\end{figure}

This phenomenon is clearly illustrated in Figure~\ref{fig:h4_linear_energies_first27}, which depicts the ansatz composition at iterations 26 and 27. 
These iterations coincide with a sudden energy drop (Figure~\ref{fig:comparison_h4}a), demonstrating how the cooperative action of multiple operators can unlock lower-energy solutions that were previously inaccessible.
Operator $\hat{A}_{1u1u}^{2u2u}$ in position 26 (and, to a lesser extent, operator $\hat{A}_{1u1u}^{2u4u}$ in position 25) initially appears with a small parameter value.
However, after an additional ADAPT iteration, specifically, upon the introduction of operator $\hat{A}_{1g1g}^{2g2g}$ in position 27, operator $\hat{A}_{1u1u}^{2u2u}$ becomes one of the most significant contributors to the ansatz.
The cooperative effect between operators $\hat{A}_{1u1u}^{2u2u}$ and $\hat{A}_{1g1g}^{2g2g}$ is evident when evaluating the energy without operator $\hat{A}_{1u1u}^{2u2u}$, which leads to an increase of 6.2~mHa. 
This energy shift exceeds the threshold for chemical accuracy,\cite{pople1999nobel} and is more than two orders of magnitude greater than the change observed when removing the poorly selected operators $\hat{A}_{1g1g}^{4u4u}$ and $\hat{A'}_{1g1u}^{4g4u}$ (in positions 19 and 20), as discussed in Section~\ref{sec:poor_selection}.
This distinction underscores the difference between truly redundant operators and those that, despite initially having small parameter values, play a crucial role by collectively unlocking key regions of the Hilbert space.

%--------------------------------------------------------
\section{Pruned-ADAPT-VQE algorithm} \label{sec:prune_algorithm}
Motivated by the various mechanisms that introduce ineffective operators into the ADAPT-VQE ansatz, and recognizing the potential for ansatz compaction, leading to shorter circuits without compromising energy accuracy, we develop an automated algorithm to remove these unnecessary contributions. 
The proposed method systematically eliminates poorly selected operators, those with fading parameter values, and redundant operators arising from reordering, while preserving operators involved in cooperative interactions. 
To achieve this, we introduce a simple yet effective routine, which we call Pruned-ADAPT-VQE.

%--------------------------------------------------------
% Decision factor
We recognize that the criterion for removing the $i$th operator from the ansatz must take into account both the magnitude of its parameter $\theta_i$ and its position within the ansatz. 
To formalize this, we introduce a decision factor $f_i$ for each operator, defined as the product of two functions: one dependent on the operator’s parameter and the other on its position:
\begin{equation} \label{eq:decision_factor}
    f_i = F_1(\theta_i)F_2(x_i)
\end{equation}
where $x_i$ is the relative position of the operator in an ansatz with $N$ operators, given by $x_i=i/N$.
We use this decision factor evaluated over each operator at each ADAPT-VQE iteration to determine which one will be removed.
This formulation ensures that both small-amplitude operators and those appearing earlier in the ansatz are systematically evaluated for potential removal.

Among the possible functions that assign a larger factor to operators with smaller absolute parameter values, we select the inverse of the squared amplitude, as it naturally disregards the sign of $\theta_i$ and strongly emphasizes operators with near-zero parameter values:
\begin{equation}\label{eq:F1_function} 
    F_1(\theta_i) = \dfrac{1}{\theta_i^2} 
\end{equation}
Testing with alternative functions ($|\theta_i^{-n}|$, with $n=0,1,3,4$) did not yield any noticeable advantage or improvement over the choice in equation~\ref{eq:F1_function} (see discussion in Section S3 of the Supporting Information).

Additionally, we prioritize the removal of operators with small parameter values that appear early in the ansatz. 
This prevents the elimination of cooperative operators and accounts for the natural decrease in amplitude of newly added operators as the ansatz approaches convergence.
To achieve this, we introduce a position-dependent function $F_2$ that decays exponentially with the operator’s position within the ansatz:
\begin{equation}\label{eq:F2_function}
    F_2(x_i) = e^{-\alpha x_i}
\end{equation}
where $\alpha$ is a positive hyperparameter controlling the influence of position on the removal criterion.
Choosing an appropriate value for $\alpha$ requires balancing competing effects: a very large $\alpha$ would give priority to the operators in the first positions, preventing the removal of the intermediate or later ones, while a value approaching zero would make the position irrelevant, potentially eliminating recent additions that still contribute to convergence.
Through preliminary testing, we find that when $\alpha$ is too small, proper convergence is not achieved, as the latest operator added is always removed in each iteration due to having the smallest coefficient. 
To prevent this, the function must assign sufficient relevance to the position. 
For the studied cases, we found that beyond a certain threshold ($\alpha\approx10$) the method provides a reasonable compromise, ensuring effective pruning while maintaining smooth ansatz optimization (see the discussion in Section S4 of the Supporting Information).

\begin{figure}[H]
    \centering
    \includegraphics[width=12cm]{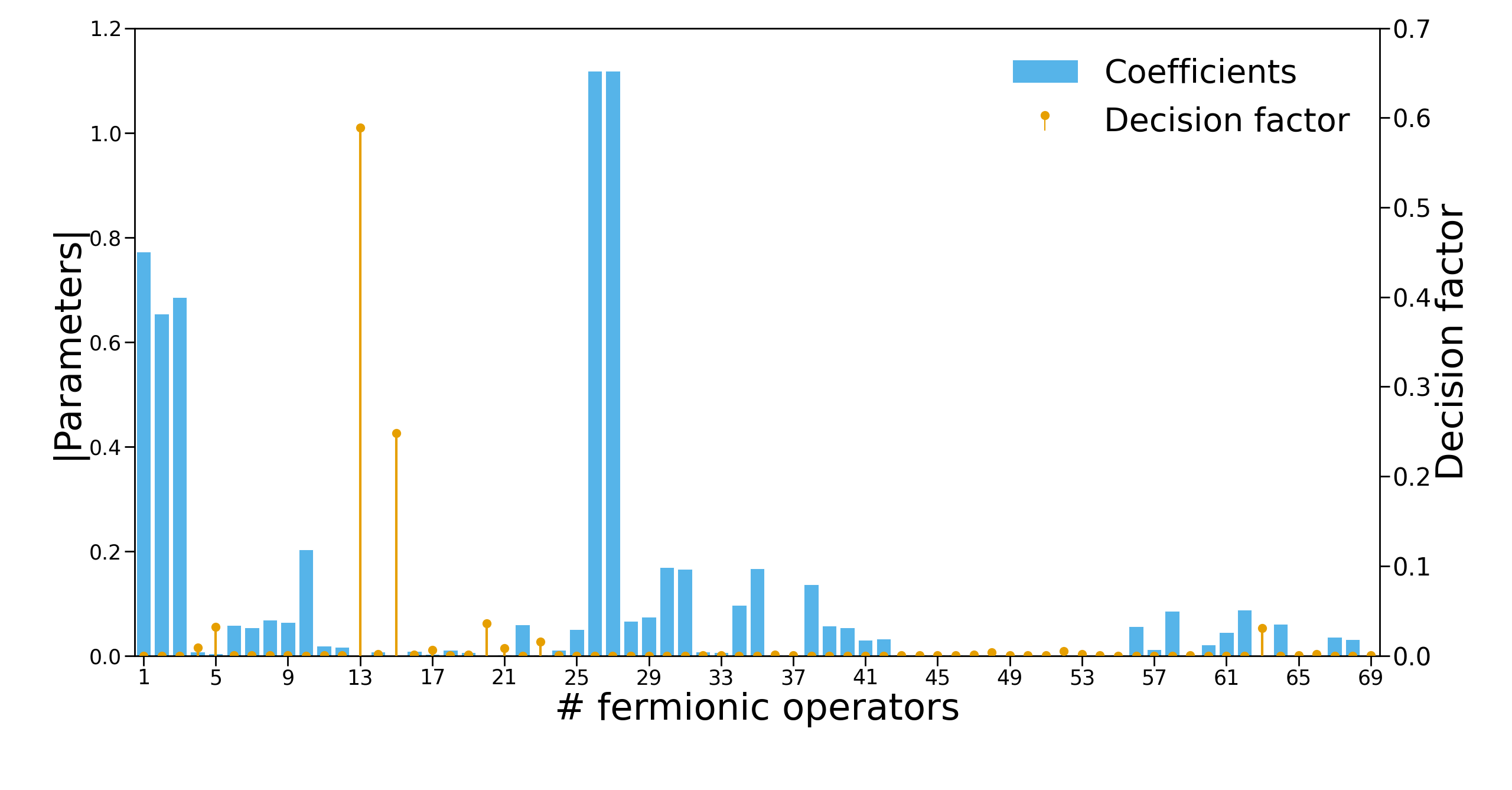}
    \caption{Absolute parameter values (blue bars) and decision factor values (orange sticks) for every ansatz operator of the linear \ce{H4} with interatomic distance of 3.0~\AA, with the 3-21G basis set.}
    \label{fig:selecting_function_profile}
\end{figure}

Figure~\ref{fig:selecting_function_profile} illustrates the set of $\{f_i\}$ values when applied to the final ansatz (with 69 excitation operators) of the linear \ce{H4} molecule.
The operator with the highest factor and selected for potential removal corresponds to operator in position 13, identified as a poorly selected operator and appearing along a flat energy region in Figure~\ref{fig:comparison_h4}a.

Note that, despite their proximity, operators 13 and 15 have coefficients on the order of $10^{-4}$, whereas operators 14 and 16–21 lie in the $10^{-3}$–$10^{-2}$ range, leading to substantially lower decision factors. 
This distinction is not apparent in the plot, but the use of a logarithmic scale was avoided to ensure consistency with the rest of plots.

%--------------------------------------------------------
% Filtering process (algorithm)

Therefore, to refine the ADAPT-VQE ansatz, we introduce a filtering process (Algorithm~\ref{alg:prune}) that evaluates operators after each optimization iteration, identifying candidates for removal while preserving essential contributions.
This approach prioritizes the removal of early-appearing operators with negligibly small parameter values while retaining recently added operators that may play a cooperative role. 
To define $\tau$, we use a fraction (0.1) of the average amplitude of the $N_L$ most recently added operators:
\begin{equation}
    \tau = \frac{0.1}{N_L}\sum_{i=0}^{N_L-1}|\theta_{N-i}|
    \label{eq:threshold}
\end{equation}
We examined the dependence of the dynamic threshold on $N_L$ in equation~\ref{eq:threshold}, varying $N_L$ from 1 to 10 (see Section S5 of the Supporting Information).
Our analysis shows that adjusting this parameter has a mild impact on the final outcome for $2\le N_L\le10$.
Therefore, as a balanced choice, we set $N_L=4$ for all subsequent analyses. 
We have also found (see Supporting Information Sections S5 and S9) that varying the fraction of the average amplitude provides a convenient way to control simulation convergence. Using a large fraction results in a more aggressive removal criterion and faster convergence, which can lead to premature termination if the fraction is too large. Conversely, a small fraction causes the simulation to behave more like the standard ADAPT-VQE.

The removal process is performed after a standard ADAPT-VQE iteration, as outlined in Algorithm~\ref{alg:prune}, and proceeds without reoptimizing parameters, as the elimination of these low-impact operators has a negligible effect on the ansatz.
As a result, this step does not increase the overall computational cost of the algorithm, as the evaluation of the decision factor and dynamic threshold is negligible. Figure S44 in Section S13 of the Supporting Information examines whether the removal of operators results in additional optimizer iterations in the classical routine, revealing a minor overhead that is mitigated by the end of the simulation.
The Pruned-ADAPT-VQE routine (Algorithm~\ref{alg:prune}) proceed as follows: (i) after each ADAPT-VQE iteration all operators in the ansatz are evaluated using equation~\ref{eq:decision_factor}; (ii) 
the parameter $\theta_j$ associated with the operator with the largest decision factor $f_j$ is compared against a dynamic threshold $\tau$, proceeding to the removal only if $\theta_j < \tau$; and (iii) the energy is calculated with the corresponding ansatz.
\begin{algorithm}[H]
\caption{Pruned-ADAPT-VQE algorithm}\label{alg:prune}
\begin{algorithmic}[1]
\State \textbf{Initialize:} Choose initial ansatz $\psi(0)$ and operator pool $\{\hat{A}_i\}$
\Repeat
    \State Compute gradients $\partial E / \partial \theta_i ~\forall~\hat{A}_i$
    \State Select operator $\hat{A}_k$ with the largest gradient
    \If{$|\partial E / \partial \theta_k| < \epsilon$}
        \State \textbf{Terminate}
    \Else
        \State Add $e^{\theta_k\hat{A}_k}$ to the ansatz and optimize parameters $\theta$
    \EndIf
    \State Compute $f_i~\forall~\hat{A}_i \in \text{ansatz}$ \Comment{equation~\ref{eq:decision_factor}} 
    \State Select operator $\hat{A}_j$ with largest $f_j$
    \State Compute threshold $\tau$ \Comment{equation~\ref{eq:threshold}}
    \If{$\theta_j<\tau$}
        \State Remove operator $\hat{A}_j$
    \EndIf
    \State Compute energy $\bra{\psi(\theta)} \hat{H} \ket{\psi(\theta)}$
\Until{Convergence}
\State \textbf{Output:} Optimized ansatz and energy 
\end{algorithmic}
\end{algorithm}

%------------------------------------------------
% Performance of Pruned-ADAPT-VQE
\section{Performance of Pruned-ADAPT-VQE} \label{sec:prune_performance}
\begin{figure}[H]
    \centering
    \includegraphics[width=11cm]{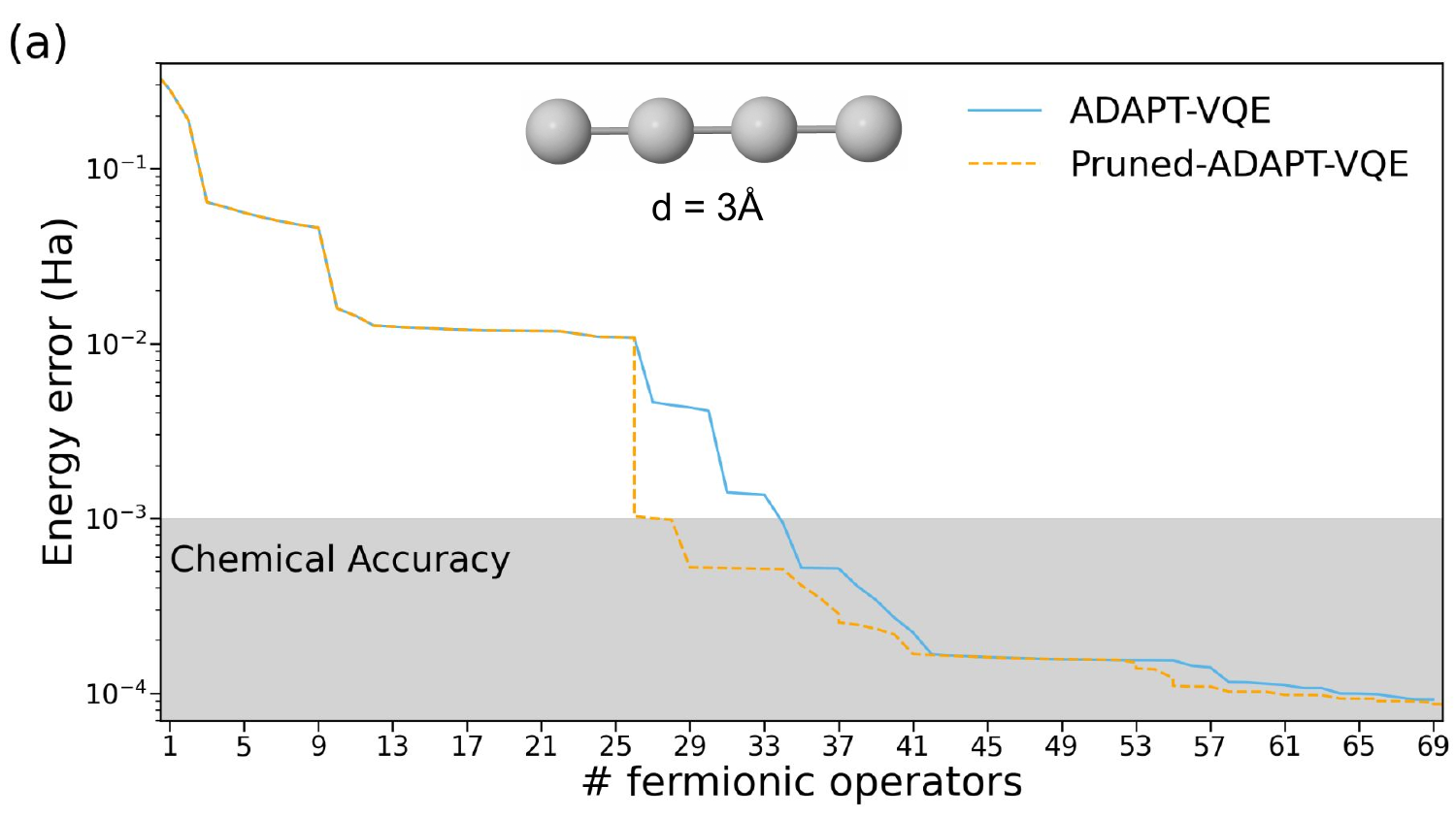}
    \includegraphics[width=10cm]{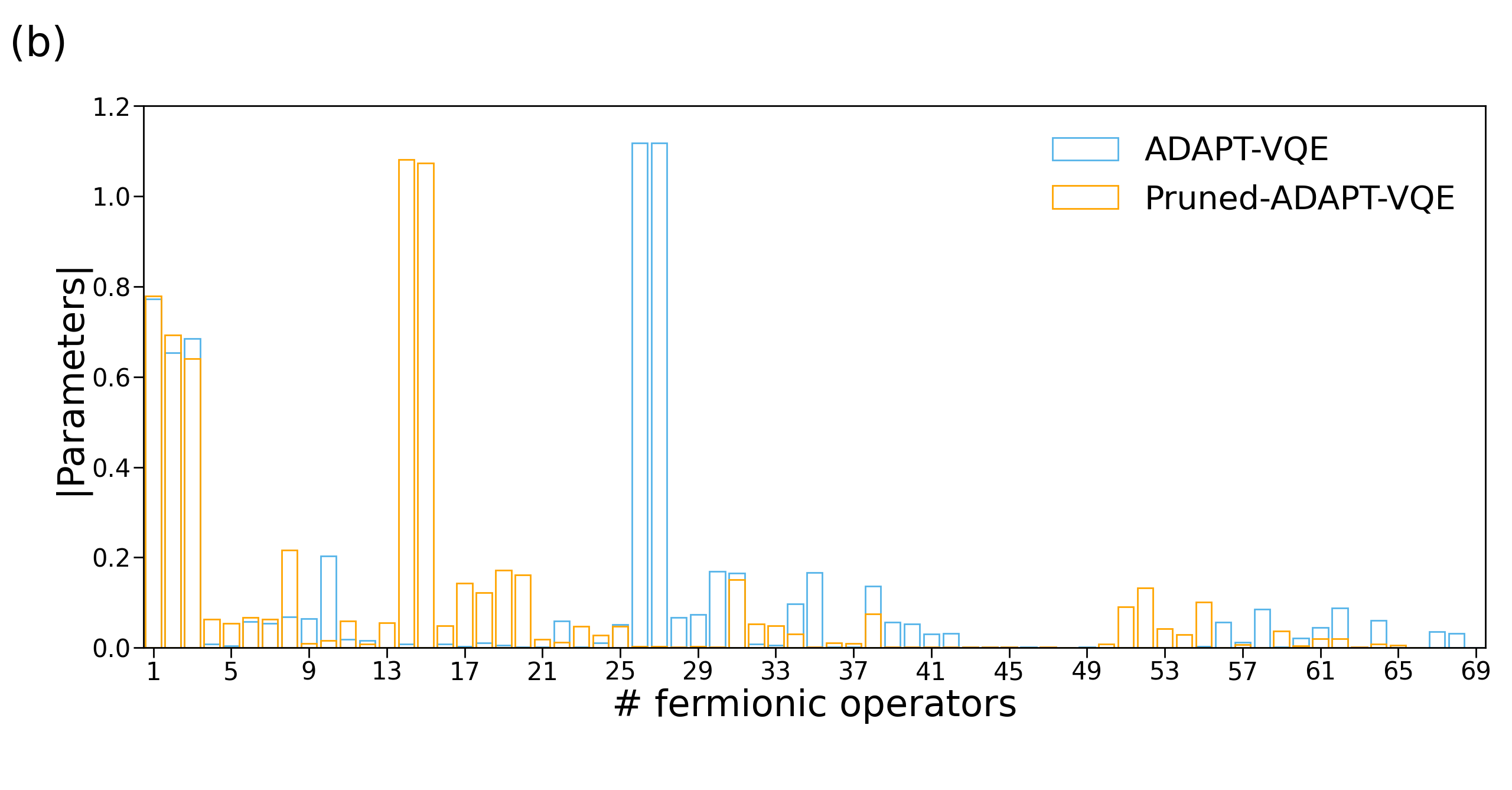}
    \caption{(a) Energy errors (in Hartree) with respect to FCI and (b) absolute parameter values for the $N=69$ ansatz obtained with ADAPT-VQE (blue) and Pruned-ADAPT-VQE (orange) for the linear \ce{H4} system with interatomic distance of 3.0~{\AA} and with the 3-21G basis set.}
    \label{fig:h4_linear_energies_ampl}
\end{figure}

Figure~\ref{fig:h4_linear_energies_ampl}a compares the performance of Pruned-ADAPT-VQE against standard ADAPT-VQE in the simulation of linear \ce{H4}. 
Without pruning, approximately 35 operators are needed to achieve chemical accuracy, whereas the refinement method reduces this number to 26. 
Initially, both approaches follow the same energy error profile. The pruning mechanism becomes active just after the flat energy region, around the 26th ADAPT iteration, ensuring that only non-cooperative, unnecessary operators are removed while maintaining accuracy.

The ansatz compaction achieved through operator removal becomes evident when comparing the amplitude distributions of Pruned-ADAPT-VQE and standard ADAPT-VQE (Figure~\ref{fig:h4_linear_energies_ampl}b). 
The elimination of irrelevant operators leads to a more compact ansatz, in which the most significant operators act earliest on the initial state.
Our pruning strategy appears to be conservative, as several operators with near-zero parameter values remain. 
Adjusting the parameters that define the decision factor (equation~\ref{eq:decision_factor}) and the dynamic threshold (equation~\ref{eq:threshold}) could enable further (or more restrained) ansatz simplification, depending on computational constraints and accuracy requirements.
For instance, increasing the 0.1 prefactor in the dynamic threshold enhances the likelihood of removing low-contributing operators, leading to more compact ans\"atze.
While this may require additional ADAPT-VQE iterations to reach convergence, the trade-off can be favorable in scenarios where reducing circuit depth is critical due to hardware constraints. 
In fact, from length 53 to the end of the simulation, all removed operators were re-added in the immediate next iteration, resulting in a significant computational overhead. Such behavior tends to appear towards the end of the simulation, when the ansatz is converging and the parameters associated to the added operators become vanishingly small. This behavior also becomes more pronounced in cases where the pruning strategy is more aggressive.

%-------------------
% Analysis of removed operators
A closer examination of the operators removed throughout the Pruned-ADAPT-VQE iterations reveals distinct cases within the category of operators with small-valued parameter.
The pruning strategy successfully eliminates wrongly selected excitations, such as operators $\hat{A}_{1g1g}^{4u4u}$ and $\hat{A'}_{1g1u}^{4g4u}$ (Figure~\ref{fig:h4_linear_badops}), as well as fading operators like $\hat{A}_{1g1u}^{2g4u}$ and $\hat{A}_{1g1u}^{4g2u}$ (Figure~\ref{fig:h4_linear_lostops}). 
Additionally, it removes redundant operators arising from reordering, such as operator $\hat{A}_{1g}^{3g}$, which was removed and reinserted several times before finding a more suitable position that balances energy accuracy and ansatz length (Figure~\ref{fig:h4_linear_reorderedops_del}).
Conversely, the algorithm effectively identifies and preserves cooperative operators, such as $\hat{A}_{1u1u}^{2u2u}$ (Figure~\ref{fig:h4_linear_energies_first27}), ensuring that essential contributions to the ansatz remain intact.

\begin{figure}[H]
    \centering
    \includegraphics[width=7cm]{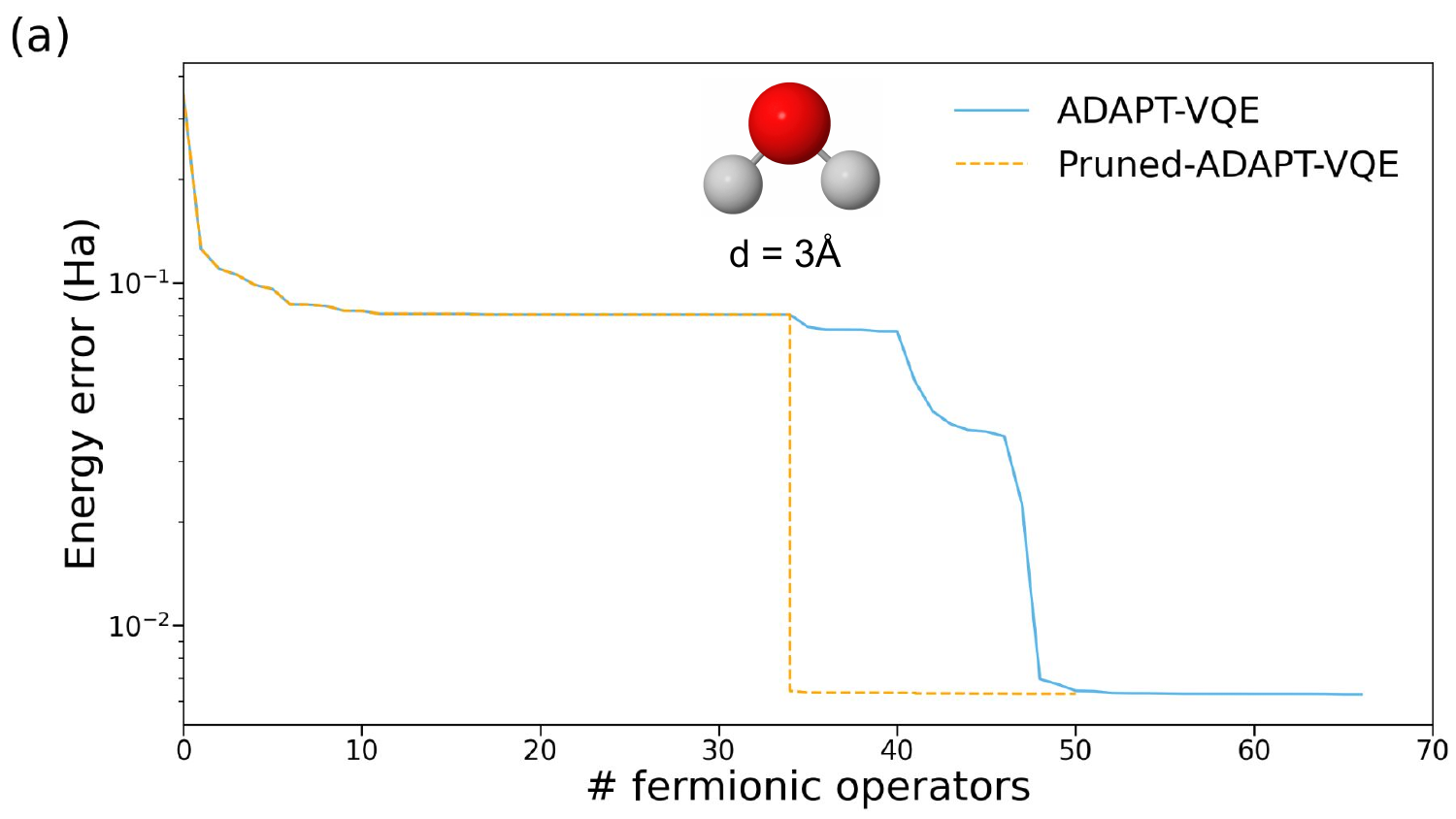}
    \includegraphics[width=7cm]{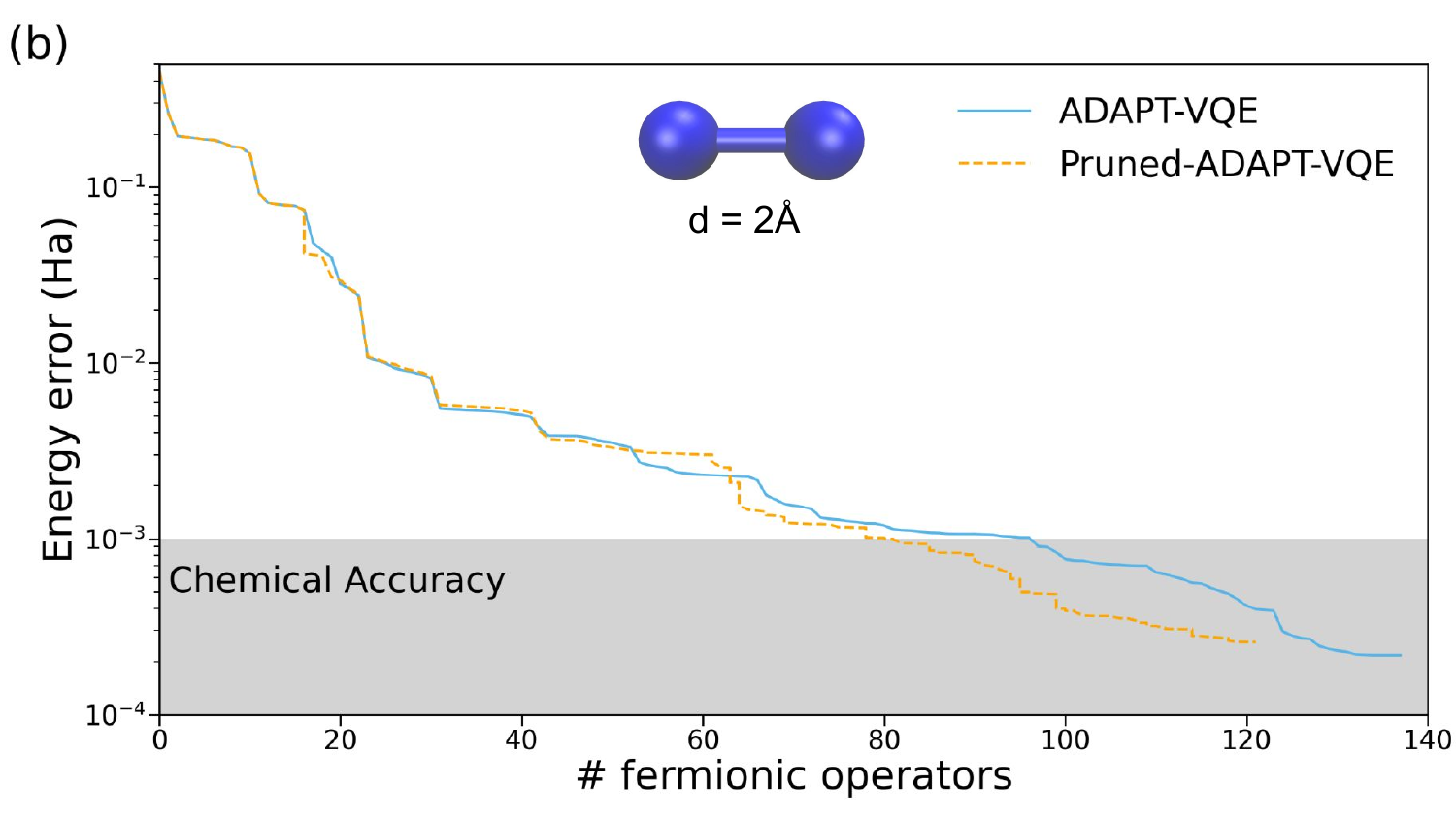}
    \caption{Energy errors (in Hartree) with respect to CAS(8,8) (\ce{H2O}) and CAS(10,8) (\ce{N2}) obtained with ADAPT-VQE (solid blue) and Pruned-ADAPT-VQE (dashed orange) ans\"atze for the (a) \ce{H2O}  with 3.0~{\AA} \ce{O-H} distance and computed in the 3-21G basis set with active space of 8 orbitals plus one frozen orbital, and (b) \ce{N2} with interatomic distance of 2.0~{\AA}, and computed with the 3-21G basis set with 8 active plus 2 frozen orbitals.}
    \label{fig:h2o_n2_energies_pruned}
\end{figure}

%-------------------
% Other system
Similar results, demonstrating a reduction in the number of fermionic operators without any loss of accuracy, have been observed in simulations of other molecular ground states.  
This includes the stretched \ce{H2O} and \ce{N2} molecules (Figure~\ref{fig:h2o_n2_energies_pruned}) as well as in other systems (see Supporting Information, Section S8).  
As seen in the case of linear \ce{H4}, ansatz reductions due to operator pruning tend to occur after flat energy regions in the ADAPT iterations. In these examples a computational overhead  reappears towards the final simulation steps. For instance, in the \ce{H2O} simulation, 5 operators are removed only to be reintroduced in the next iteration.

It is worth noting that, in some cases, Pruned-ADAPT-VQE does not offer a substantial improvement over standard ADAPT-VQE in terms of ansatz reduction. 
For example, when computing ground-state energies using the minimal STO-3G basis set, the size of the operator pool is significantly reduced (see Supporting Information, Section S10), and Pruned-ADAPT-VQE exhibits a convergence profile nearly identical to that of ADAPT-VQE.
These observations suggest that the benefits of pruning become more relevant when larger operator pools are involved, as typically encountered with larger basis sets in more realistic simulations. 
Additional comparisons of Pruned-ADAPT-VQE and ADAPT-VQE using various operator pools are presented in Supporting Information, Section S9.
Notably, even in cases where operator pruning does not yield substantial ansatz reductions, the removal of excitation operators using the proposed approach never compromises the accuracy achieved by ADAPT-VQE.

\section{Conclusions} \label{sec:conclusions}
The adaptive nature of ADAPT-VQE allows the ansatz to iteratively evolve, refining itself to better capture the problem landscape as the algorithm progresses.  
While the gradient-driven approach for ansatz growth offers a clear advantage over other strategies, it is not infallible.  
In some cases, the wavefunction may contain non-contributing operators, those with near-zero parameter values, which unnecessarily increase circuit depth without improving accuracy.  
In this work, we have identified three distinct mechanisms responsible for the presence of such redundant operators:  
(i) suboptimal operator selection due to limitations in the gradient criterion,  
(ii) operator reordering effects, where previously selected operators are reintroduced while the parameter values of their earlier instances diminish, leading to unnecessary redundancy in the ansatz, and  
(iii) fading operators, which become irrelevant as the ansatz evolves.  
Recognizing and addressing these issues enables the simplification and compression of ADAPT-VQE ans\"atze, leading to shorter quantum circuits without compromising accuracy.  

Our approach systematically evaluates all ansatz operators after each optimization step, using a function that accounts for both operator position and amplitude.  
This strategy removes operators with minimal contributions while preserving those that may have cooperative effects in future iterations, which results in shallower circuits better suited to current NISQ hardware.
By integrating this selection function with a dynamic threshold, adapted based on the parameter values of recently added operators, we effectively eliminate redundancies while ensuring algorithmic convergence.  
We validated this approach by computing the ground-state energies of several molecular systems.
Our results demonstrate that this method potentially improves the efficiency of ADAPT-VQE simulations, particularly in scenarios with long energy plateaus, at the expense of a greater number of iterations.
While in some instances the method does not lead to significant improvements, it consistently performs at least as well as standard ADAPT-VQE.  
Given its benefits in reducing ansatz complexity without introducing additional quantum computational overhead, we propose that this pruning strategy to be incorporated into ADAPT-VQE when circuit compactness is prioritized over iteration minimization.

%%%%%%%%%%%%%%%%%%%%%%%%%%%%%%%%%%%%%%%%%%%%%%%%%%%%%%%%%%%%%%%%%%%%%
\section*{Data Availability Statement}
The data that supports the findings of this study are available within the article and its supplementary material. 
The code developed for the numerical simulations is openly available on GitHub.\cite{VQEmulti}

\begin{acknowledgement}

We acknowledge financial support from MICIU/AEI/10.13039/501100011033 (project PID2022-136231NB-I00) and by FEDER, UE.
We also thank the support by the IKUR Strategy under the collaboration agreement between Ikerbasque Foundation and DIPC on behalf of the Department of Education of the Basque Government.
The authors are thankful for the technical and human support provided by the Donostia International Physics Center (DIPC) Computer Center. 

\end{acknowledgement}

\begin{suppinfo}
In the supportive information section further details about the theory and methodology are provided. These include the definition of the notation of the fermionic excitations in the pool (Section S1) and a description of the molecular orbitals of linear \ce{H4} (Section S2). Details of both $F_{1}(\theta_{i})$ (Section S3) and $F_{2}(x_{i})$ (Section S4) functions, part of the decision factor, are provided, as well as from the dynamical threshold (Section S5). 
Analysis of the Pruned-ADAPT-VQE performance tested on several molecules (Section S6). 
Analysis of the gradient-based selection criterion (Section S7).
Performance of Pruned-ADAPT-VQE with UCCSD (Section S8) and qubit-excitation-based (Section S9) pools.
Results with the STO-3G basis set (Section S10).
Dependence of parameter values (Section S11).
Gate counts (Section S12).
\end{suppinfo}

\bibliography{abbr, references} % Produces the bibliography via BibTeX.

\providecommand{\latin}[1]{#1}
\makeatletter
\providecommand{\doi}
  {\begingroup\let\do\@makeother\dospecials
  \catcode`\{=1 \catcode`\}=2 \doi@aux}
\providecommand{\doi@aux}[1]{\endgroup\texttt{#1}}
\makeatother
\providecommand*\mcitethebibliography{\thebibliography}
\csname @ifundefined\endcsname{endmcitethebibliography}  {\let\endmcitethebibliography\endthebibliography}{}
\begin{mcitethebibliography}{36}
\providecommand*\natexlab[1]{#1}
\providecommand*\mciteSetBstSublistMode[1]{}
\providecommand*\mciteSetBstMaxWidthForm[2]{}
\providecommand*\mciteBstWouldAddEndPuncttrue
  {\def\EndOfBibitem{\unskip.}}
\providecommand*\mciteBstWouldAddEndPunctfalse
  {\let\EndOfBibitem\relax}
\providecommand*\mciteSetBstMidEndSepPunct[3]{}
\providecommand*\mciteSetBstSublistLabelBeginEnd[3]{}
\providecommand*\EndOfBibitem{}
\mciteSetBstSublistMode{f}
\mciteSetBstMaxWidthForm{subitem}{(\alph{mcitesubitemcount})}
\mciteSetBstSublistLabelBeginEnd
  {\mcitemaxwidthsubitemform\space}
  {\relax}
  {\relax}

\bibitem[Bharti \latin{et~al.}(2022)Bharti, Cervera-Lierta, Kyaw, Haug, Alperin-Lea, Anand, Degroote, Heimonen, Kottmann, Menke, Mok, Sim, Kwek, and Aspuru-Guzik]{bharti2022}
Bharti,~K.; Cervera-Lierta,~A.; Kyaw,~T.~H.; Haug,~T.; Alperin-Lea,~S.; Anand,~A.; Degroote,~M.; Heimonen,~H.; Kottmann,~J.~S.; Menke,~T.; Mok,~W.-K.; Sim,~S.; Kwek,~L.-C.; Aspuru-Guzik,~A. Noisy intermediate-scale quantum algorithms. \emph{Rev. Mod. Phys.} \textbf{2022}, \emph{94}, 015004\relax
\mciteBstWouldAddEndPuncttrue
\mciteSetBstMidEndSepPunct{\mcitedefaultmidpunct}
{\mcitedefaultendpunct}{\mcitedefaultseppunct}\relax
\EndOfBibitem
\bibitem[McArdle \latin{et~al.}(2020)McArdle, Endo, Aspuru-Guzik, Benjamin, and Yuan]{McArdle2020}
McArdle,~S.; Endo,~S.; Aspuru-Guzik,~A.; Benjamin,~S.~C.; Yuan,~X. Quantum computational chemistry. \emph{Rev. Mod. Phys.} \textbf{2020}, \emph{92}, 015003\relax
\mciteBstWouldAddEndPuncttrue
\mciteSetBstMidEndSepPunct{\mcitedefaultmidpunct}
{\mcitedefaultendpunct}{\mcitedefaultseppunct}\relax
\EndOfBibitem
\bibitem[Aspuru-Guzik \latin{et~al.}(2005)Aspuru-Guzik, Dutoi, Love, and Head-Gordon]{Aspuru-Guzik2005}
Aspuru-Guzik,~A.; Dutoi,~A.~D.; Love,~P.~J.; Head-Gordon,~M. Simulated Quantum Computation of Molecular Energies. \emph{Science} \textbf{2005}, \emph{309}, 1704--1707\relax
\mciteBstWouldAddEndPuncttrue
\mciteSetBstMidEndSepPunct{\mcitedefaultmidpunct}
{\mcitedefaultendpunct}{\mcitedefaultseppunct}\relax
\EndOfBibitem
\bibitem[Peruzzo \latin{et~al.}(2014)Peruzzo, McClean, Shadbolt, Yung, Zhou, Love, Aspuru-Guzik, and O'Brien]{Peruzzo:2014}
Peruzzo,~A.; McClean,~J.; Shadbolt,~P.; Yung,~M.-H.; Zhou,~X.-Q.; Love,~P.~J.; Aspuru-Guzik,~A.; O'Brien,~J.~L. A variational eigenvalue solver on a photonic quantum processor. \emph{Nat. Commun.} \textbf{2014}, \emph{5}, 4213\relax
\mciteBstWouldAddEndPuncttrue
\mciteSetBstMidEndSepPunct{\mcitedefaultmidpunct}
{\mcitedefaultendpunct}{\mcitedefaultseppunct}\relax
\EndOfBibitem
\bibitem[McClean \latin{et~al.}(2016)McClean, Romero, Babbush, and Aspuru-Guzik]{McClean_2016}
McClean,~J.~R.; Romero,~J.; Babbush,~R.; Aspuru-Guzik,~A. The theory of variational hybrid quantum-classical algorithms. \emph{New J. Phys.} \textbf{2016}, \emph{18}, 023023\relax
\mciteBstWouldAddEndPuncttrue
\mciteSetBstMidEndSepPunct{\mcitedefaultmidpunct}
{\mcitedefaultendpunct}{\mcitedefaultseppunct}\relax
\EndOfBibitem
\bibitem[Arute \latin{et~al.}(2019)Arute, Arya, Babbush, Bacon, Bardin, Barends, Biswas, Boixo, Brandao, Buell, and et~al.]{Arute_Arya_Babbush_Bacon_Bardin_Barends_Biswas_Boixo_Brandao_Buell_etal._2019}
Arute,~F.; Arya,~K.; Babbush,~R.; Bacon,~D.; Bardin,~J.~C.; Barends,~R.; Biswas,~R.; Boixo,~S.; Brandao,~F.~G.; Buell,~D.~A.; et~al. Quantum supremacy using a programmable superconducting processor. \emph{Nature} \textbf{2019}, \emph{574}, 505–510\relax
\mciteBstWouldAddEndPuncttrue
\mciteSetBstMidEndSepPunct{\mcitedefaultmidpunct}
{\mcitedefaultendpunct}{\mcitedefaultseppunct}\relax
\EndOfBibitem
\bibitem[Preskill(2018)]{Preskill2018quantumcomputingi}
Preskill,~J. Quantum {C}omputing in the {NISQ} era and beyond. \emph{{Quantum}} \textbf{2018}, \emph{2}, 79\relax
\mciteBstWouldAddEndPuncttrue
\mciteSetBstMidEndSepPunct{\mcitedefaultmidpunct}
{\mcitedefaultendpunct}{\mcitedefaultseppunct}\relax
\EndOfBibitem
\bibitem[Anschuetz and Kiani(2022)Anschuetz, and Kiani]{Anschuetz2022}
Anschuetz,~E.~R.; Kiani,~B.~T. Quantum variational algorithms are swamped with traps. \emph{Nat. Commun.} \textbf{2022}, \emph{13}, 7760\relax
\mciteBstWouldAddEndPuncttrue
\mciteSetBstMidEndSepPunct{\mcitedefaultmidpunct}
{\mcitedefaultendpunct}{\mcitedefaultseppunct}\relax
\EndOfBibitem
\bibitem[Bittel and Kliesch(2021)Bittel, and Kliesch]{Bittel2021}
Bittel,~L.; Kliesch,~M. Training Variational Quantum Algorithms Is NP-Hard. \emph{Phys. Rev. Lett.} \textbf{2021}, \emph{127}, 120502\relax
\mciteBstWouldAddEndPuncttrue
\mciteSetBstMidEndSepPunct{\mcitedefaultmidpunct}
{\mcitedefaultendpunct}{\mcitedefaultseppunct}\relax
\EndOfBibitem
\bibitem[McClean \latin{et~al.}(2018)McClean, Boixo, Smelyanskiy, Babbush, and Neven]{McClean2018}
McClean,~J.~R.; Boixo,~S.; Smelyanskiy,~V.~N.; Babbush,~R.; Neven,~H. Barren plateaus in quantum neural network training landscapes. \emph{Nat. Commun.} \textbf{2018}, \emph{9}, 4812\relax
\mciteBstWouldAddEndPuncttrue
\mciteSetBstMidEndSepPunct{\mcitedefaultmidpunct}
{\mcitedefaultendpunct}{\mcitedefaultseppunct}\relax
\EndOfBibitem
\bibitem[Arrasmith \latin{et~al.}(2022)Arrasmith, Holmes, Cerezo, and Coles]{Arrasmith_2022}
Arrasmith,~A.; Holmes,~Z.; Cerezo,~M.; Coles,~P.~J. Equivalence of quantum barren plateaus to cost concentration and narrow gorges. \emph{Quantum Sci. Technol.} \textbf{2022}, \emph{7}, 045015\relax
\mciteBstWouldAddEndPuncttrue
\mciteSetBstMidEndSepPunct{\mcitedefaultmidpunct}
{\mcitedefaultendpunct}{\mcitedefaultseppunct}\relax
\EndOfBibitem
\bibitem[Romero \latin{et~al.}(2018)Romero, Babbush, McClean, Hempel, Love, and Aspuru-Guzik]{Romero2018-zr}
Romero,~J.; Babbush,~R.; McClean,~J.~R.; Hempel,~C.; Love,~P.~J.; Aspuru-Guzik,~A. Strategies for quantum computing molecular energies using the unitary coupled cluster ansatz. \emph{Quantum Sci. Technol.} \textbf{2018}, \emph{4}, 014008\relax
\mciteBstWouldAddEndPuncttrue
\mciteSetBstMidEndSepPunct{\mcitedefaultmidpunct}
{\mcitedefaultendpunct}{\mcitedefaultseppunct}\relax
\EndOfBibitem
\bibitem[Lee \latin{et~al.}(2019)Lee, Huggins, Head-Gordon, and Whaley]{Lee2019}
Lee,~J.; Huggins,~W.~J.; Head-Gordon,~M.; Whaley,~K.~B. {Generalized Unitary Coupled Cluster Wave functions for Quantum Computation}. \emph{J. Chem. Theory Comput.} \textbf{2019}, \emph{15}, 311--324\relax
\mciteBstWouldAddEndPuncttrue
\mciteSetBstMidEndSepPunct{\mcitedefaultmidpunct}
{\mcitedefaultendpunct}{\mcitedefaultseppunct}\relax
\EndOfBibitem
\bibitem[Yordanov \latin{et~al.}(2020)Yordanov, Arvidsson-Shukur, and Barnes]{Yordanov2020}
Yordanov,~Y.~S.; Arvidsson-Shukur,~D. R.~M.; Barnes,~C. H.~W. Efficient quantum circuits for quantum computational chemistry. \emph{Phys. Rev. A} \textbf{2020}, \emph{102}, 062612\relax
\mciteBstWouldAddEndPuncttrue
\mciteSetBstMidEndSepPunct{\mcitedefaultmidpunct}
{\mcitedefaultendpunct}{\mcitedefaultseppunct}\relax
\EndOfBibitem
\bibitem[Grimsley \latin{et~al.}(2019)Grimsley, Economou, Barnes, and Mayhall]{Grimsley2019}
Grimsley,~H.~R.; Economou,~S.~E.; Barnes,~E.; Mayhall,~N.~J. An adaptive variational algorithm for exact molecular simulations on a quantum computer. \emph{Nat. Commun.} \textbf{2019}, \emph{10}, 3007\relax
\mciteBstWouldAddEndPuncttrue
\mciteSetBstMidEndSepPunct{\mcitedefaultmidpunct}
{\mcitedefaultendpunct}{\mcitedefaultseppunct}\relax
\EndOfBibitem
\bibitem[Grimsley \latin{et~al.}(2023)Grimsley, Barron, Barnes, Economou, and Mayhall]{Grimsley2023}
Grimsley,~H.~R.; Barron,~G.~S.; Barnes,~E.; Economou,~S.~E.; Mayhall,~N.~J. {Adaptive, problem-tailored variational quantum eigensolver mitigates rough parameter landscapes and barren plateaus}. \emph{npj Quantum Inf.} \textbf{2023}, \emph{9}, 19\relax
\mciteBstWouldAddEndPuncttrue
\mciteSetBstMidEndSepPunct{\mcitedefaultmidpunct}
{\mcitedefaultendpunct}{\mcitedefaultseppunct}\relax
\EndOfBibitem
\bibitem[Puig \latin{et~al.}(2025)Puig, Drudis, Thanasilp, and Holmes]{Puig2025}
Puig,~R.; Drudis,~M.; Thanasilp,~S.; Holmes,~Z. Variational Quantum Simulation: A Case Study for Understanding Warm Starts. \emph{PRX Quantum} \textbf{2025}, \emph{6}, 010317\relax
\mciteBstWouldAddEndPuncttrue
\mciteSetBstMidEndSepPunct{\mcitedefaultmidpunct}
{\mcitedefaultendpunct}{\mcitedefaultseppunct}\relax
\EndOfBibitem
\bibitem[Carreras \latin{et~al.}(2025)Carreras, Casanova, and Orús]{carreras2025limitationsquantumhardwaremolecular}
Carreras,~A.; Casanova,~D.; Orús,~R. Limitations of Quantum Hardware for Molecular Energy Estimation Using VQE. 2025; \url{https://arxiv.org/abs/2506.03995}\relax
\mciteBstWouldAddEndPuncttrue
\mciteSetBstMidEndSepPunct{\mcitedefaultmidpunct}
{\mcitedefaultendpunct}{\mcitedefaultseppunct}\relax
\EndOfBibitem
\bibitem[Dalton \latin{et~al.}(2024)Dalton, Long, Yordanov, Smith, Barnes, Mertig, and Arvidsson-Shukur]{Dalton_Long_Yordanov_Smith_Barnes_Mertig_Arvidsson-Shukur_2024b}
Dalton,~K.; Long,~C.~K.; Yordanov,~Y.~S.; Smith,~C.~G.; Barnes,~C.~H.; Mertig,~N.; Arvidsson-Shukur,~D.~R. Quantifying the effect of gate errors on variational quantum eigensolvers for Quantum Chemistry. \emph{npj Quantum Inf.} \textbf{2024}, \emph{10}, 18\relax
\mciteBstWouldAddEndPuncttrue
\mciteSetBstMidEndSepPunct{\mcitedefaultmidpunct}
{\mcitedefaultendpunct}{\mcitedefaultseppunct}\relax
\EndOfBibitem
\bibitem[Kim \latin{et~al.}(2023)Kim, Eddins, Anand, Wei, van~den Berg, Rosenblatt, Nayfeh, Wu, Zaletel, Temme, and et~al.]{Kim_Eddins_Anand_Wei_van_den_Berg_Rosenblatt_Nayfeh_Wu_Zaletel_Temme_etal._2023}
Kim,~Y.; Eddins,~A.; Anand,~S.; Wei,~K.~X.; van~den Berg,~E.; Rosenblatt,~S.; Nayfeh,~H.; Wu,~Y.; Zaletel,~M.; Temme,~K.; et~al. Evidence for the utility of quantum computing before Fault Tolerance. \emph{Nature} \textbf{2023}, \emph{618}, 500–505\relax
\mciteBstWouldAddEndPuncttrue
\mciteSetBstMidEndSepPunct{\mcitedefaultmidpunct}
{\mcitedefaultendpunct}{\mcitedefaultseppunct}\relax
\EndOfBibitem
\bibitem[Fedorov \latin{et~al.}(2022)Fedorov, Peng, Govind, and Alexeev]{Fedorov_Peng_Govind_Alexeev_2022a}
Fedorov,~D.~A.; Peng,~B.; Govind,~N.; Alexeev,~Y. VQE method: A short survey and recent developments. \emph{Mater. Theory} \textbf{2022}, \emph{6}, 2\relax
\mciteBstWouldAddEndPuncttrue
\mciteSetBstMidEndSepPunct{\mcitedefaultmidpunct}
{\mcitedefaultendpunct}{\mcitedefaultseppunct}\relax
\EndOfBibitem
\bibitem[Yordanov \latin{et~al.}(2021)Yordanov, Armaos, Barnes, and Arvidsson-Shukur]{Yordanov2021}
Yordanov,~Y.~S.; Armaos,~V.; Barnes,~C. H.~W.; Arvidsson-Shukur,~D. R.~M. {Qubit-excitation-based adaptive variational quantum eigensolver}. \emph{Commun. Phys.} \textbf{2021}, \emph{4}, 228\relax
\mciteBstWouldAddEndPuncttrue
\mciteSetBstMidEndSepPunct{\mcitedefaultmidpunct}
{\mcitedefaultendpunct}{\mcitedefaultseppunct}\relax
\EndOfBibitem
\bibitem[Anastasiou \latin{et~al.}(2024)Anastasiou, Chen, Mayhall, Barnes, and Economou]{Anastasiou:tetris-adapt:2024}
Anastasiou,~P.~G.; Chen,~Y.; Mayhall,~N.~J.; Barnes,~E.; Economou,~S.~E. TETRIS-ADAPT-VQE: An adaptive algorithm that yields shallower, denser circuit Ans\"atze. \emph{Phys. Rev. Res.} \textbf{2024}, \emph{6}, 013254\relax
\mciteBstWouldAddEndPuncttrue
\mciteSetBstMidEndSepPunct{\mcitedefaultmidpunct}
{\mcitedefaultendpunct}{\mcitedefaultseppunct}\relax
\EndOfBibitem
\bibitem[Tang \latin{et~al.}(2021)Tang, Shkolnikov, Barron, Grimsley, Mayhall, Barnes, and Economou]{Tang:adapt:2021}
Tang,~H.~L.; Shkolnikov,~V.; Barron,~G.~S.; Grimsley,~H.~R.; Mayhall,~N.~J.; Barnes,~E.; Economou,~S.~E. {Qubit-ADAPT-VQE: An Adaptive Algorithm for Constructing Hardware-Efficient Ans{\"{a}}tze on a Quantum Processor}. \emph{PRX Quantum} \textbf{2021}, \emph{2}, 020310\relax
\mciteBstWouldAddEndPuncttrue
\mciteSetBstMidEndSepPunct{\mcitedefaultmidpunct}
{\mcitedefaultendpunct}{\mcitedefaultseppunct}\relax
\EndOfBibitem
\bibitem[Vaquero-Sabater \latin{et~al.}(2024)Vaquero-Sabater, Carreras, Orús, Mayhall, and Casanova]{Vaquero-Sabater_Carreras_Orús_Mayhall_Casanova_2024}
Vaquero-Sabater,~N.; Carreras,~A.; Orús,~R.; Mayhall,~N.~J.; Casanova,~D. Physically motivated improvements of variational quantum eigensolvers. \emph{J. Chem. Theory Comput.} \textbf{2024}, \emph{20}, 5133–5144\relax
\mciteBstWouldAddEndPuncttrue
\mciteSetBstMidEndSepPunct{\mcitedefaultmidpunct}
{\mcitedefaultendpunct}{\mcitedefaultseppunct}\relax
\EndOfBibitem
\bibitem[Alves(2022)]{Ramoa_thesis}
Alves,~M. F. R. d.~C. Ans\"atze for Noisy Variational Quantum Eigensolvers. Ph.D.\ thesis, University do Minho, 2022\relax
\mciteBstWouldAddEndPuncttrue
\mciteSetBstMidEndSepPunct{\mcitedefaultmidpunct}
{\mcitedefaultendpunct}{\mcitedefaultseppunct}\relax
\EndOfBibitem
\bibitem[Jordan and Wigner(1928)Jordan, and Wigner]{JW:mapping:1928}
Jordan,~P.; Wigner,~E. {\"U}ber das Paulische {\"A}quivalenzverbot. \emph{Z. Phys.} \textbf{1928}, \emph{47}, 631--651\relax
\mciteBstWouldAddEndPuncttrue
\mciteSetBstMidEndSepPunct{\mcitedefaultmidpunct}
{\mcitedefaultendpunct}{\mcitedefaultseppunct}\relax
\EndOfBibitem
\bibitem[Nocedal and Wright(1994)Nocedal, and Wright]{BFGS}
Nocedal,~J.; Wright,~S.~J. \emph{Numerical Optimization}; Springer New York, NY, 1994; pp XXII, 664\relax
\mciteBstWouldAddEndPuncttrue
\mciteSetBstMidEndSepPunct{\mcitedefaultmidpunct}
{\mcitedefaultendpunct}{\mcitedefaultseppunct}\relax
\EndOfBibitem
\bibitem[Carreras and Vaquero-Sabater(2023)Carreras, and Vaquero-Sabater]{VQEmulti}
Carreras,~A.; Vaquero-Sabater,~N. VQEmulti. 2023; \url{https://github.com/abelcarreras/VQEmulti}\relax
\mciteBstWouldAddEndPuncttrue
\mciteSetBstMidEndSepPunct{\mcitedefaultmidpunct}
{\mcitedefaultendpunct}{\mcitedefaultseppunct}\relax
\EndOfBibitem
\bibitem[Harris \latin{et~al.}(2020)Harris, Millman, van~der Walt, Gommers, Virtanen, Cournapeau, Wieser, Taylor, Berg, Smith, Kern, Picus, Hoyer, van Kerkwijk, Brett, Haldane, del R{\'{i}}o, Wiebe, Peterson, G{\'{e}}rard-Marchant, Sheppard, Reddy, Weckesser, Abbasi, Gohlke, and Oliphant]{numpy}
Harris,~C.~R. \latin{et~al.}  Array programming with {NumPy}. \emph{Nature} \textbf{2020}, \emph{585}, 357--362\relax
\mciteBstWouldAddEndPuncttrue
\mciteSetBstMidEndSepPunct{\mcitedefaultmidpunct}
{\mcitedefaultendpunct}{\mcitedefaultseppunct}\relax
\EndOfBibitem
\bibitem[Virtanen \latin{et~al.}(2020)Virtanen, Gommers, Oliphant, Haberland, Reddy, Cournapeau, Burovski, Peterson, Weckesser, Bright, {van der Walt}, Brett, Wilson, Millman, Mayorov, Nelson, Jones, Kern, Larson, Carey, Polat, Feng, Moore, {VanderPlas}, Laxalde, Perktold, Cimrman, Henriksen, Quintero, Harris, Archibald, Ribeiro, Pedregosa, {van Mulbregt}, and {SciPy 1.0 Contributors}]{2020SciPy}
Virtanen,~P. \latin{et~al.}  {SciPy} 1.0: Fundamental Algorithms for Scientific Computing in Python. \emph{Nat. Methods} \textbf{2020}, \emph{17}, 261--272\relax
\mciteBstWouldAddEndPuncttrue
\mciteSetBstMidEndSepPunct{\mcitedefaultmidpunct}
{\mcitedefaultendpunct}{\mcitedefaultseppunct}\relax
\EndOfBibitem
\bibitem[McClean \latin{et~al.}(2019)McClean, Sung, Kivlichan, Cao, Dai, Fried, Gidney, Gimby, Gokhale, Häner, Hardikar, Havlíček, Higgott, Huang, Izaac, Jiang, Liu, McArdle, Neeley, O'Brien, O'Gorman, Ozfidan, Radin, Romero, Rubin, Sawaya, Setia, Sim, Steiger, Steudtner, Sun, Sun, Wang, Zhang, and Babbush]{mcclean2019openfermion}
McClean,~J.~R. \latin{et~al.}  OpenFermion: The Electronic Structure Package for Quantum Computers. 2019\relax
\mciteBstWouldAddEndPuncttrue
\mciteSetBstMidEndSepPunct{\mcitedefaultmidpunct}
{\mcitedefaultendpunct}{\mcitedefaultseppunct}\relax
\EndOfBibitem
\bibitem[Grimsley \latin{et~al.}(2023)Grimsley, Barron, Barnes, Economou, and Mayhall]{Grimsley:adapt:2023}
Grimsley,~H.~R.; Barron,~G.~S.; Barnes,~E.; Economou,~S.~E.; Mayhall,~N.~J. {Adaptive, problem-tailored variational quantum eigensolver mitigates rough parameter landscapes and barren plateaus}. \emph{npj Quantum Inf.} \textbf{2023}, \emph{9}, 19\relax
\mciteBstWouldAddEndPuncttrue
\mciteSetBstMidEndSepPunct{\mcitedefaultmidpunct}
{\mcitedefaultendpunct}{\mcitedefaultseppunct}\relax
\EndOfBibitem
\bibitem[Grimsley \latin{et~al.}(2020)Grimsley, Claudino, Economou, Barnes, and Mayhall]{Grimsley2020}
Grimsley,~H.~R.; Claudino,~D.; Economou,~S.~E.; Barnes,~E.; Mayhall,~N.~J. Is the Trotterized UCCSD Ansatz Chemically Well-Defined? \emph{J. Chem. Theory Comput.} \textbf{2020}, \emph{16}, 1--6\relax
\mciteBstWouldAddEndPuncttrue
\mciteSetBstMidEndSepPunct{\mcitedefaultmidpunct}
{\mcitedefaultendpunct}{\mcitedefaultseppunct}\relax
\EndOfBibitem
\bibitem[Pople(1999)]{pople1999nobel}
Pople,~J.~A. Nobel lecture: Quantum chemical models. \emph{Rev. Mod. Phys.} \textbf{1999}, \emph{71}, 1267\relax
\mciteBstWouldAddEndPuncttrue
\mciteSetBstMidEndSepPunct{\mcitedefaultmidpunct}
{\mcitedefaultendpunct}{\mcitedefaultseppunct}\relax
\EndOfBibitem
\end{mcitethebibliography}


\providecommand{\latin}[1]{#1}
\makeatletter
\providecommand{\doi}
  {\begingroup\let\do\@makeother\dospecials
  \catcode`\{=1 \catcode`\}=2 \doi@aux}
\providecommand{\doi@aux}[1]{\endgroup\texttt{#1}}
\makeatother
\providecommand*\mcitethebibliography{\thebibliography}
\csname @ifundefined\endcsname{endmcitethebibliography}  {\let\endmcitethebibliography\endthebibliography}{}
\begin{mcitethebibliography}{2}
\providecommand*\natexlab[1]{#1}
\providecommand*\mciteSetBstSublistMode[1]{}
\providecommand*\mciteSetBstMaxWidthForm[2]{}
\providecommand*\mciteBstWouldAddEndPuncttrue
  {\def\EndOfBibitem{\unskip.}}
\providecommand*\mciteBstWouldAddEndPunctfalse
  {\let\EndOfBibitem\relax}
\providecommand*\mciteSetBstMidEndSepPunct[3]{}
\providecommand*\mciteSetBstSublistLabelBeginEnd[3]{}
\providecommand*\EndOfBibitem{}
\mciteSetBstSublistMode{f}
\mciteSetBstMaxWidthForm{subitem}{(\alph{mcitesubitemcount})}
\mciteSetBstSublistLabelBeginEnd
  {\mcitemaxwidthsubitemform\space}
  {\relax}
  {\relax}

\bibitem[Yordanov \latin{et~al.}(2021)Yordanov, Armaos, Barnes, and Arvidsson-Shukur]{Yordanov2021}
Yordanov,~Y.~S.; Armaos,~V.; Barnes,~C. H.~W.; Arvidsson-Shukur,~D. R.~M. {Qubit-excitation-based adaptive variational quantum eigensolver}. \emph{Commun. Phys.} \textbf{2021}, \emph{4}, 228\relax
\mciteBstWouldAddEndPuncttrue
\mciteSetBstMidEndSepPunct{\mcitedefaultmidpunct}
{\mcitedefaultendpunct}{\mcitedefaultseppunct}\relax
\EndOfBibitem
\end{mcitethebibliography}

\end{document}

\clearpage
\tableofcontents
%###########################################################################

\clearpage
\section{Excitation operators in the pool}\label{sec:operators_SI}
Each individual excitation operator in the pool consists of a UCC operator restricted to occupied-to-virtual spin-singlet adapted single and double excitations.
Occupied (virtual) orbitals in the HF determinant are indicated with $i,j$ ($a,b$) indices.
Single and double excitations take the form: 
\begin{eqnarray}
% A_i^a
    \hat{A}_i^a = \frac{1}{2} \left[\hat{\tau}_i^a +\hat{\tau}_{\bar{i}}^{\bar{a}} - \text{h.c.} \right] \\
% A_ii^aa
    \hat{A}_{ii}^{aa} = \frac{1}{\sqrt{2}} \left[\hat{\tau}_{i\bar{i}}^{a\bar{a}} - \text{h.c.} \right] \\
% A_ii^ab
    \hat{A}_{ii}^{ab} = \frac{1}{2} \left[\hat{\tau}_{i\bar{i}}^{a\bar{b}} +\hat{\tau}_{\bar{i}i}^{\bar{a}b} - \text{h.c.} \right]\\
% A_ij^aa
    \hat{A}_{ij}^{aa} = \frac{1}{2} \left[\hat{\tau}_{i\bar{j}}^{a\bar{a}} +\hat{\tau}_{\bar{i}j}^{\bar{a}a} - \text{h.c.} \right]\\
% A_ij^ab (1)
    \hat{A}_{ij}^{ab} = \frac{1}{2\sqrt{6}} \left[2(\hat{\tau}_{ij}^{ab}+\hat{\tau}_{\bar{i}\bar{j}}^{\bar{a}\bar{b}}) + \hat{\tau}_{i\bar{j}}^{a\bar{b}} +\hat{\tau}_{\bar{i}j}^{\bar{a}b} +\hat{\tau}_{\bar{i}j}^{a\bar{b}} +\hat{\tau}_{i\bar{j}}^{\bar{a}b} - \text{h.c.} \right]\\
% A_ij^ab (2)
    \hat{A'}_{ij}^{ab} = \frac{1}{2\sqrt{2}} \left[\hat{\tau}_{i\bar{j}}^{a\bar{b}} +\hat{\tau}_{\bar{i}j}^{\bar{a}b} -\hat{\tau}_{\bar{i}j}^{a\bar{b}} -\hat{\tau}_{i\bar{j}}^{\bar{a}b} - \text{h.c.} \right]\\
\end{eqnarray}
where h.c. indicates hermitian conjugate, the bar over orbital indices refer to $\beta$-spin orbitals, and $\hat{\tau}_i^a$ and $\hat{\tau}_{ij}^{ab}$ are expressed in terms of creation and annihilation operators as: 
\begin{equation}\label{eq:label_ops_single} 
  \hat{\tau}^{a}_{i} = \hat{a}^\dagger_a \hat{a}_i 
\end{equation} 

\begin{equation}\label{eq:label_ops_double} 
  \hat{\tau}^{ab}_{ij} = \hat{a}^\dagger_a \hat{a}^\dagger_b \hat{a}_j \hat{a}_i 
\end{equation}

\section{Molecular orbitals of \ce{H4} and \ce{H2O}}

\begin{figure}[H]
    \centering
    \includegraphics[width=5cm]{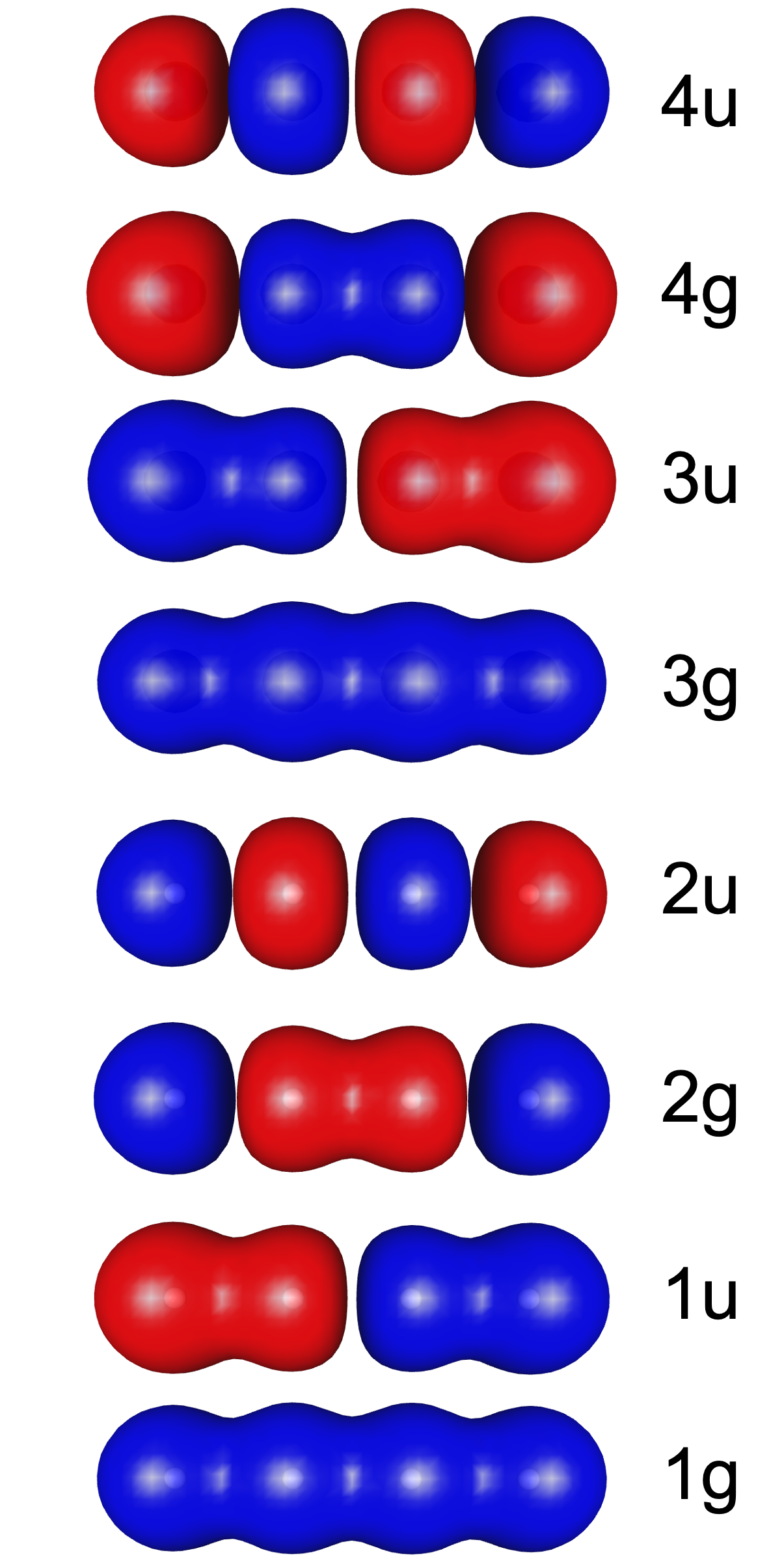}
    \caption{Molecular orbitals computed at the HF/3-21G level of linear \ce{H4} with 3.0~\AA~intermolecular separation}
    \label{fig:h4_mos}
\end{figure}

\begin{figure}[H]
    \centering
    \includegraphics[width=10cm]{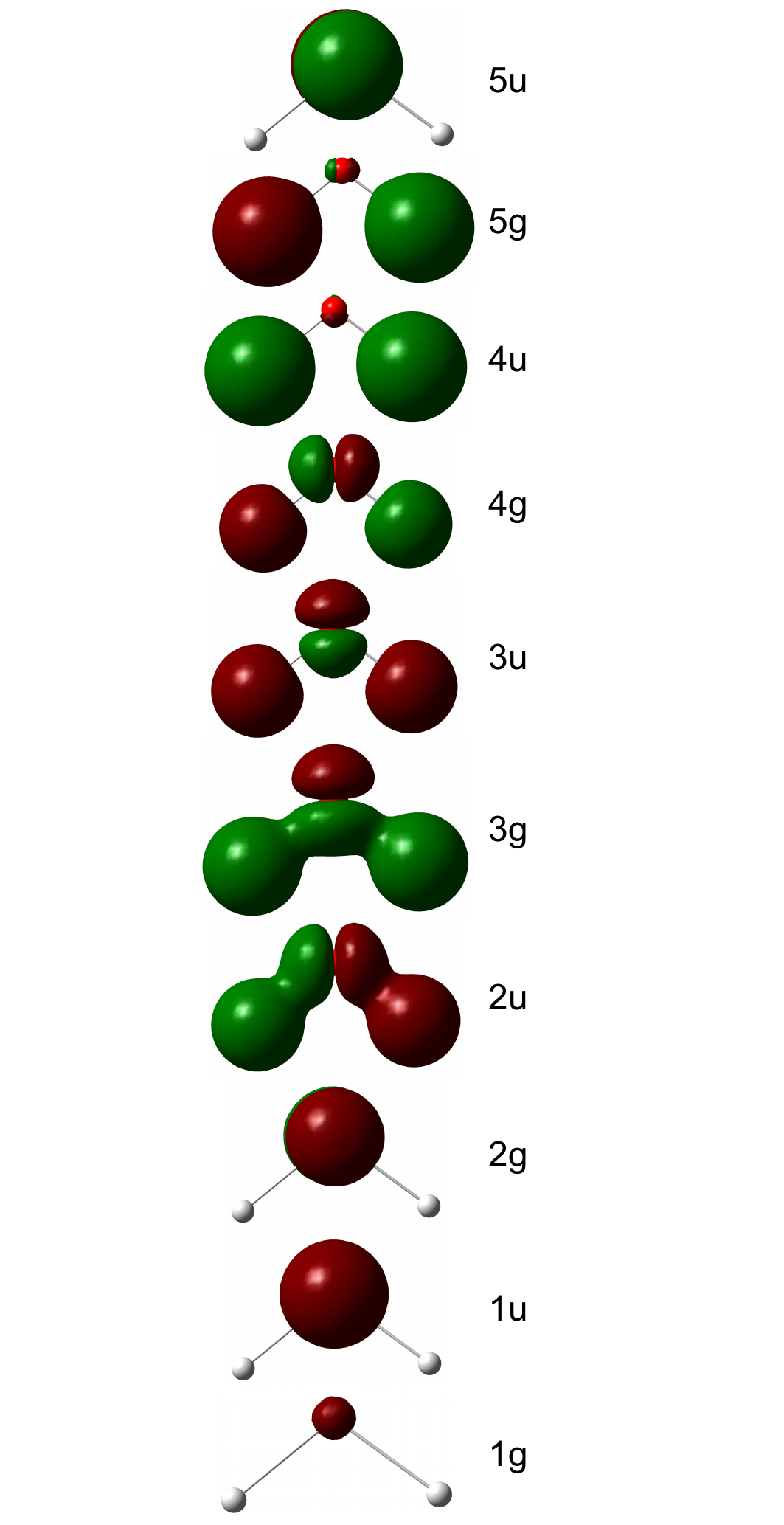}
    \caption{First 10 molecular orbitals computed at the HF/3-21G level of \ce{H2O} with 3.0~\AA~intermolecular \ce{H-O} separation}
    \label{fig:h2o_mos}
\end{figure}

In the orbital labels, the subscripts $g$ (gerade) and $u$ (ungerade) denote the symmetry of the molecular orbitals with respect to inversion through the center of mass. A gerade orbital is symmetric under inversion (i.e., its wavefunction remains unchanged), while an ungerade orbital is antisymmetric (i.e., the wavefunction changes its sign).

\section{Alternative functions for $F_1(\theta_i)$} \label{sec:F1_SI}
The objective of this term is to account for the amplitude of each operator, prioritizing those with the smallest contributions. In this section, several values of $n$ ($n=0,1,3,4$) in equation \ref{eq:F1_function_SI} have been tested.
\begin{equation}\label{eq:F1_function_SI} 
    F_1(\theta_i) = \dfrac{1}{|\theta_i^{-n}|} 
\end{equation}

\begin{figure}[H]
    \centering
    \includegraphics[width=10cm]{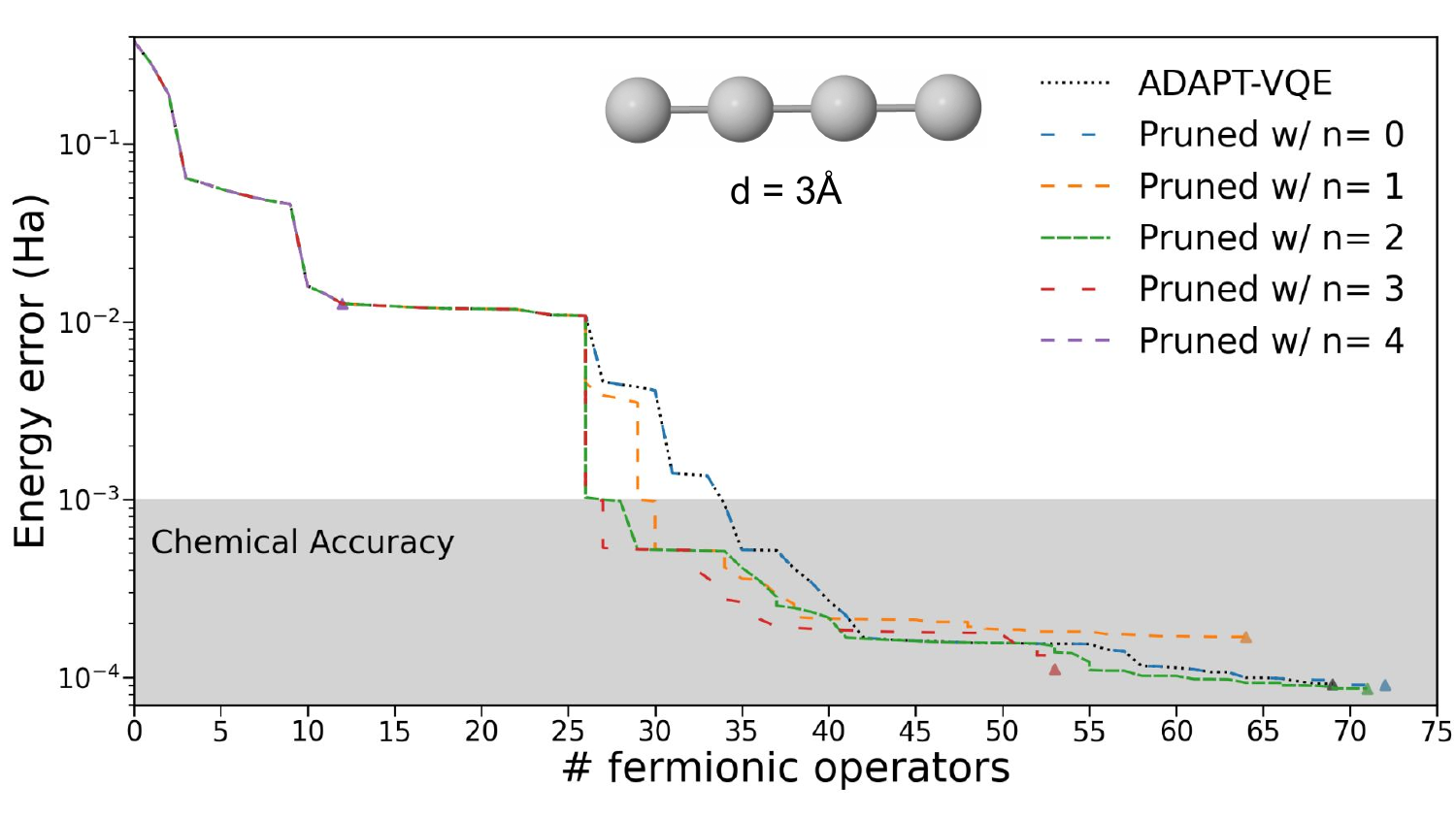}
    \caption{Energy error (in Hartree) with respect to FCI of ADAPT-VQE (black dotted line) and Pruned-ADAPT-VQE with different amplitude weights for the linear \ce{H4} with interatomic distance of 3.0~{\AA}, and computed with the 3-21G basis set and an active space of 8 orbitals. The triangle indicates where the simulation ended.}
    \label{si:fig:linh4_weight_test}
\end{figure}

\begin{figure}[H]
    \centering
    \includegraphics[width=10cm]{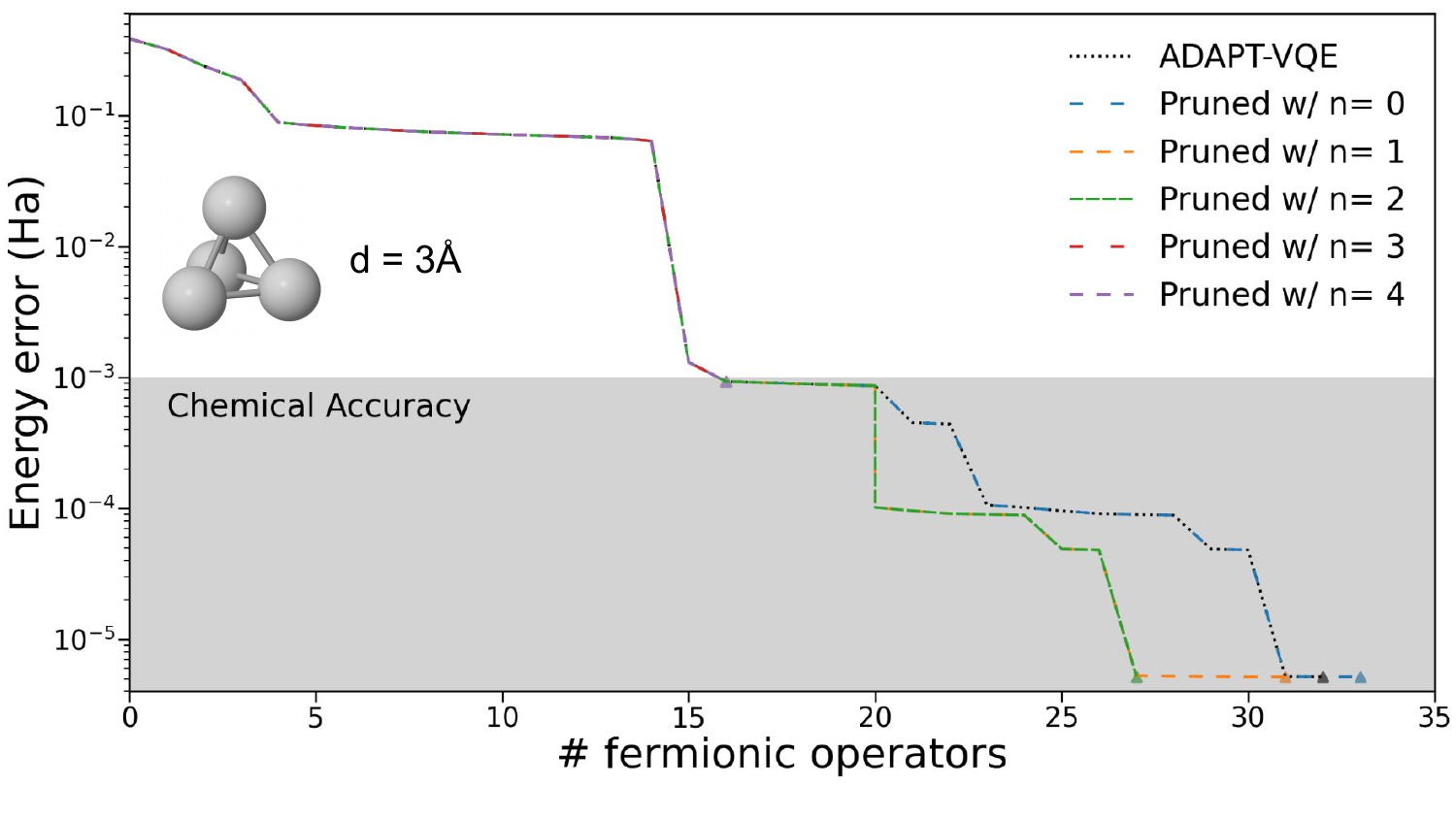}
    \caption{Energy error (in Hartree) with respect to FCI of ADAPT-VQE (black dotted line) and Pruned-ADAPT-VQE with different amplitude weights for the tetrahedral \ce{H4} with interatomic distance of 3.0~{\AA}, and computed with the 3-21G basis set and an active space of 8 orbitals. The triangle indicates where the simulation ended.}
    \label{si:fig:tetrah4_weight_test}
\end{figure}

\begin{figure}[H]
    \centering
    \includegraphics[width=10cm]{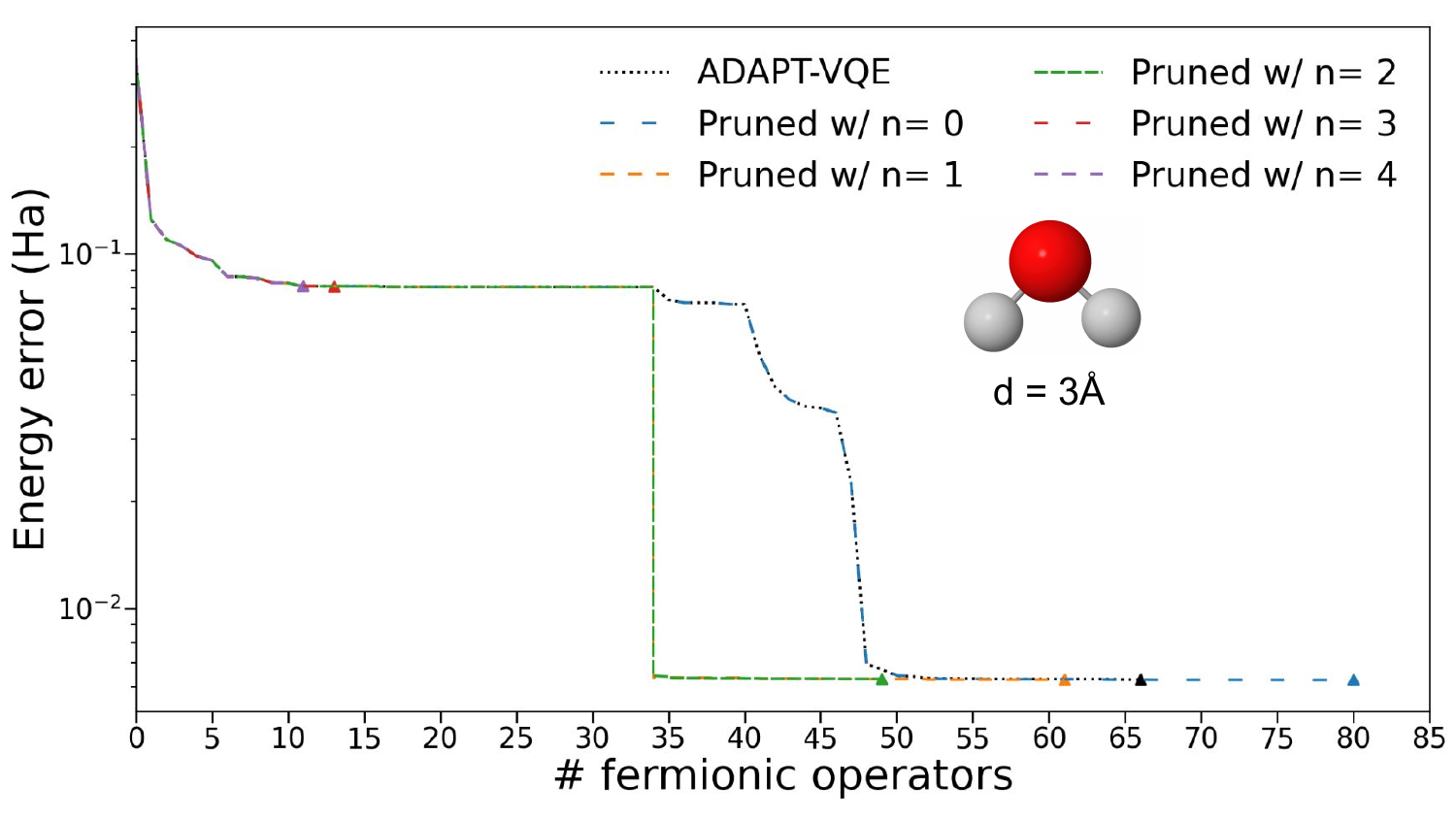}
    \caption{Energy error (in Hartree) with respect to CAS(8,8) of ADAPT-VQE (black dotted line) and Pruned-ADAPT-VQE with different amplitude weights for the \ce{H2O} molecule with 3.0~{\AA} \ce{O-H} distance, and computed with the 3-21G basis set with active space of 8 orbitals plus one frozen orbital. The triangle indicates where the simulation ended.}
    \label{si:fig:h2o_weight_test}
\end{figure}

Each value of $n$ represents a different weight assigned to the amplitude within the total function.\\
$\theta_i^{-0}$ means that no weight is assigned to the amplitude, giving no preference to operators with small coefficients. This function is expected to yield the same result as ADAPT-VQE. On the other hand, $\theta_i^{-4}$ assigns a very high weight to the amplitude. While the position is still considered, the influence of the amplitude becomes dominant, leading to the selection of operators with the smallest coefficients over those in earlier positions. Since operator amplitudes tend to decrease as the simulation progresses, this would result in the last operator being repeatedly selected, preventing the simulation from converging (see $n=4$ in Figure~\ref{si:fig:linh4_weight_test} and $n=3,4$ in Figures~\ref{si:fig:tetrah4_weight_test} and \ref{si:fig:h2o_weight_test}).

\section{Dependence on $\alpha$ value in $F_2(x_i)$} \label{sec:alpha_SI}

The objective of this term is to take into account the position of each operator. 
The weight of the position is modulated by the $\alpha$ value in equation~\ref{eq:si:alpha_function}.
\begin{equation}\label{eq:si:alpha_function}
    F_2(x_i) = e^{-\alpha x_i}
\end{equation}

\begin{figure}[H]
    \centering
    \includegraphics[width=10cm]{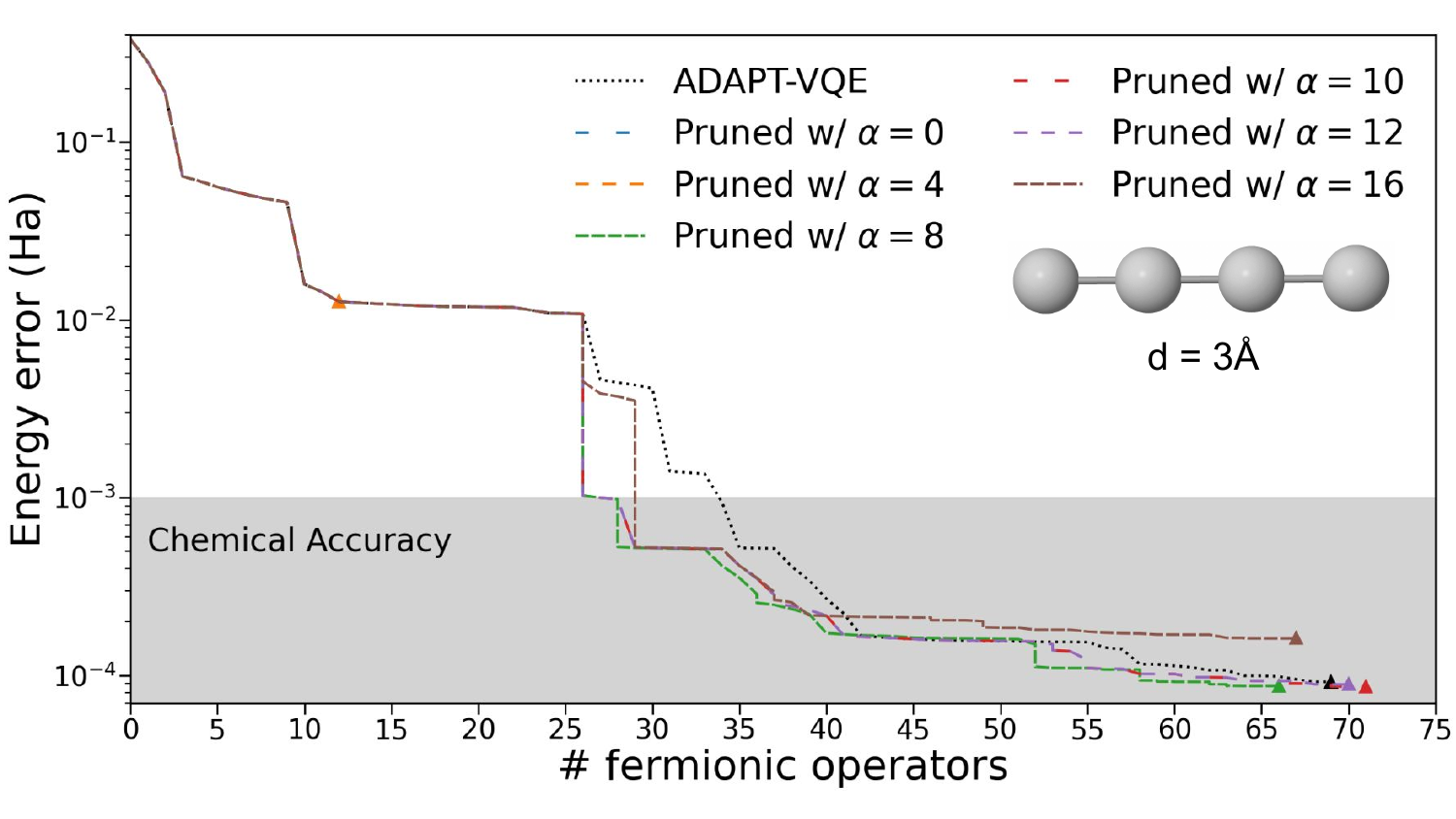}
    \caption{Energy error (in Hartree) with respect to FCI of ADAPT-VQE (black dotted line) and Pruned-ADAPT-VQE with different $\alpha$ values for the linear \ce{H4} with interatomic distance of 3.0~{\AA}, and computed with the 3-21G basis set and an active space of 8 orbitals. The triangle indicates where the simulation ended.}
    \label{si:fig:linh4_alpha_test}
\end{figure}

\begin{figure}[H]
    \centering
    \includegraphics[width=10cm]{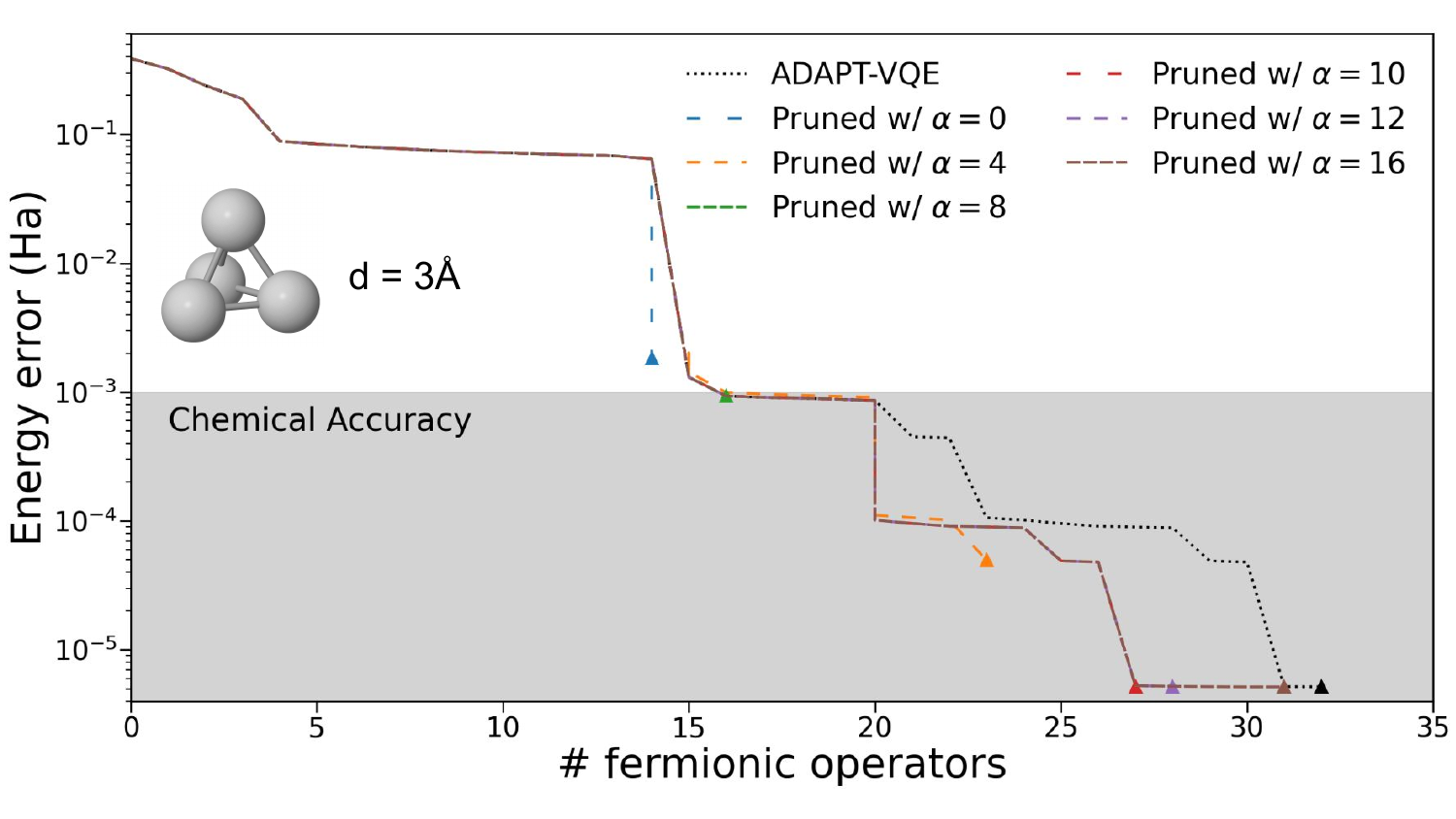}
    \caption{Energy error (in Hartree) with respect to FCI of ADAPT-VQE (black dotted line) and Pruned-ADAPT-VQE with different $\alpha$ values for the tetrahedral \ce{H4} with interatomic distance of 3.0~{\AA}, and computed with the 3-21G basis set and an active space of 8 orbitals. The triangle indicates where the simulation ended.}
    \label{si:fig:tetrah4_alpha_test}
\end{figure}

\begin{figure}[H]
    \centering
    \includegraphics[width=10cm]{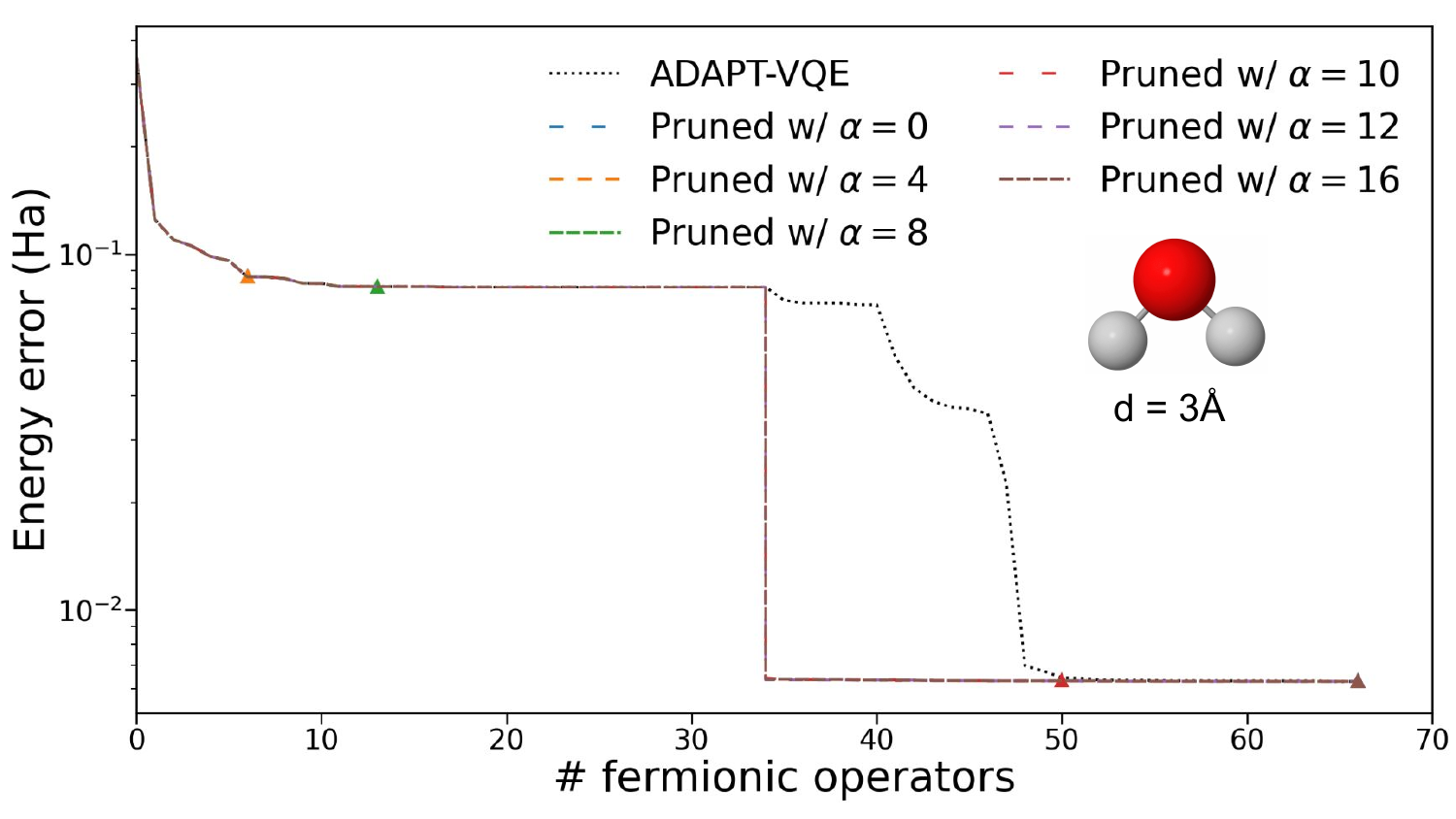}
    \caption{Energy error (in Hartree) with respect to CAS(8,8) of ADAPT-VQE (black dotted line) and Pruned-ADAPT-VQE with $\alpha$ values weights for the \ce{H2O} molecule with 3.0~{\AA} \ce{O-H} distance, and computed with the 3-21G basis set with active space of 8 orbitals plus one frozen orbital. The triangle indicates where the simulation ended.}
    \label{si:fig:h2o_alpha_test}
\end{figure}

Setting $\alpha = 0$ gives no importance to position, meaning the total function would depend only on the coefficient. This would cause the latter operators to always be removed, as the algorithm tends to add operators with smaller coefficients over time. A high value of $\alpha$ may overly prioritize the operators at the beginning of the ansatz, potentially preventing some intermediate operators from being removed. 
This effect can be visualized in Figures~{\ref{si:fig:linh4_alpha_test}}, {\ref{si:fig:tetrah4_alpha_test}}, and~{\ref{si:fig:h2o_alpha_test}}, where the smallest $\alpha$ values make the simulation end prematurely while the biggest ones lead to longer ans\"atze.

\section{Threshold analysis} \label{sec:threshold_SI}
Once an operator is selected using the decision factor, we must determine whether to remove it based on a threshold. As the simulation progresses, amplitudes tend to decrease, so operators with smaller coefficients must be allowed into the ansatz, this is taken into account with a dynamic threshold. 
Equation~\ref{eq:threshold_SI} shows that the threshold value depends on the coefficients of the most $N_{L}$ recently added operators in the ansatz. 
\begin{equation}
        \tau = \frac{0.1}{N_L}\sum_{i=0}^{N_L-1}|\theta_{N-i}|
    \label{eq:threshold_SI}
\end{equation}
where $N$ is to total number of operators in the ansatz.
This adaptive approach facilitates convergence, as the decreasing coefficients lead to a progressively lower threshold, resulting in fewer operators being removed over time. Additionally, it enables the elimination of operators in flat regions, where subsequent operators tend to have larger amplitudes, causing the threshold to rise. Several tests have been conducted, considering the last $N_{L}$ operators, with $N_{L}$ ranging from 1 to 10.

\begin{figure}[H]
    \centering
    \includegraphics[width=10cm]{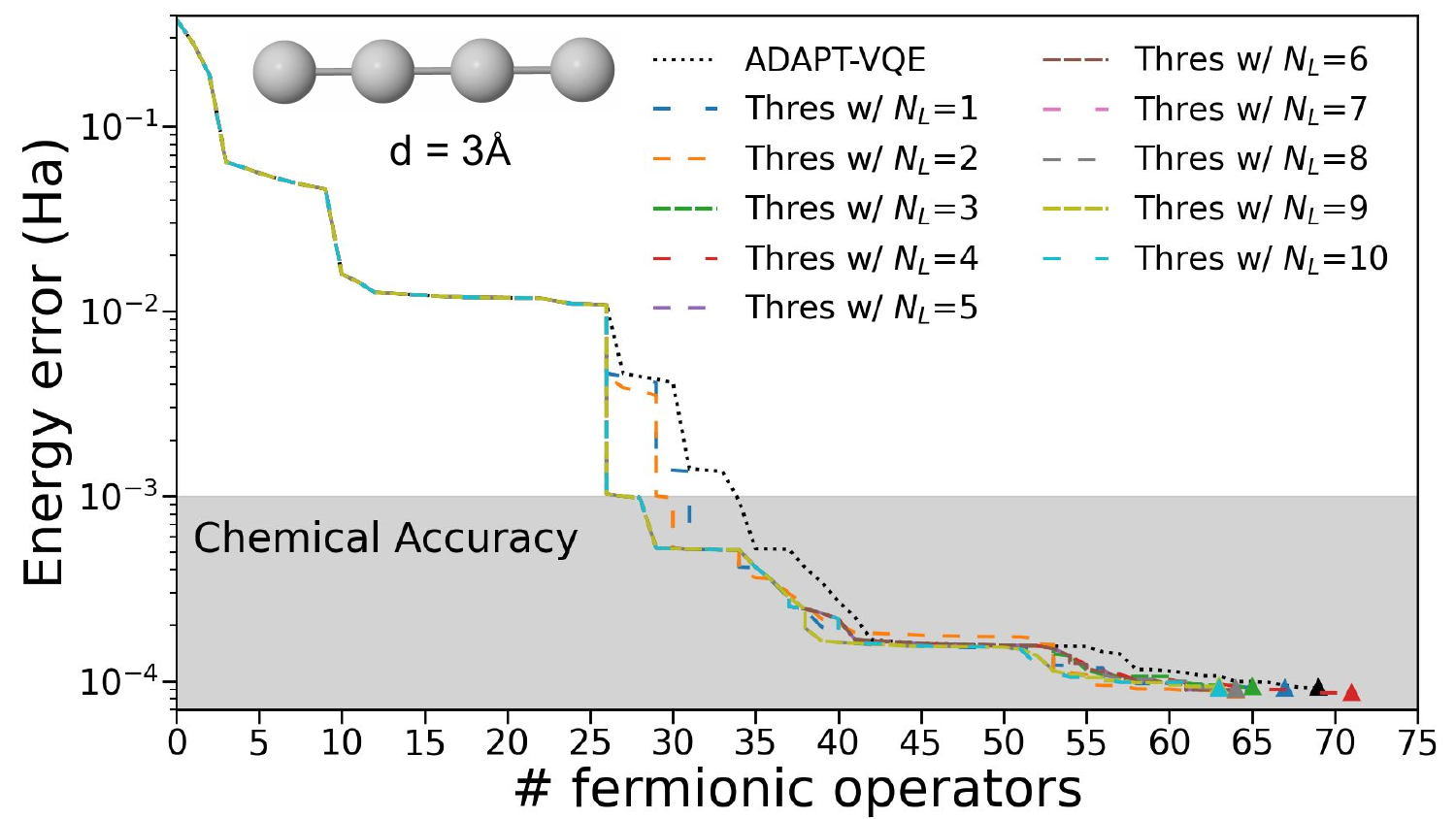}
    \caption{Energy error (in Hartree) with respect to FCI of ADAPT-VQE (black dotted line) and Pruned-ADAPT-VQE with different $N_{L}$ values for the linear \ce{H4} with interatomic distance of 3.0~{\AA}, and computed with the 3-21G basis set and an active space of 8 orbitals. The triangle indicates where the simulation ended.}
    \label{si:fig:linh4_thres_test}
\end{figure}

\begin{figure}[H]
    \centering
    \includegraphics[width=10cm]{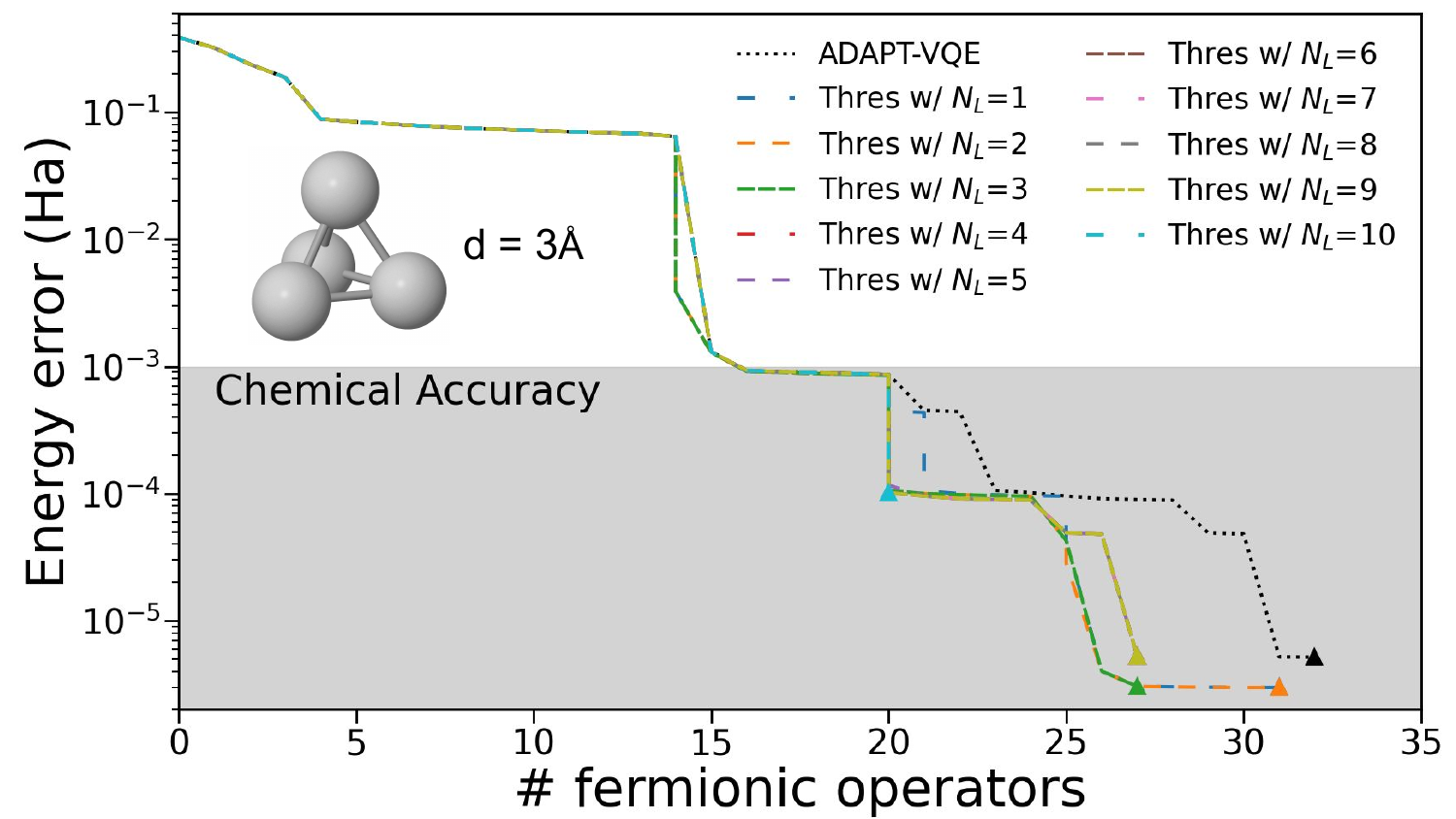}
    \caption{Energy error (in Hartree) with respect to FCI of ADAPT-VQE (black dotted line) and Pruned-ADAPT-VQE with different $N_{L}$ values for the tetrahedral \ce{H4} with interatomic distance of 3.0~{\AA}, and computed with the 3-21G basis set and an active space of 8 orbitals. The triangle indicates where the simulation ended.}
    \label{si:fig:tetrah4_thres_test}
\end{figure}

\begin{figure}[H]
    \centering
    \includegraphics[width=10cm]{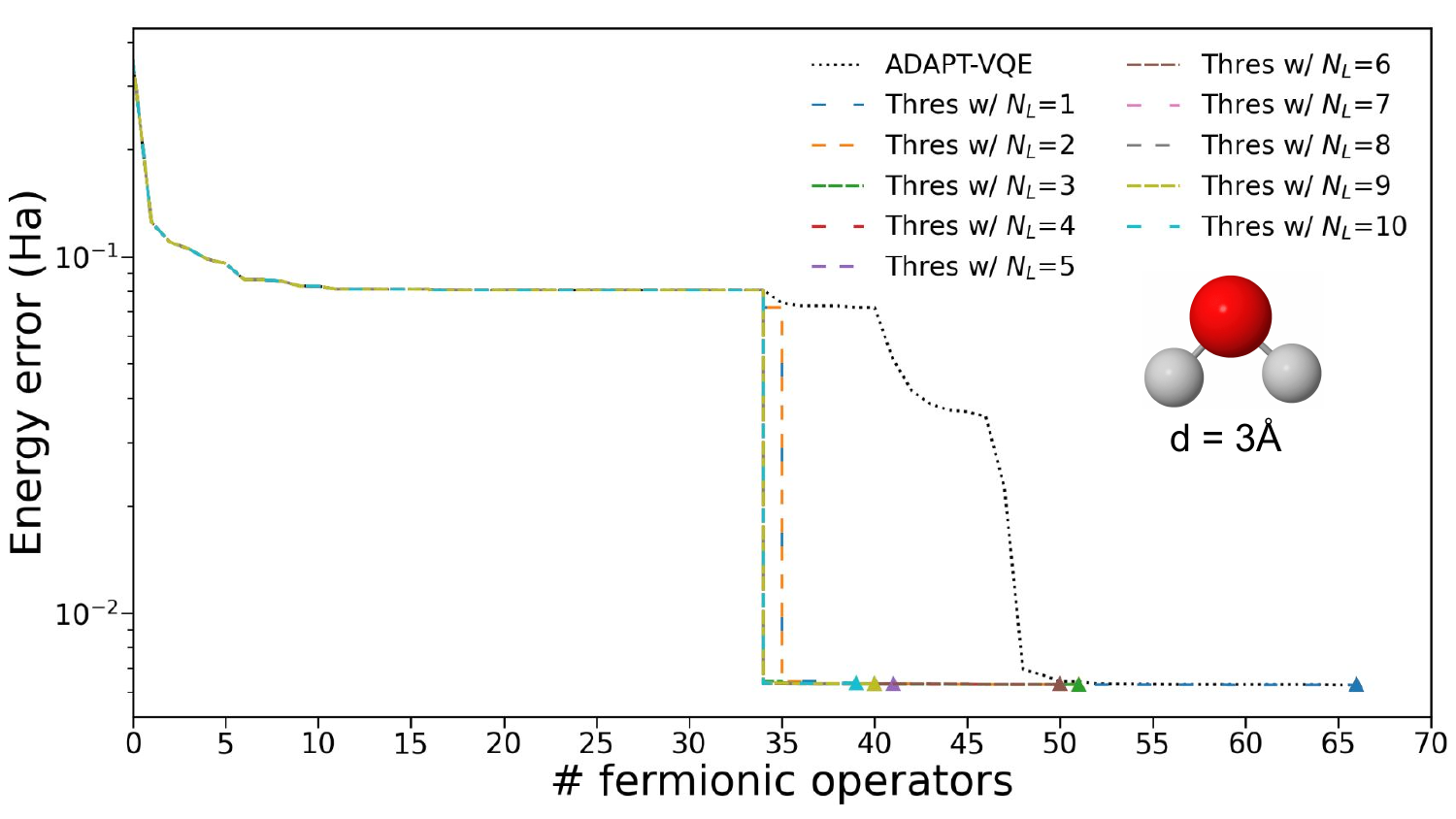}
    \caption{Energy error (in Hartree) with respect to CASCI(8,8) of ADAPT-VQE (black dotted line) and Pruned-ADAPT-VQE with $N_{L}$ values weights for the \ce{H2O} molecule with 3.0~{\AA} \ce{O-H} distance, and computed with the 3-21G basis set with active space of 8 orbitals plus one frozen orbital. The triangle indicates where the simulation ended.}
    \label{si:fig:h2o_thres_test}
\end{figure}

\begin{figure}
    \centering
    \includegraphics[width=10cm]{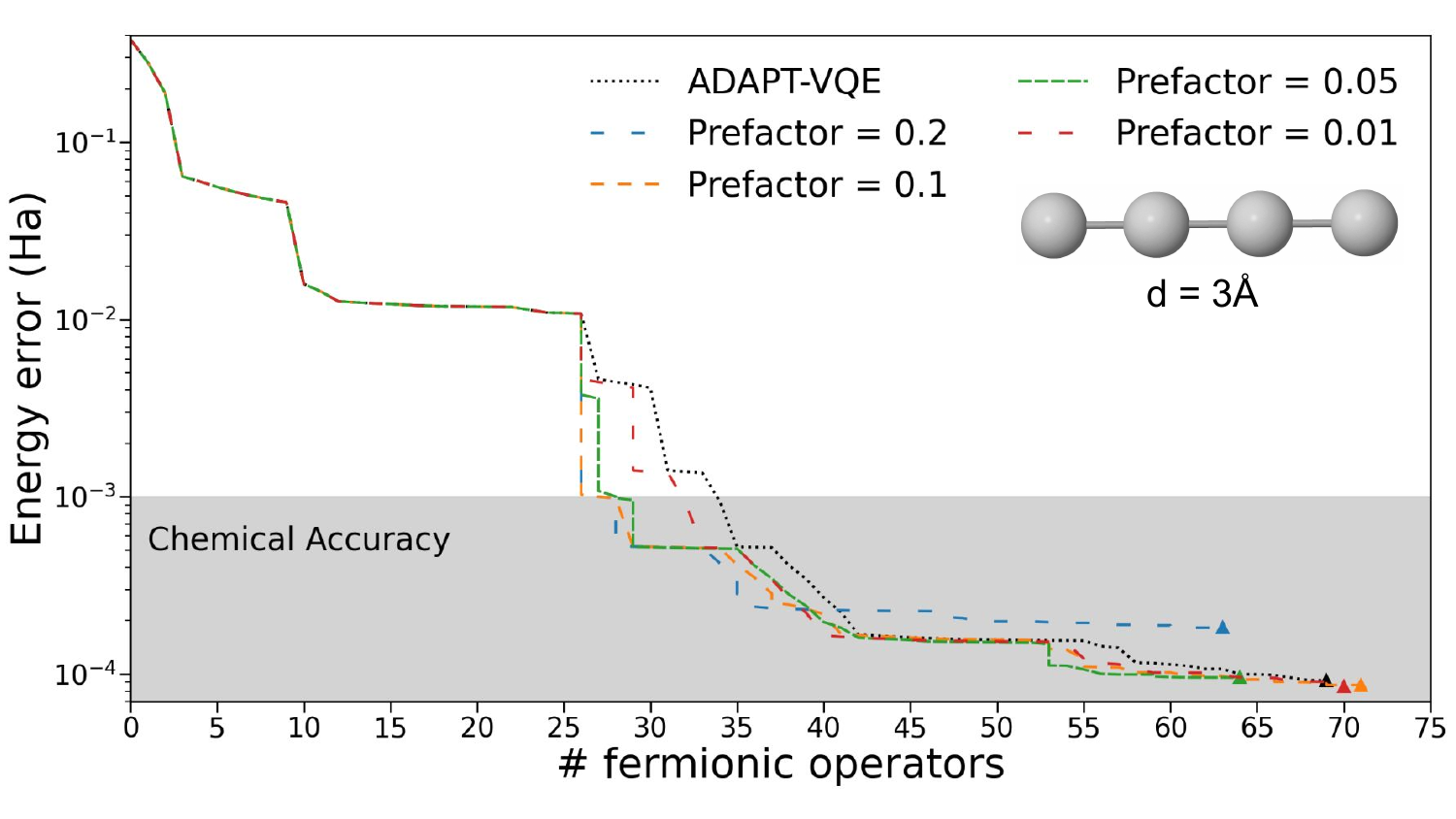}
    \caption{Energy error (in Hartree) with respect to FCI of ADAPT-VQE (black dotted line) and Pruned-ADAPT-VQE with different prefactor values for the linear \ce{H4} with interatomic distance of 3.0~{\AA}, and computed with the 3-21G basis set and an active space of 8 orbitals. The triangle indicates where the simulation ended.}
    \label{si:fig:linh4_prefactor_test}
\end{figure}

Setting the prefactor to 0.1 in equation~{\ref{eq:threshold_SI}} balances the likelihood of removing low-contributing operators. Decreasing it allow smaller operators appear in the ansatz and reach smaller errors, while increasing it creates more compact ans\"atze that may not reach the lowest energies. 
A large prefactor can also trigger premature operator removals, causing operators to be reintroduced in subsequent iterations and resulting in non-useful iterations. 
Figures~{\ref{si:fig:linh4_prefactor_test}}-{\ref{si:fig:h2o_prefactor_test}} study this effect by varying the prefactor we denominate from 0.2 to 0.01.

\begin{figure}[H]
    \centering
    \includegraphics[width=10cm]{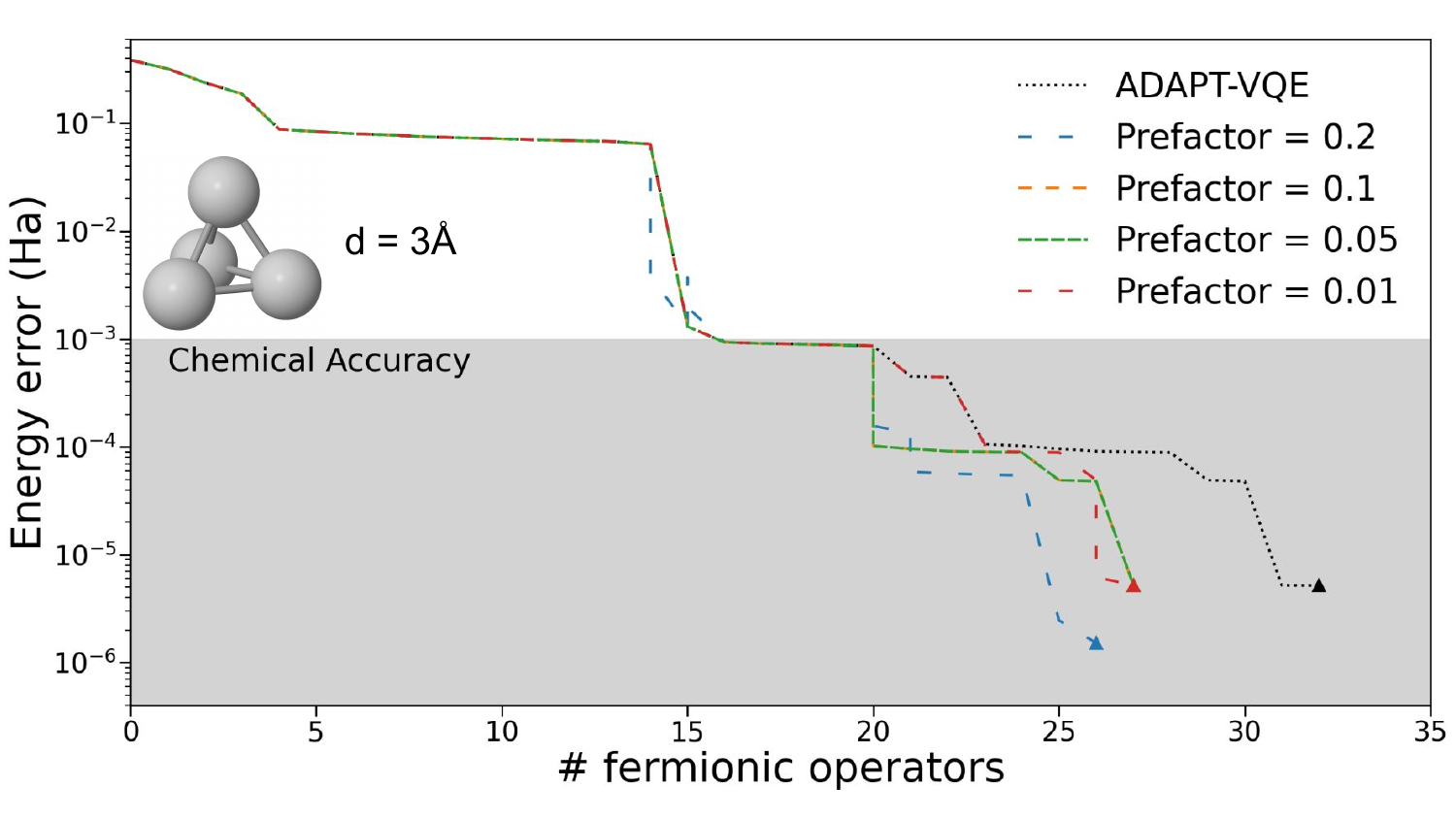}
    \caption{Energy error (in Hartree) with respect to FCI of ADAPT-VQE (black dotted line) and Pruned-ADAPT-VQE with different prefactor values for the tetrahedral \ce{H4} with interatomic distance of 3.0~{\AA}, and computed with the 3-21G basis set and an active space of 8 orbitals. The triangle indicates where the simulation ended.}
    \label{si:fig:tetrah4_prefactor_test}
\end{figure}

\begin{figure}[H]
    \centering
    \includegraphics[width=0.5\linewidth]{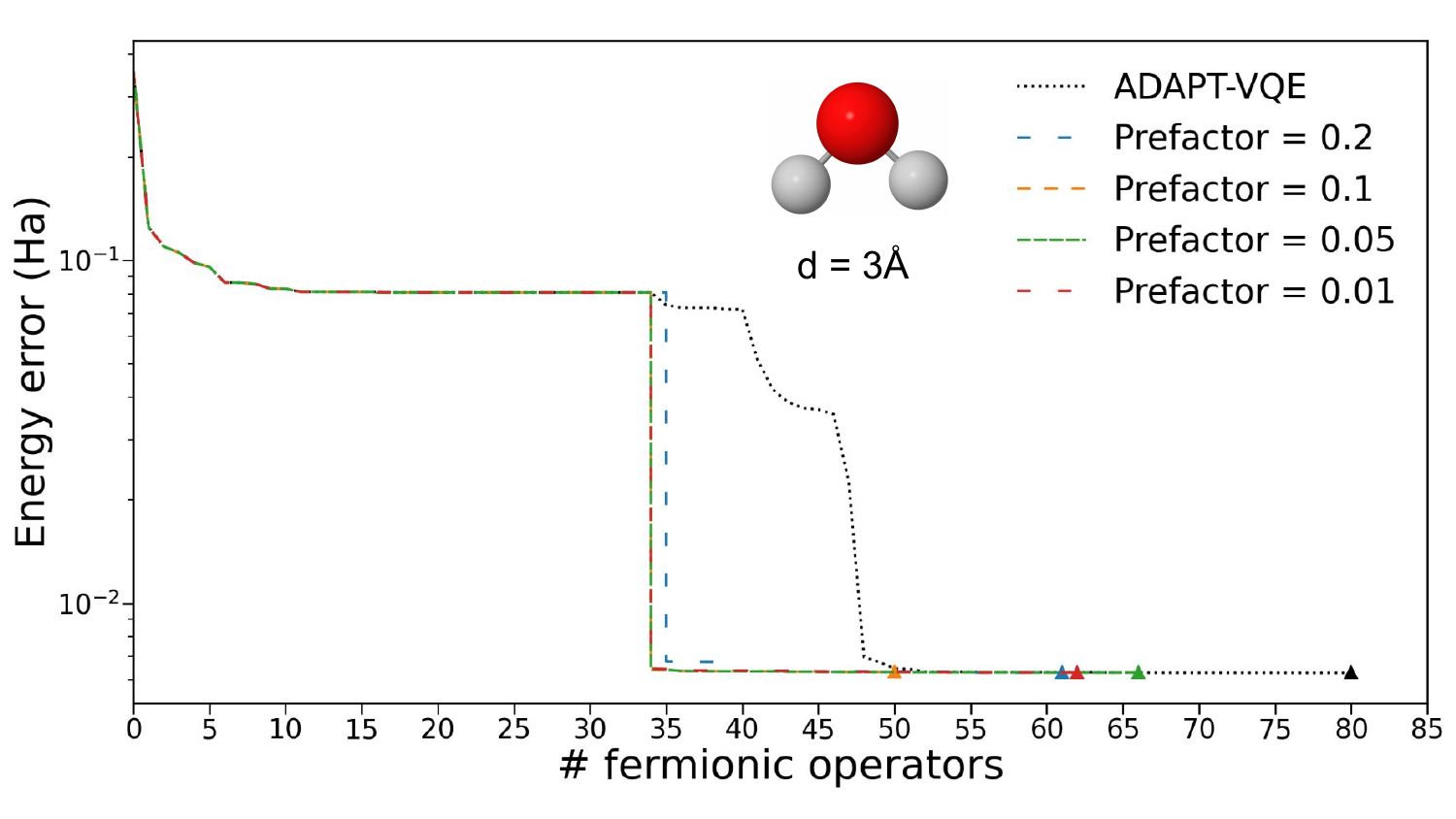}
    \caption{Energy error (in Hartree) with respect to CASCI(8,8) of ADAPT-VQE (black dotted line) and Pruned-ADAPT-VQE with different prefactor values for the \ce{H2O} molecule with 3.0~{\AA} \ce{O-H} distance, and computed with the 3-21G basis set with active space of 8 orbitals plus one frozen orbital. The triangle indicates where the simulation ended.}
    \label{si:fig:h2o_prefactor_test}
\end{figure}

We conclude that setting the prefactor to 0.1 achieves, in general, a good compromise between accuracy and ansatz size.
However, as observed in Figure~{\ref{si:fig:tetrah4_prefactor_test}}, in some cases an aggressive removal criteria may lead to more efficient ans\"atze.

\section{Additional molecule analysis}

\begin{figure}[H]
    \centering
    \includegraphics[width=10cm]{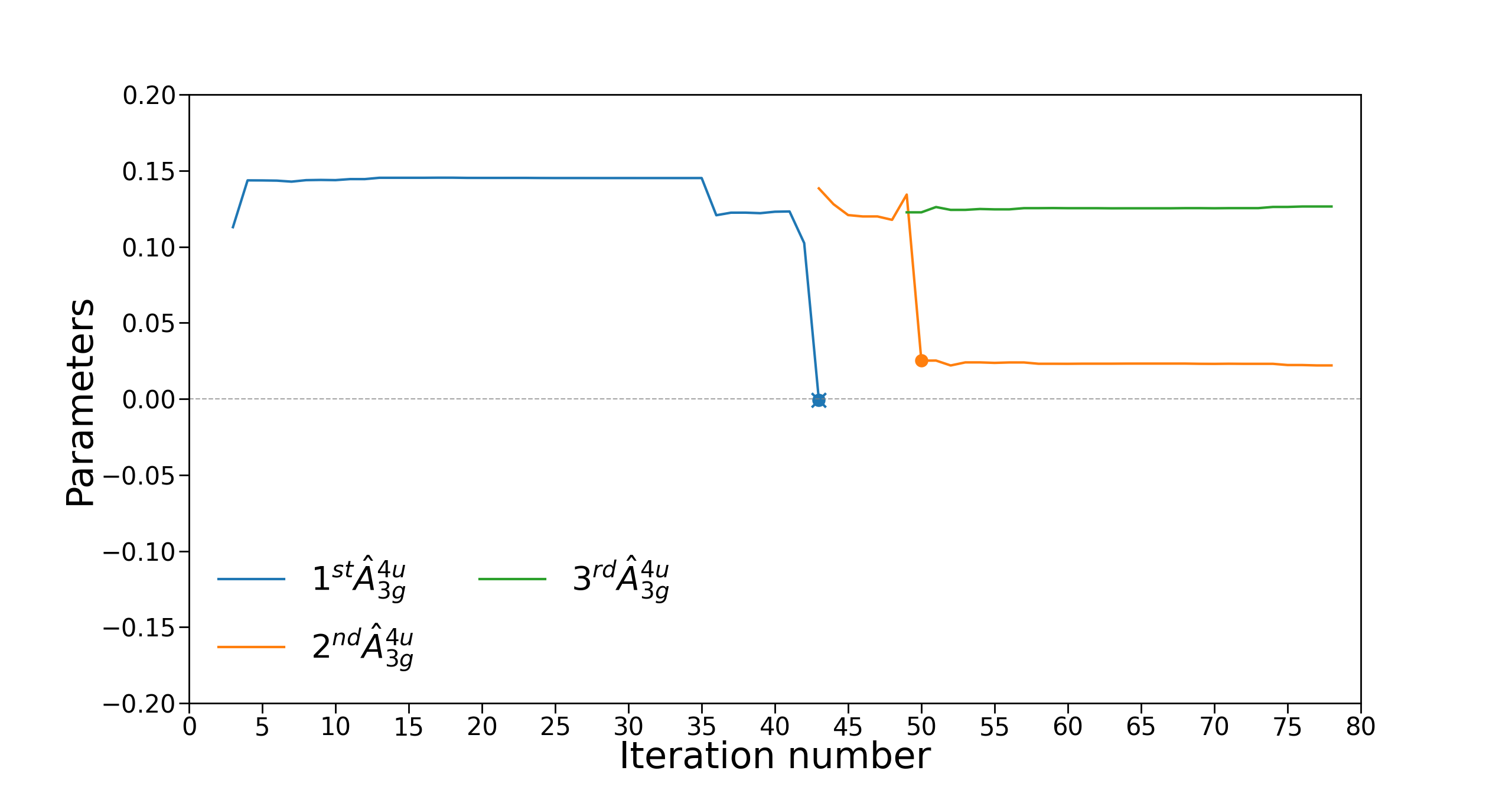}
    \caption{Parameter values of various operators from their introduction to iteration 80 in the simulation of \ce{H2O} with interatomic distance \ce{H-O} of 3.0 \AA~and the 3-21G basis set, using an active space of 8 orbitals plus 1 frozen orbital. Full circles indicate introduction of another instance of the operator. Cross markers indicate when the operator is removed by Pruned-ADAPT-VQE.}
    \label{fig:si:reorder_water14}
\end{figure}

In this section we analyze the removed operators from the {\ce{H2O}} molecule.
We observe again the operator reordering phenomena in some of the removed operators. 
This is the case of operator $\hat{A}_{3g}^{4u}$, Figure~{\ref{fig:si:reorder_water14}}, where the addition of the same operator in iteration 43 causes a decrease in the parameter value of the first added operator, leading to its removal. The same happens with the second $\hat{A}_{3g}^{4u}$ added, whose coefficient decays after a third addition of the same instance. 
In this case, however, the weight is the weight is distributed between the two instances of the operator, and neither is excluded from the ansatz. 
This illustrates another instance of local quasi-commutation between the operator and this segment of the ansatz.

\begin{figure}[H]
    \centering
    \includegraphics[width=10cm]{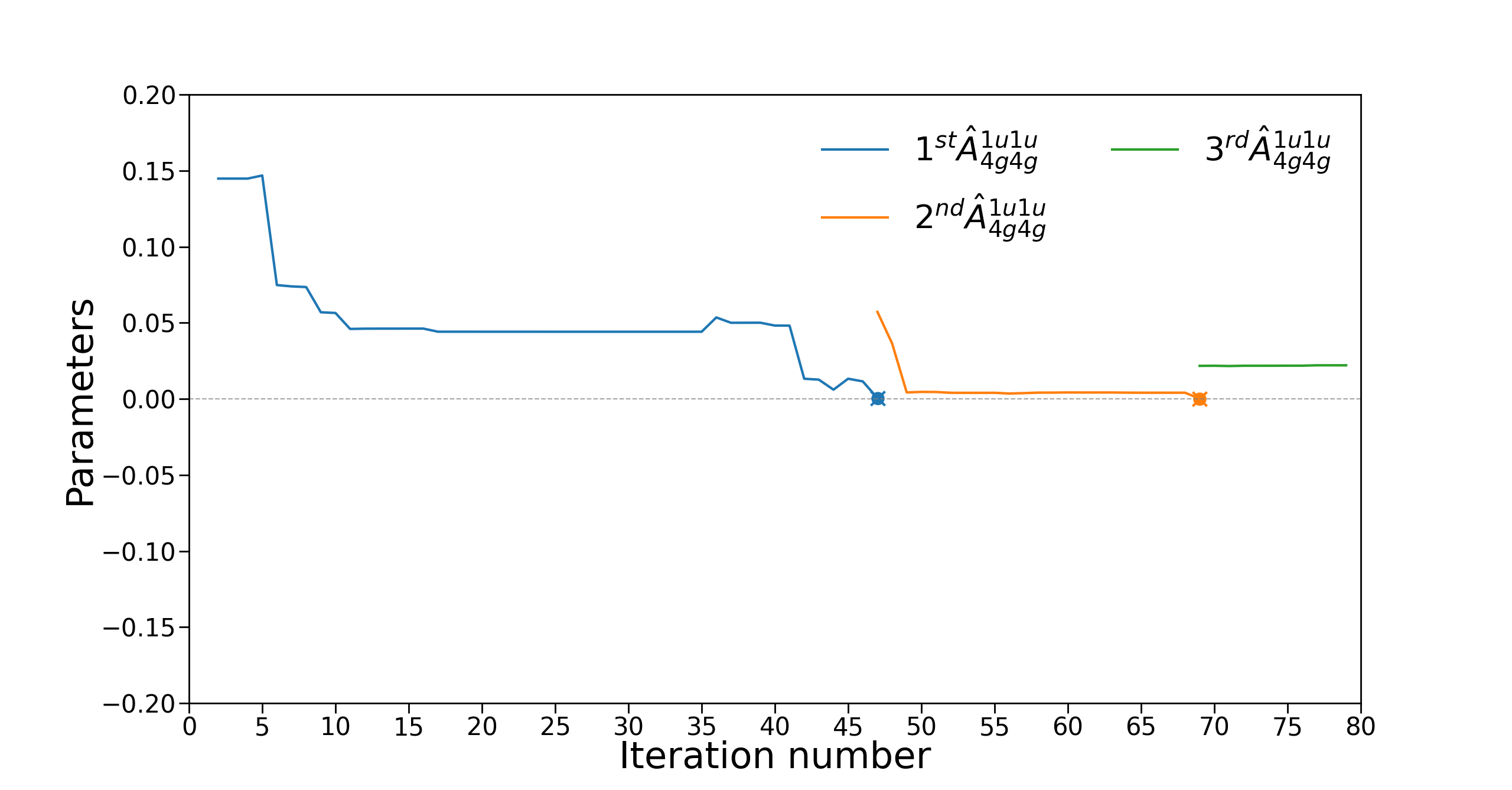}
    \caption{Parameter values of various operators from their introduction to iteration 80 in the simulation of \ce{H2O} with interatomic distance \ce{H-O} of 3.0 \AA~and the 3-21G basis set, using an active space of 8 orbitals plus 1 frozen orbital. Full circles indicate introduction of another instance of the operator. Cross markers indicate when the operator is removed by Pruned-ADAPT-VQE.}
    \label{fig:si:reorder_water20}
\end{figure}

Another example of operator reordering is $\hat{A}^{1u1u}_{4g4g}$, shown in Figure \ref{fig:si:reorder_water20}.

\begin{figure}[H]
    \centering
    \includegraphics[width=10cm]{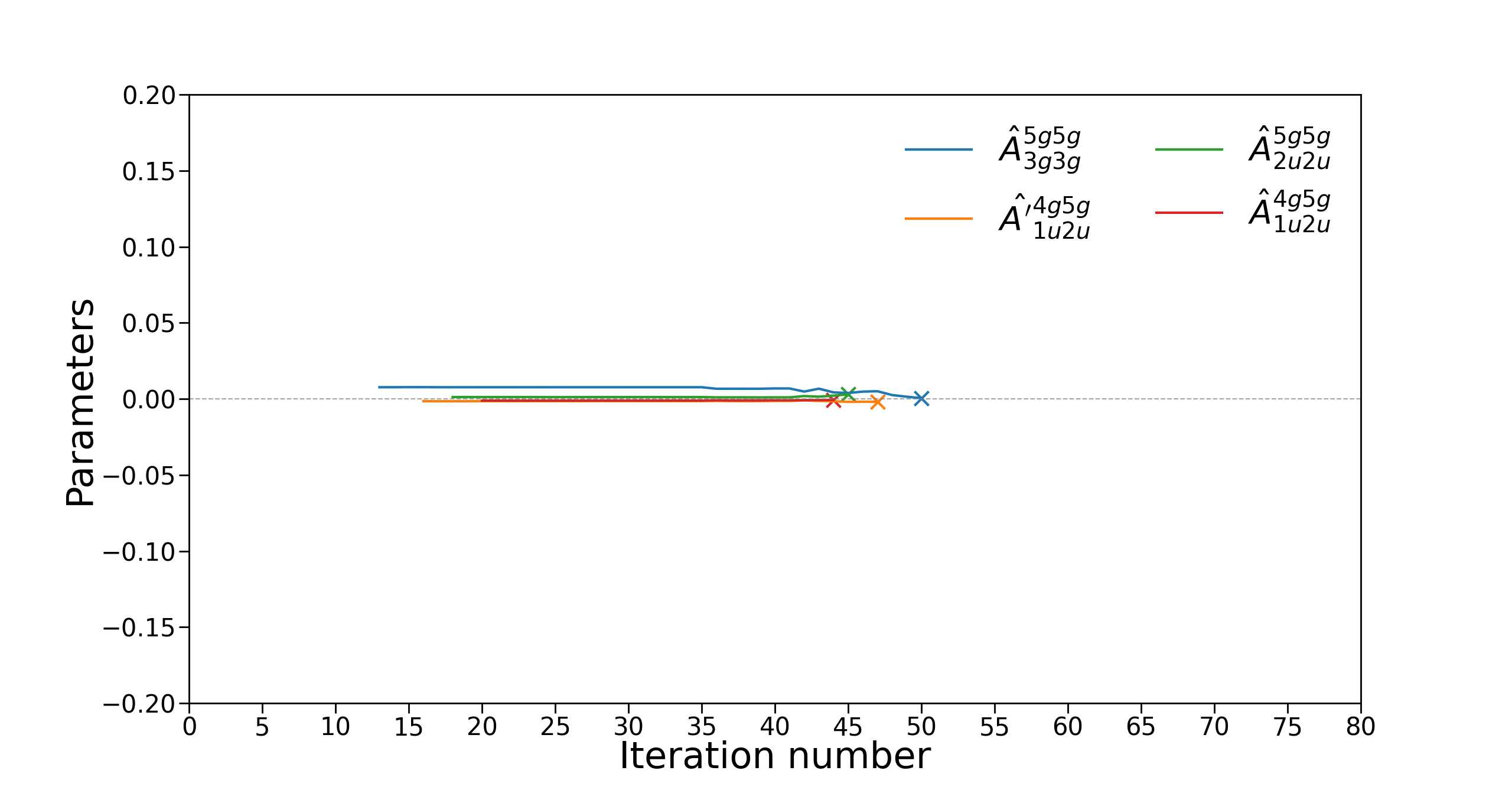}
    \caption{Parameter values of various operators from their introduction to iteration 80 in the simulation of \ce{H2O} with interatomic distance \ce{H-O} of 3.0 \AA and the 3-21G basis set, using an active space of 8 orbitals plus 1 frozen orbital. Full circles indicate introduction of another instance of the operator. Cross markers indicate when the operator is removed by Pruned-ADAPT-VQE.}
    \label{fig:si:badops_water}
\end{figure}

Almost all of the deleted operators in this simulation are due to what we defined as bad operator selection. Some examples are shown in Figure \ref{fig:si:badops_water}.

\begin{figure}[H]
    \centering
    \includegraphics[width=10cm]{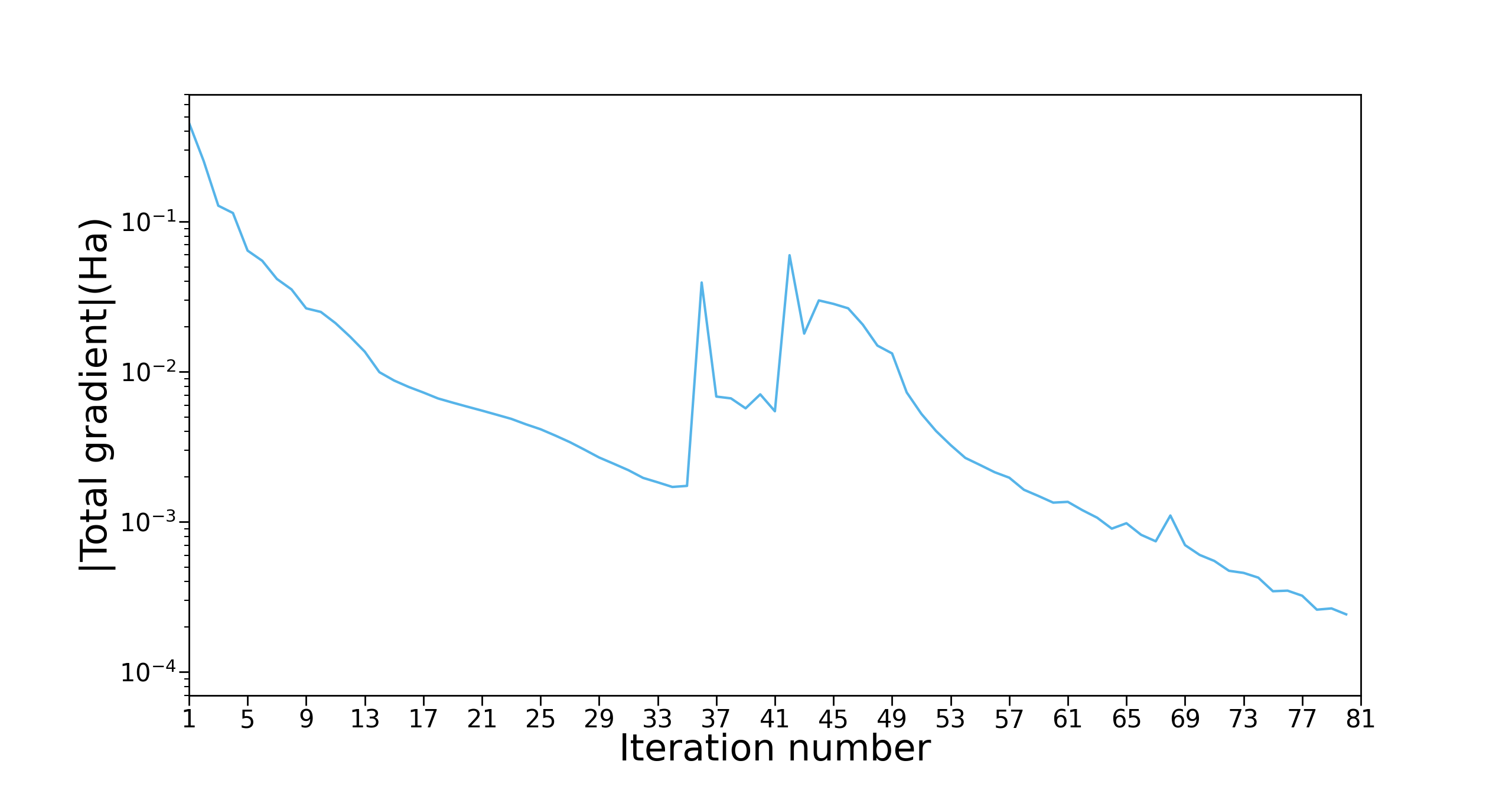}
    \caption{Total gradient of the UCCSD operators in ADAPT-VQE simulation of \ce{H2O} molecule with 3.0~{\AA} \ce{O-H} distance, and computed with the 3-21G basis set with active space of 8 orbitals plus one frozen orbital.}
    \label{fig:si:gradient_water}
\end{figure}

The simulation results are shown in Figure 8, where a large flat area is observed. 
Figure~\ref{fig:si:gradient_water} illustrates the evolution of the total gradient throughout the simulation, linking this flat region to a gradient trough. 
We attribute this phenomenon to the large number of poor operators that were selected during this simulation.

\begin{figure}[H]
    \centering
    \includegraphics[width=0.7\linewidth]{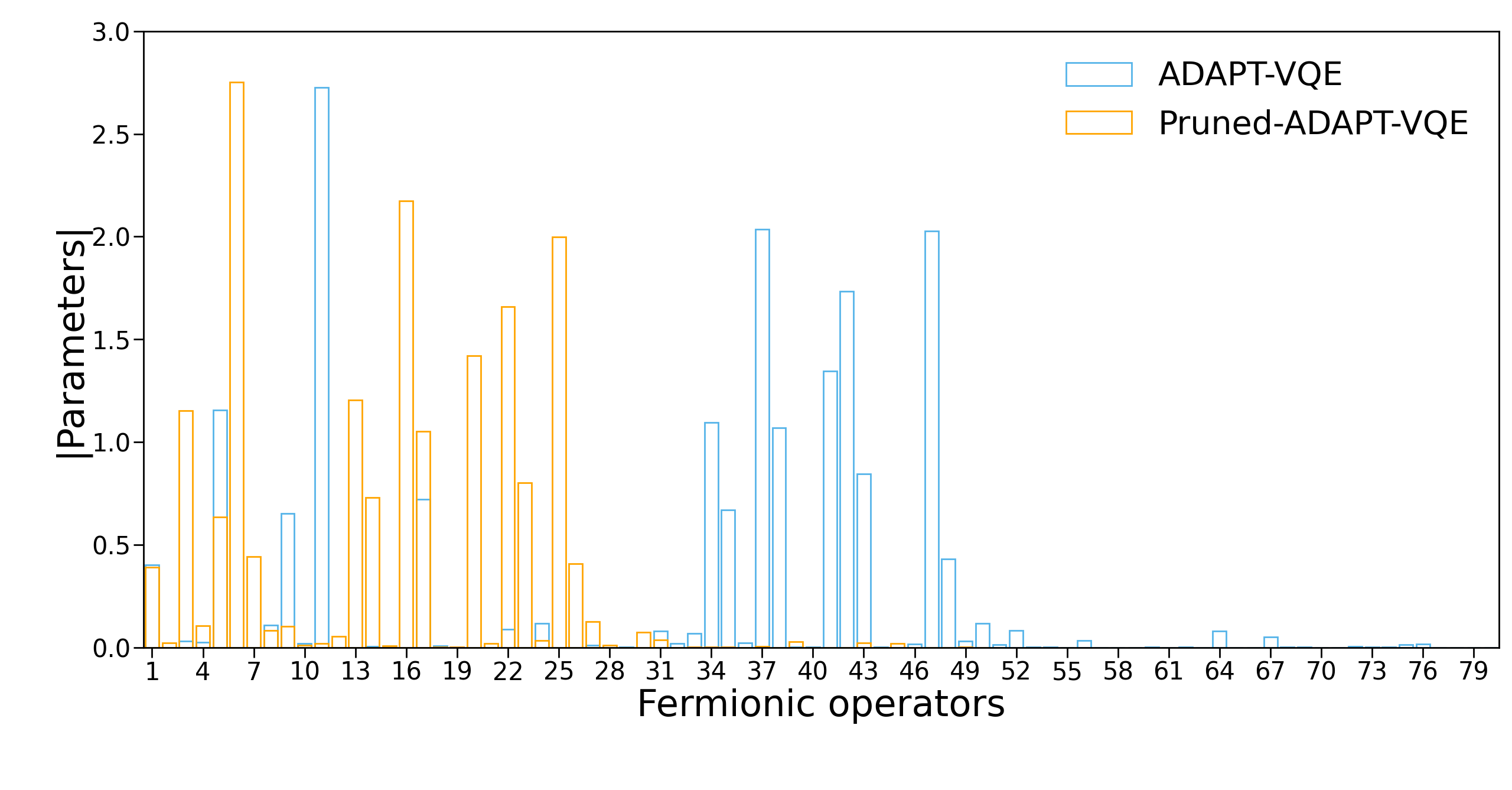}
    \caption{Absolute parameter values for the N = 80 ansatz obtained with ADAPT-VQE (blue) and Pruned-ADAPT-VQE (orange) for the \ce{H2O} system with interatomic distance \ce{H-O} of 3.0 \AA~and with the 3-21G basis set using 8 active orbitals plus one frozen orbital.}
    \label{fig:si:amplitudes_water}
\end{figure}

The comparison between the final ans\"atze of ADAPT-VQE and Pruned-ADAPT-VQE is shown in Figure~\ref{fig:si:amplitudes_water}.

\section{The gradient-based selection criterion}

\begin{figure}[H]
    \centering
    \includegraphics[width=10cm]{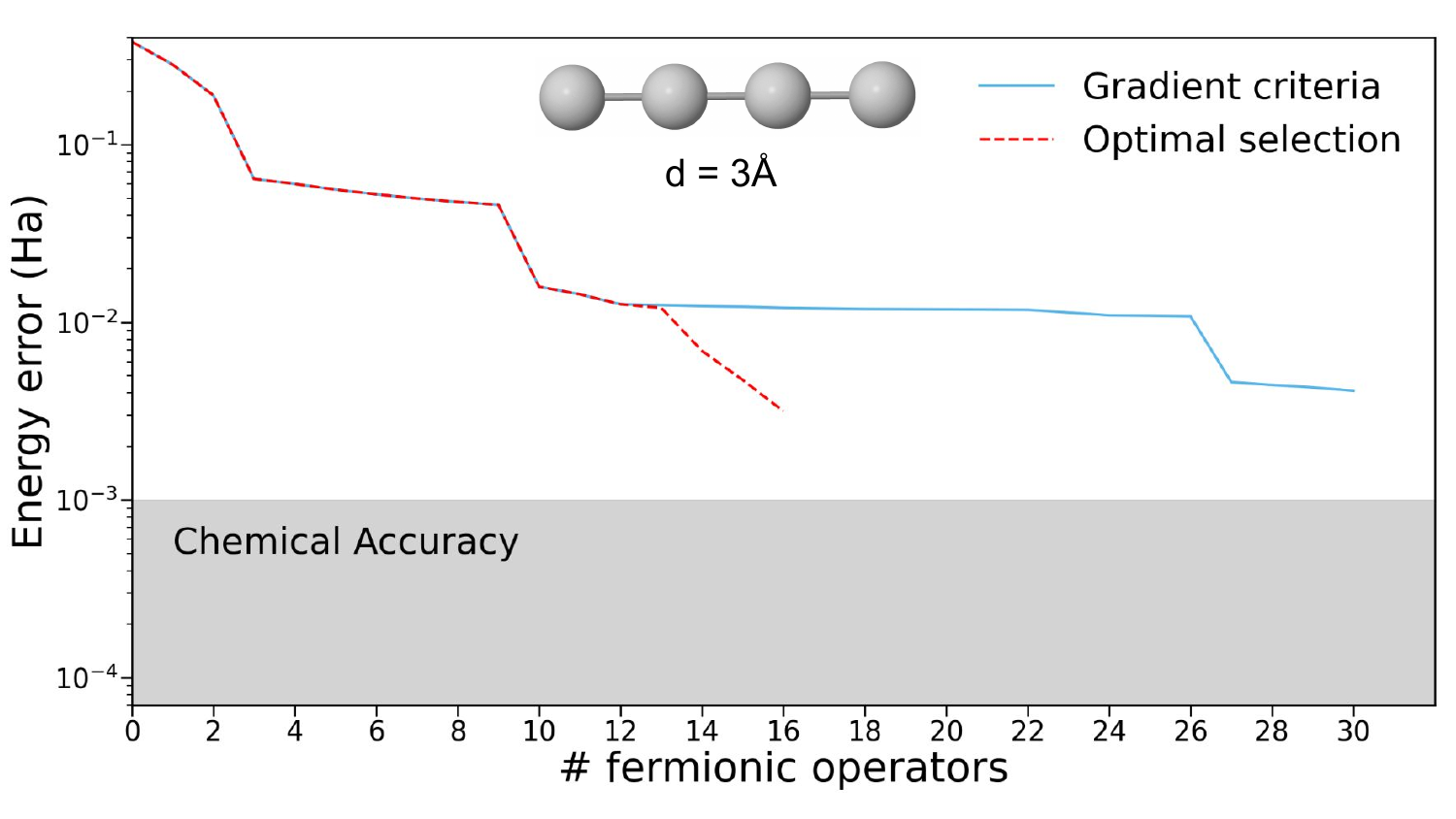}
    \caption{Energy error (in Hartree) with respect to FCI of ADAPT-VQE using the gradient-based criteria (blue line) and trying all possible options and keeping the best one (dotted red line) for the linear {\ce{H4}} with interatomic distance of 3.0 \AA, and computed with the 3-21G basis set and active space of 8 orbitals.}
    \label{fig:si:nogradient_path}
\end{figure}

This section tests wether the operator with the largest gradient is always the best choice. 
We use the first 12 operators from the fermionic ADAPT-VQE ansatz for the linear \ce{H4} system with interatomic distance of 3.0 \AA~(example from Figure 1). We take the first 12 operators because a gradient trough appears starting from that point, as can be appreciated in Figure 1c, which leads to a sequence of iterations where the energy error barely improves, this is what we call flat area. After taking the 12-operators ansatz from ADAPT-VQE, instead of selecting the next one to be added with the gradient criteria, we try every possible operator in the pool and optimize all the parameters. Repeating this process three times, led to a different ansatz that was able to avoid the flat area, as is shown in Figure~\ref{fig:si:nogradient_path}.

\section{Pruned-ADAPT-VQE with UCCSD pool}
In this section we test Pruned-ADAPT-VQE on a variety of molecular systems. The hyperparameters used are the same as those employed in the main text examples: a coefficient weight of 2, a prefactor $\alpha = 10$, and a dynamic threshold set to $10\%$ of the average amplitude of the last four added operators.
\begin{figure}[H]
    \centering
    \includegraphics[width=10cm]{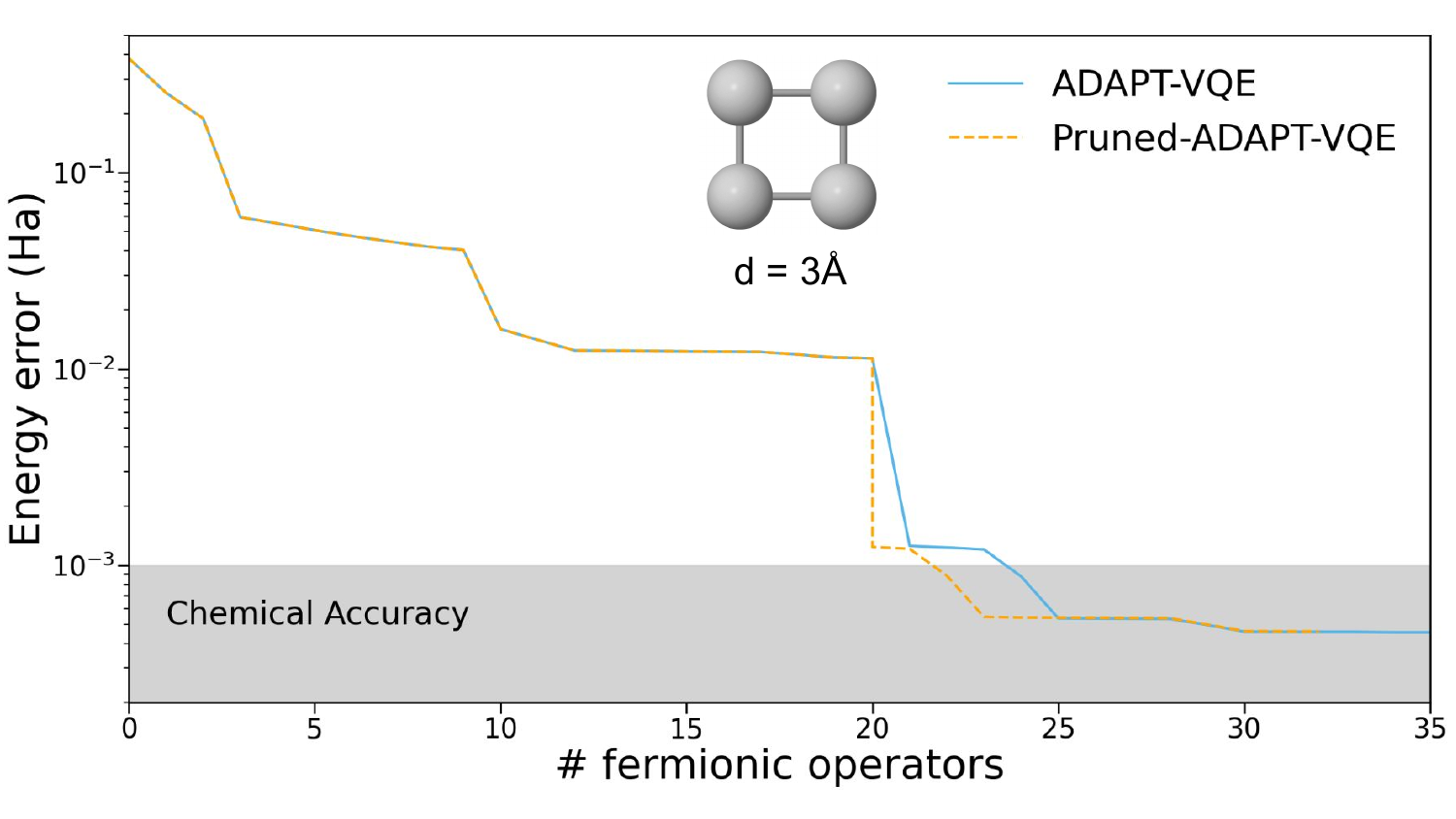}
    \caption{Energy errors (in Hartree) with respect to FCI obtained with ADAPT-VQE (solid blue) and Pruned-ADAPT-VQE (dashed orange) ans\"atze for the squared {\ce{H4}} with interatomic distance of 3.0~{\AA}, and computed with the 3-21G basis set and an active space of 8 orbitals.}
    \label{fig:sq_h4_energies_pruned}
\end{figure}

\begin{figure}[H]
    \centering
    \includegraphics[width=10cm]{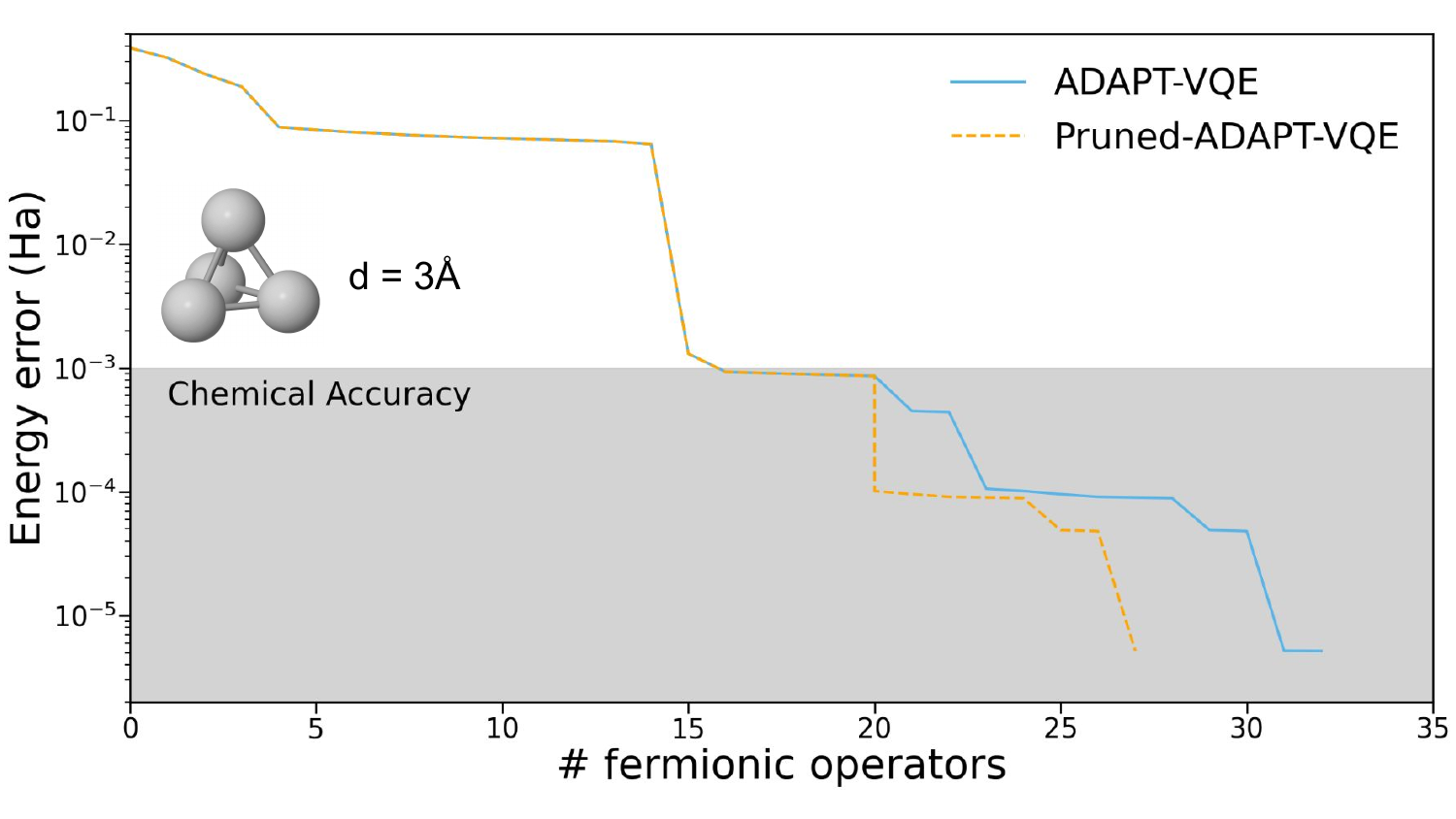}
    \caption{Energy errors (in Hartree) with respect to FCI obtained with ADAPT-VQE (solid blue) and Pruned-ADAPT-VQE (dashed orange) ans\"atze for the tetrahedral {\ce{H4}} with interatomic distance of 3.0~{\AA}, and computed with the 3-21G basis set and an active space of 8 orbitals.}
    \label{fig:td_h4_energies_pruned}
\end{figure}

\begin{figure}[H]
    \centering
    \includegraphics[width=10cm]{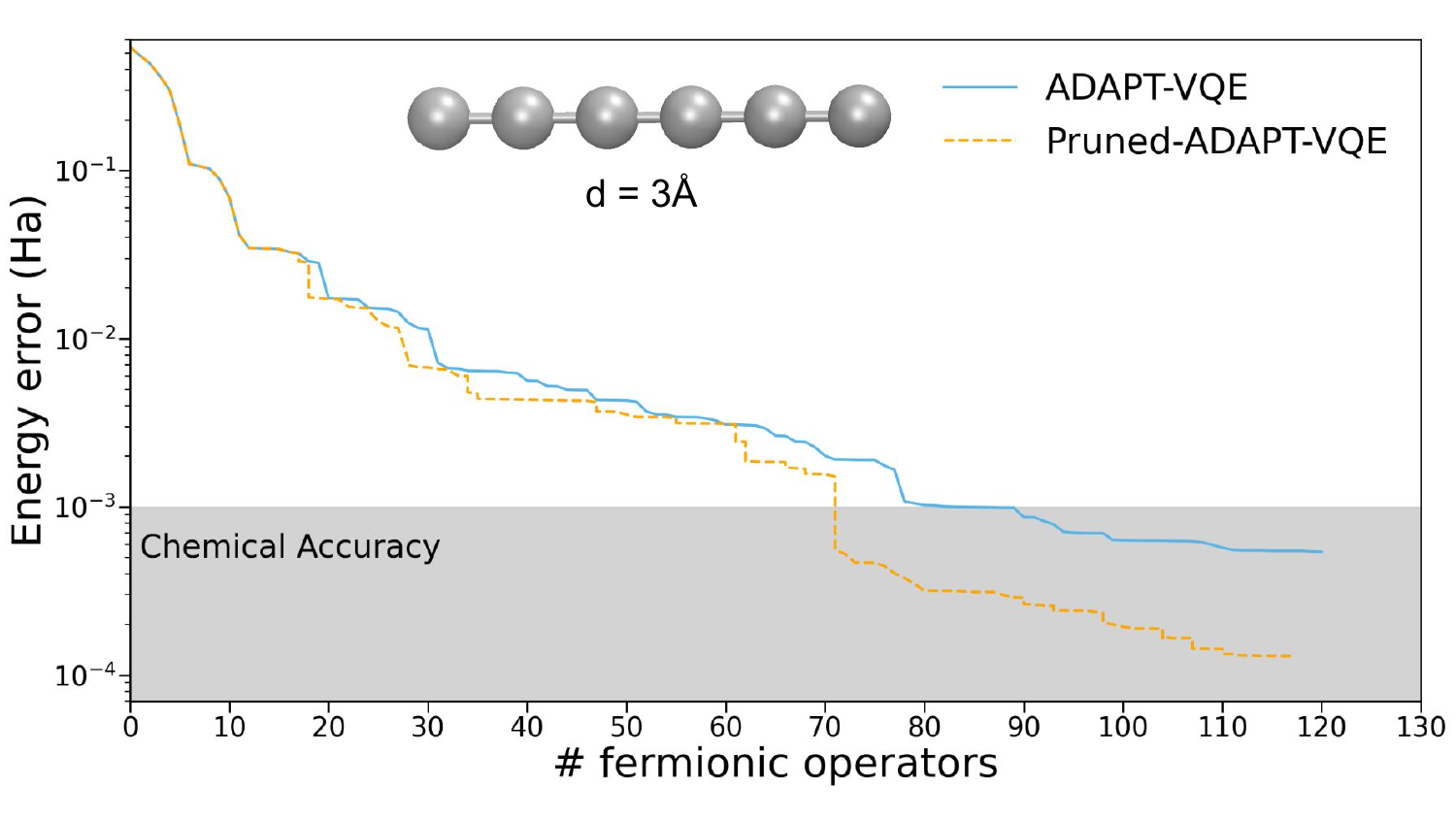}
    \caption{Energy errors (in Hartree) with respect to CASCI(6,8) obtained with ADAPT-VQE (solid blue) and Pruned-ADAPT-VQE (dashed orange) ans\"atze for the linear {\ce{H6}} system with interatomic distance of 3.0~{\AA}, and computed with the 3-21G basis set and an active space of 8 orbitals.}
    \label{fig:H6_energies_pruned}
\end{figure}

In the case of \ce{BeH2} we found that using the same parameters led to a premature convergence of the algorithm. This situation was resolved by taking a more relaxed removal strategy, specifically by reducing the prefactor in the threshold definition.

\begin{figure}[H]
    \centering
    \includegraphics[width=10cm]{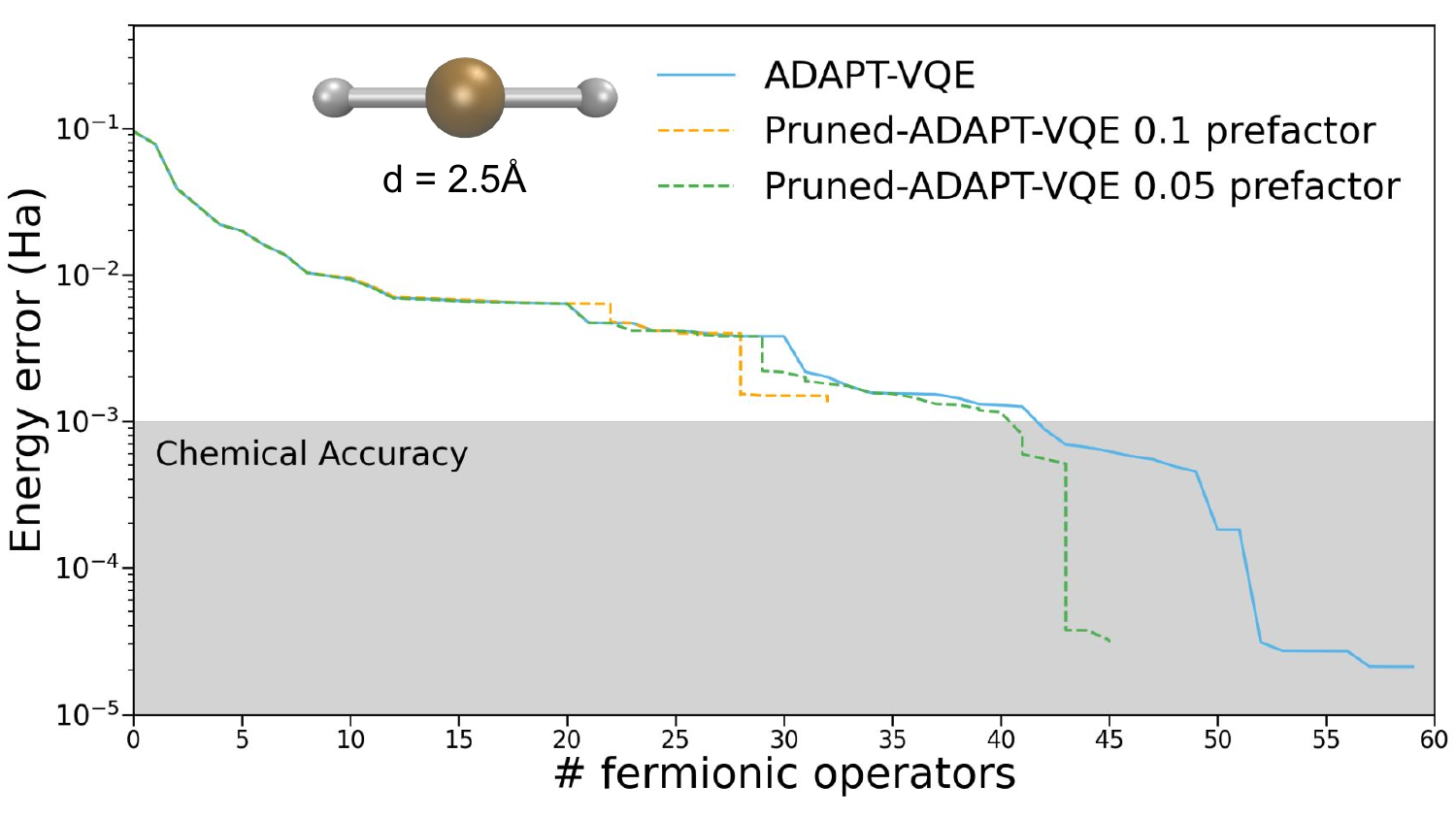}
    \caption{Energy errors (in Hartree) with respect to CASCI(4,8) obtained with ADAPT-VQE (solid blue) and Pruned-ADAPT-VQE (dashed orange for prefactor 0.1 and dashed green for prefactor 0.05, see Section 5) ans\"atze for the {\ce{BeH2}} molecule with interatomic distance of 2.5~{\AA}, and computed with the 3-21G basis set and an active space of 8 orbitals plus one frozen orbital.}
    \label{fig:BeH2_energies_pruned}
\end{figure}

\section{Pruned-ADAPT-VQE with qubit-excitation-based pool}

In this section, we evaluate the performance of Pruned-ADAPT-VQE using the qubit-excitation-based pool.{\cite{Yordanov2021}}
Unless otherwise specified, we employ the same hyperparameters as those used with the fermionic pool: a coefficient weight of 2, $\alpha = 10$, and a threshold defined as $10\%$ of the average amplitude of the last 4 added operators.
\begin{figure}[H]
    \centering
    \includegraphics[width=10cm]{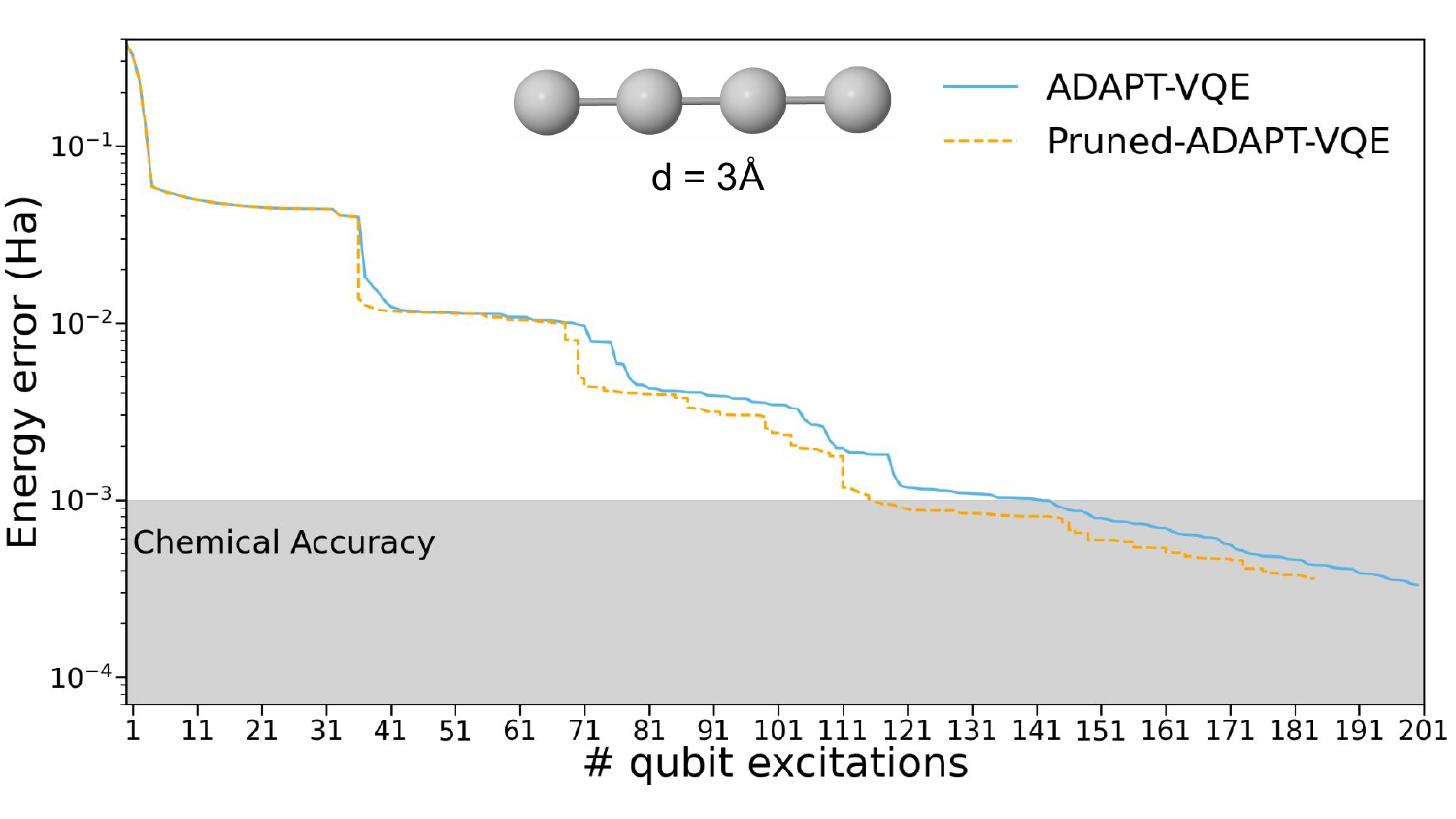}
    \caption{Energy errors (in Hartree) with respect to FCI obtained with ADAPT-VQE (solid blue) and Pruned-ADAPT-VQE (dashed orange) ans\"atze with qubit-excitation-based pool for the linear {\ce{H4}} molecule with interatomic distance of 3.0~{\AA}, and computed with the 3-21G basis set and an active space of 8 orbitals.}
    \label{fig:h4_linear_qubit}
\end{figure}

\begin{figure}[H]
    \centering
    \includegraphics[width=10cm]{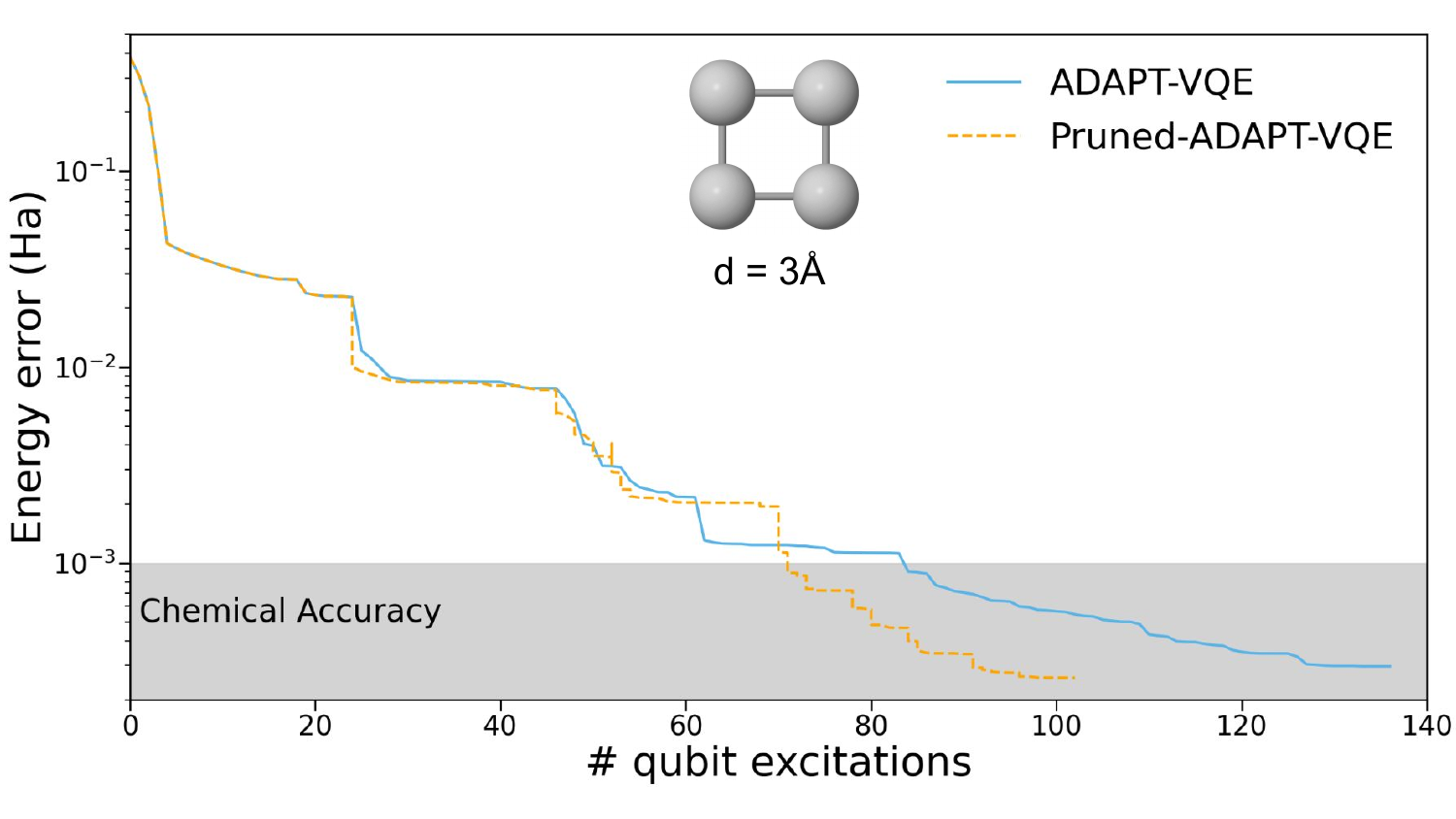}
    \caption{Energy errors (in Hartree) with respect to FCI obtained with ADAPT-VQE (solid blue) and Pruned-ADAPT-VQE (dashed orange) ans\"atze with qubit-excitation-based pool for the squared {\ce{H4}} molecule with interatomic distance of 3.0~{\AA}, and computed with the 3-21G basis set and an active space of 8 orbitals with one frozen orbital.}
    \label{fig:h4_sq_qubit}
\end{figure}

\begin{figure}[H]
    \centering
    \includegraphics[width=10cm]{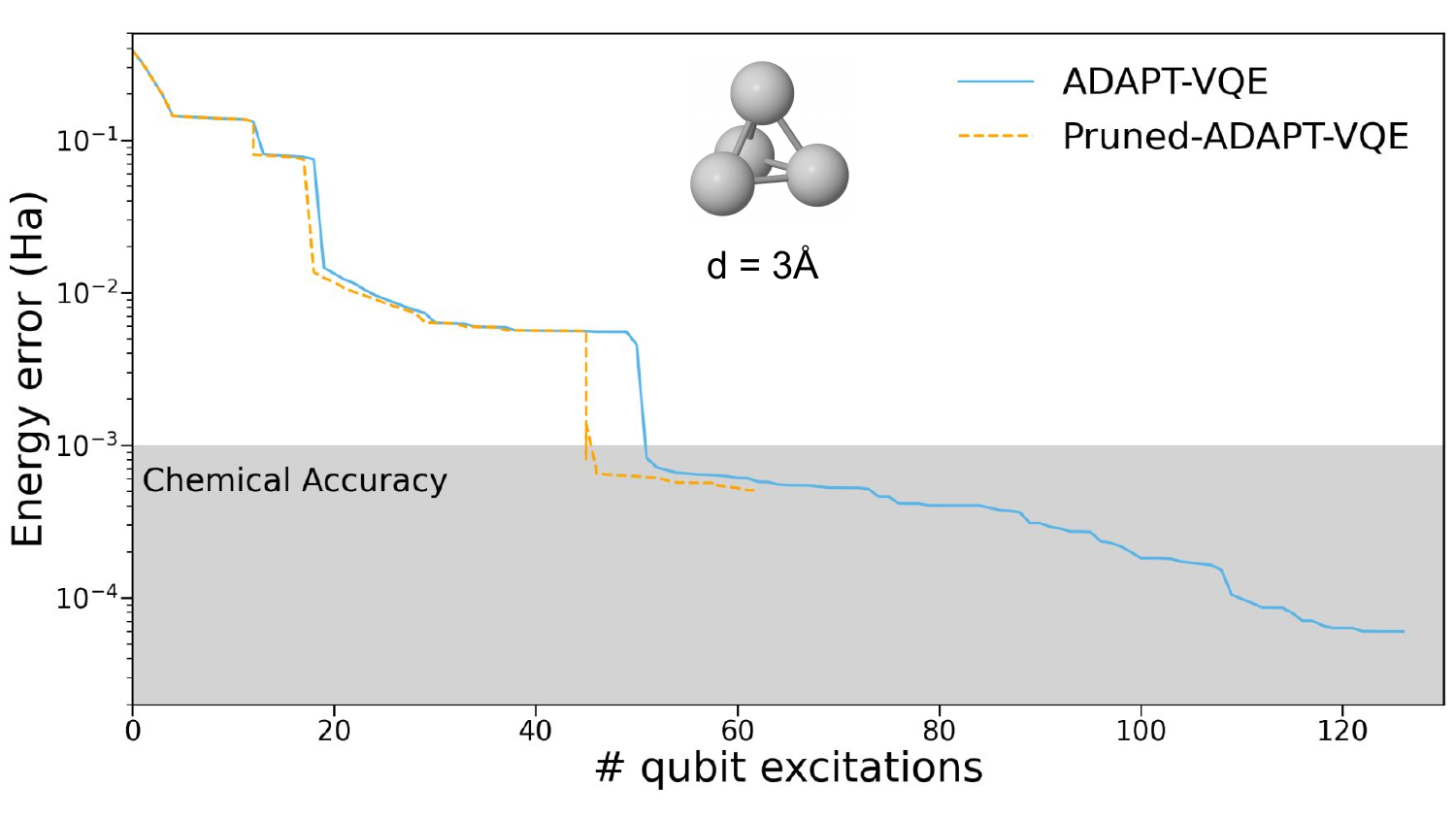}
    \caption{Energy errors (in Hartree) with respect to FCI obtained with ADAPT-VQE (solid blue) and Pruned-ADAPT-VQE (dashed orange) ans\"atze with qubit-excitation-based pool for the tetrahedral {\ce{H4}} molecule with interatomic distance of 3.0~{\AA}, and computed with the 3-21G basis set and an active space of 8 orbitals with one frozen orbital.}
    \label{fig:h4_td_qubit}
\end{figure}

\begin{figure}[H]
    \centering
    \includegraphics[width=10cm]{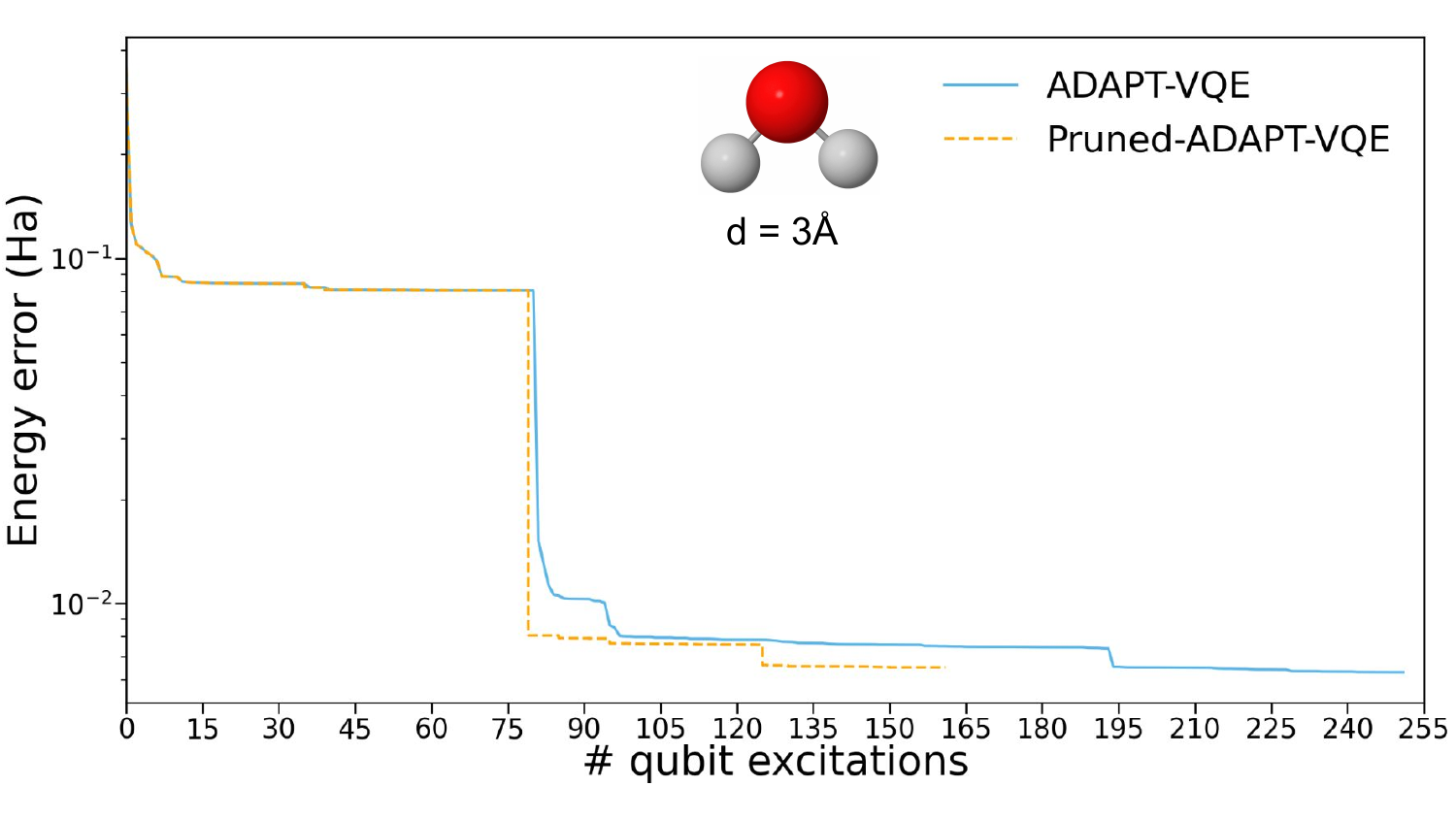}
    \caption{Energy errors (in Hartree) with respect to CASCI obtained with ADAPT-VQE (solid blue) and Pruned-ADAPT-VQE (dashed orange) ans\"atze with qubit-excitation-based pool for the {\ce{H2O}}  with 3.0~{\AA} {\ce{O-H}} distance and computed in the 3-21G basis set with active space of 8 orbitals plus one frozen orbital.}
    \label{fig:h2o_qubit}
\end{figure}

\begin{figure}[H]
    \centering
    \includegraphics[width=10cm]{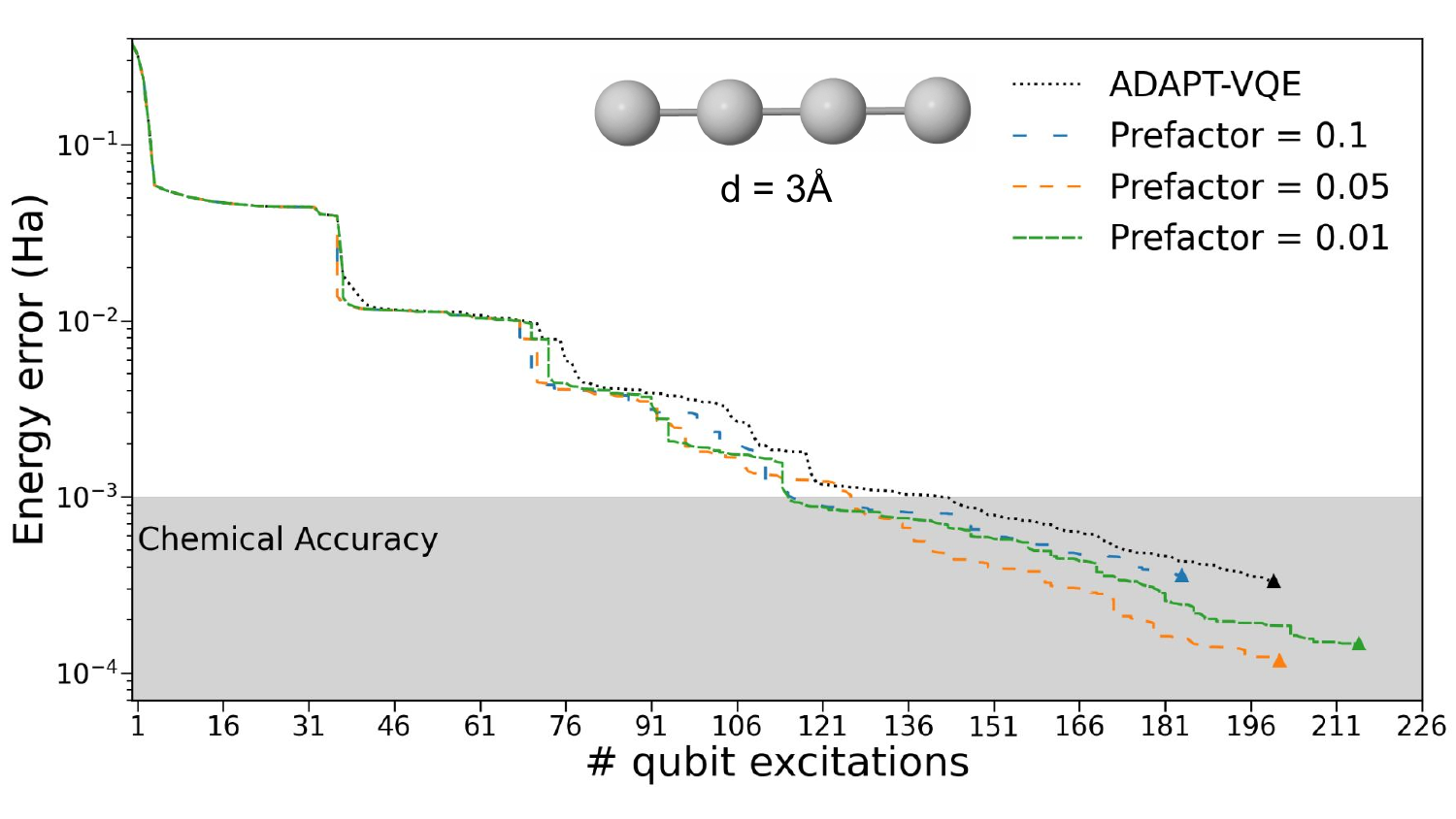}
    \caption{Energy errors (in Hartree) with respect to FCI obtained with ADAPT-VQE (solid blue) and Pruned-ADAPT-VQE (dashed orange) ans\"atze with qubit-excitation-based pool and different prefactor values for the linear {\ce{H4}} molecule with interatomic distance of 3.0~{\AA}, and computed with the 3-21G basis set and an active space of 8 orbitals.}
    \label{si:fig:pref_linh4_qubit}
\end{figure}

In the case of the tetrahedral {\ce{H4}} molecule, using a prefactor of 0.1 leads to premature convergence of the simulation. This suggest that in this case 0.1 may be a bit too aggressive. Decreasing the threshold allows the simulation to reach lower energy values, preventing the premature removal of relevant operators. 
In Figures {\ref{si:fig:pref_linh4_qubit}}-{\ref{si:fig:pref_teth4_qubit}} we test different prefactor values for equation S12.

\begin{figure}[H]
    \centering
    \includegraphics[width=10cm]{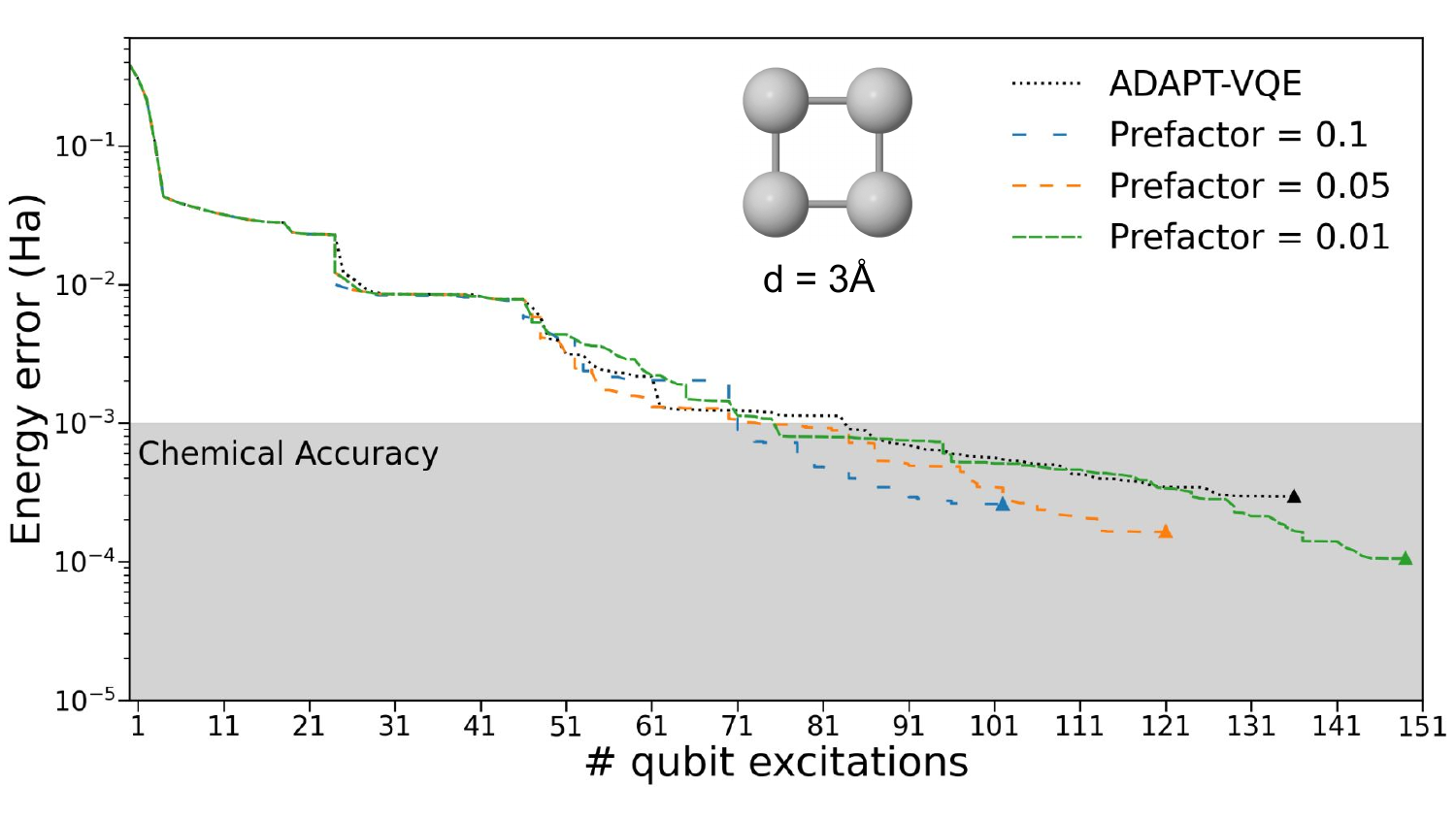}
    \caption{Energy errors (in Hartree) with respect to FCI obtained with ADAPT-VQE (solid blue) and Pruned-ADAPT-VQE (dashed orange) ans\"atze with qubit-excitation-based pool and different prefactor values for the squared {\ce{H4}} molecule with interatomic distance of 3.0~{\AA}, and computed with the 3-21G basis set and an active space of 8 orbitals.}
    \label{si:fig:pref_sqh4_qubit}
\end{figure}

\begin{figure}[H]
    \centering
    \includegraphics[width=10cm]{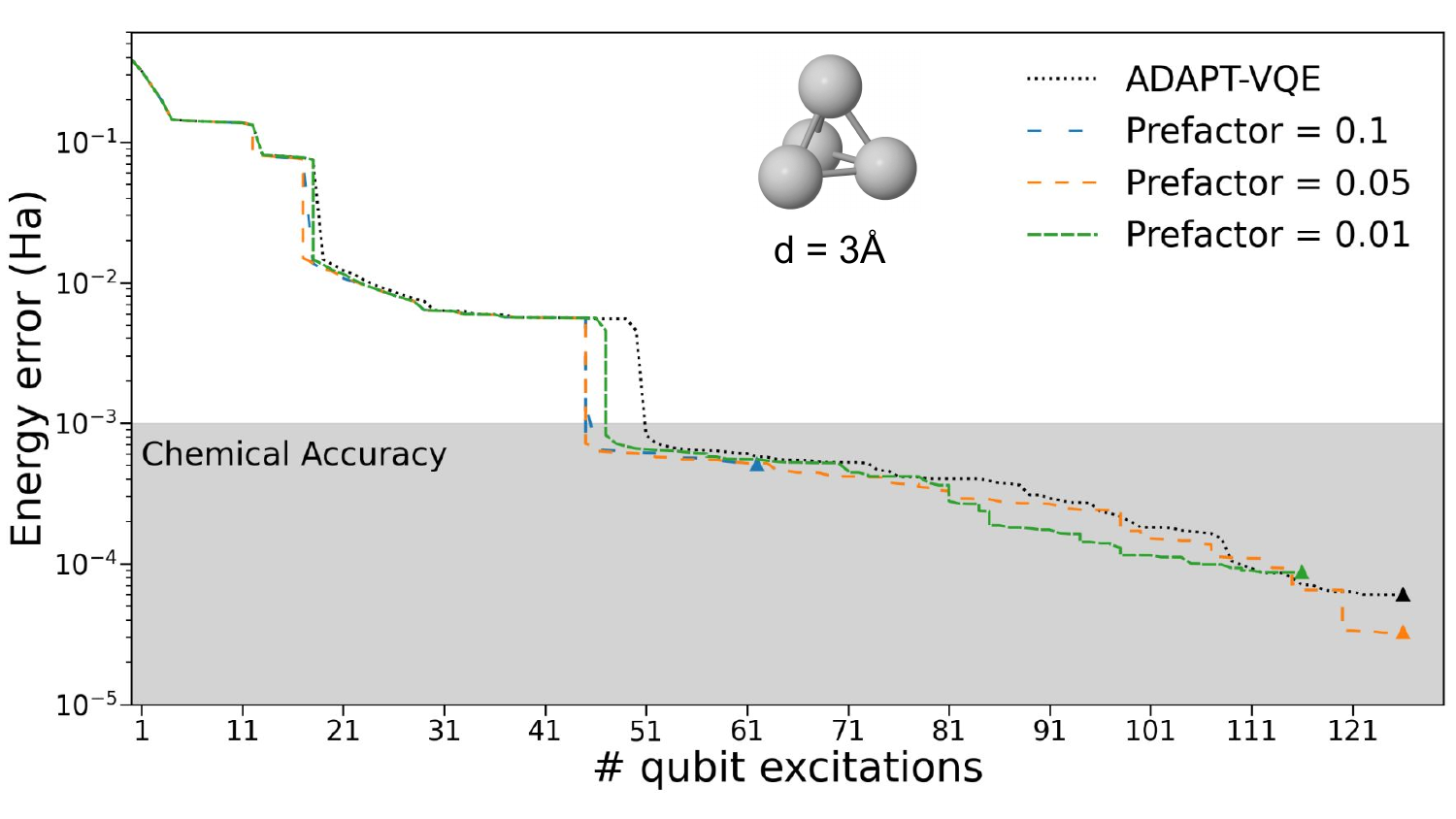}
    \caption{Energy errors (in Hartree) with respect to FCI obtained with ADAPT-VQE (solid blue) and Pruned-ADAPT-VQE (dashed orange) ans\"atze with qubit-excitation-based pool and different prefactor values for the tetrahedral {\ce{H4}} molecule with interatomic distance of 3.0~{\AA}, and computed with the 3-21G basis set and an active space of 8 orbitals.}
    \label{si:fig:pref_teth4_qubit}
\end{figure}

\begin{figure}[H]
    \centering
    \includegraphics[width=10cm]{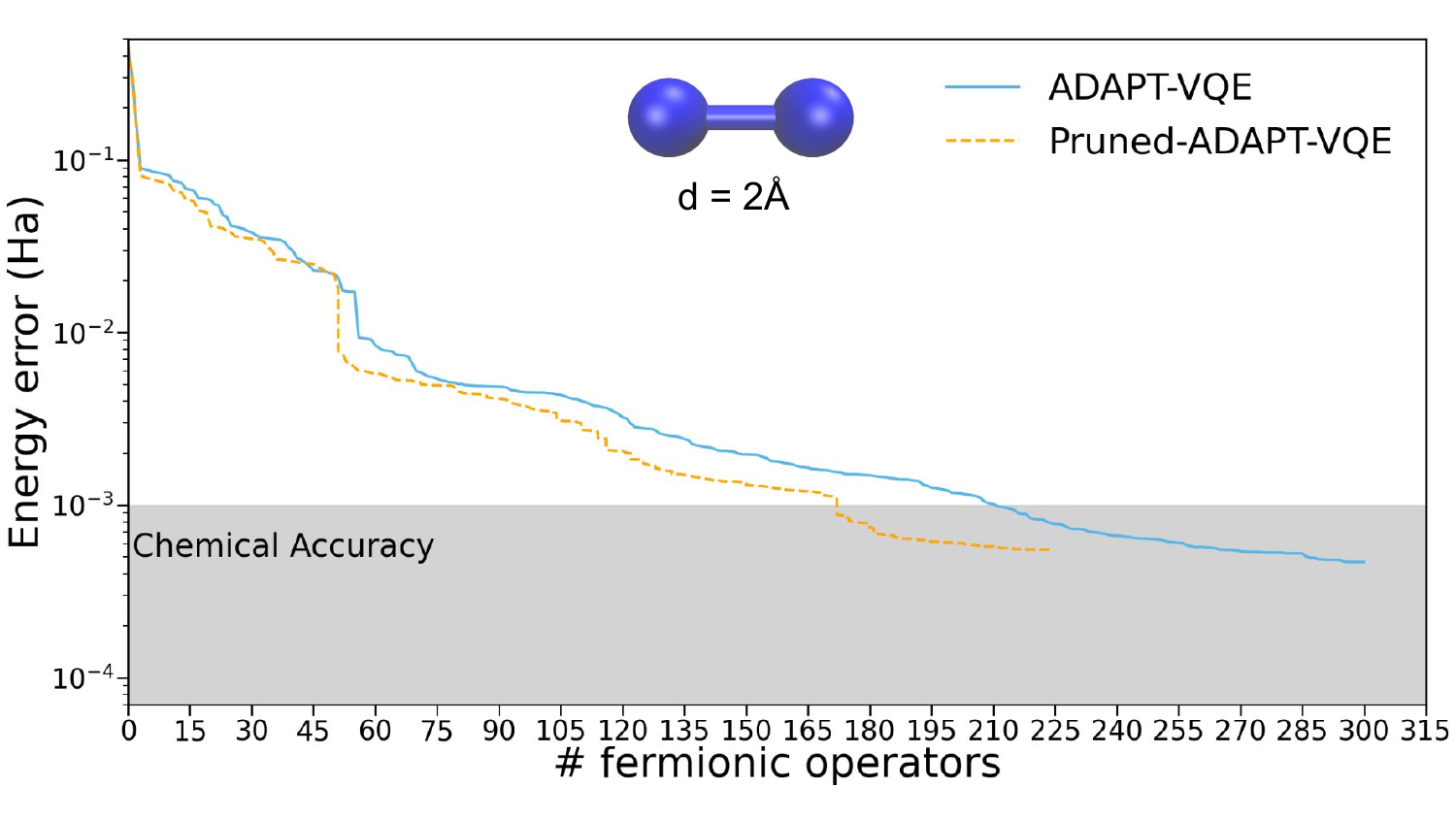}
    \caption{Energy errors (in Hartree) with respect to CASCI(10,8) obtained with ADAPT-VQE (solid blue) and Pruned-ADAPT-VQE (dashed orange) ans\"atze with qubit-excitation-based pool and prefactor = 0.05 for the {\ce{N2}} molecule with interatomic distance of 2.0~{\AA}, and computed with the 3-21G basis set and an active space of 8 orbitals with 2 frozen orbitals.}
    \label{fig:si:n2_exc}
\end{figure}

We find that when using the qubit-excitation-based pool, setting the threshold prefactor to $5\%$ of the average amplitude of the most recently added operators yields a better compromise between accuracy and circuit depth than the $10\%$ value used in the main text. This suggests that the prefactor may need to be adjusted depending on both the target accuracy and the choice of operator pool. The example in Figure~{\ref{fig:si:n2_exc}} have been carried out using a prefactor of 0.05.

\begin{figure}[H]
    \centering
    \includegraphics[width=10cm]{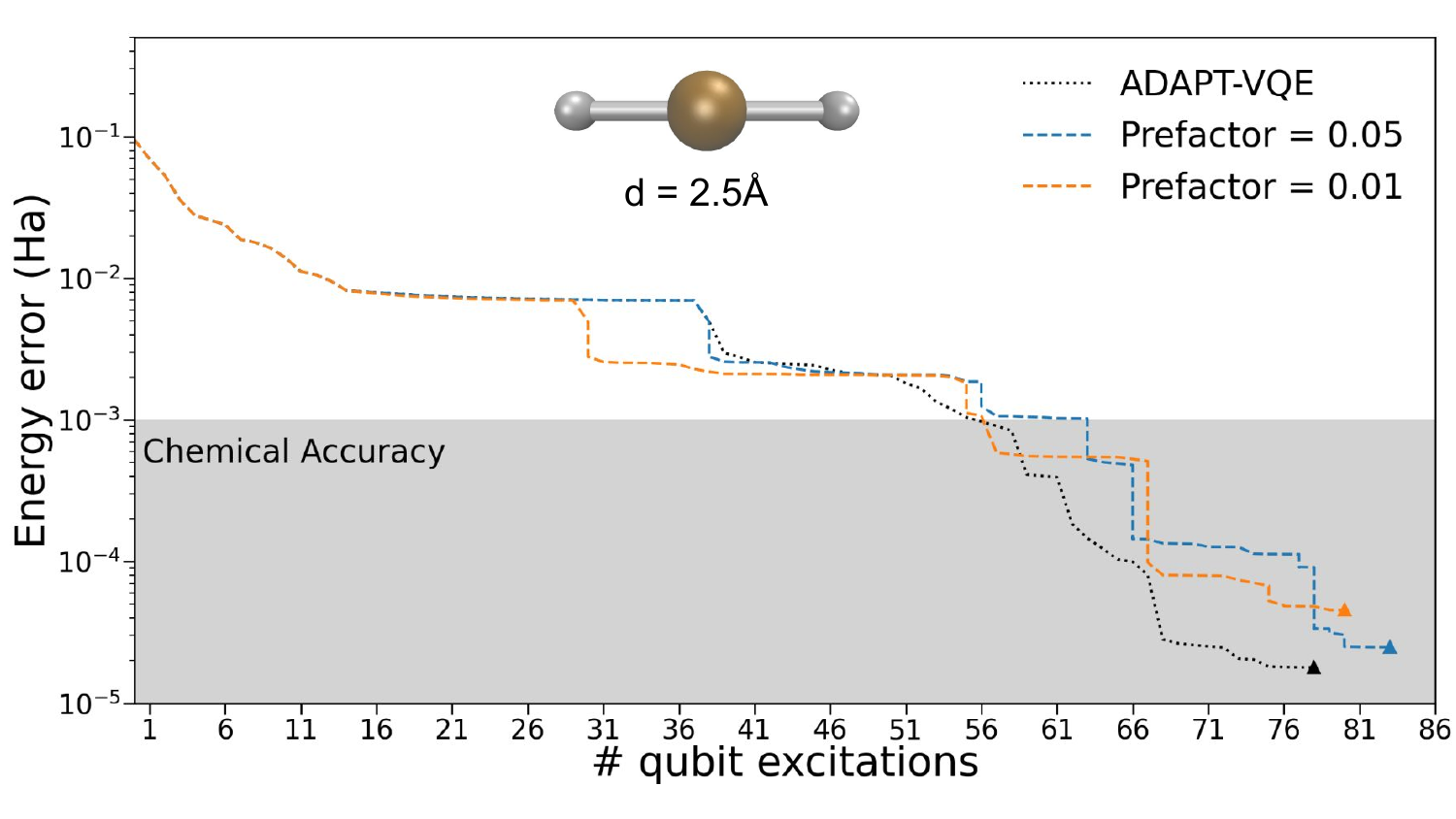}
    \caption{Energy errors (in Hartree) with respect to CASCI(4,8) obtained with ADAPT-VQE (dotted black) and Pruned-ADAPT-VQE (dashed blue for prefactor 0.05 and dashed orange for prefactor 0.01, see Section 5) ans\"atze for the {\ce{BeH2}} molecule with interatomic distance of 2.5~{\AA}, and computed with the 3-21G basis set and an active space of 8 orbitals plus one frozen orbital.}
    \label{fig:si:beh2_exc}
\end{figure}

Figure~\ref{fig:si:beh2_exc} shows an example of a system where the pruning strategy leads to longer ans\"atze than the ADAPT-VQE, suggesting that for some exceptional cases, the hyperparameters should be further adjusted.

\section{~STO-3G basis testing}
This section includes simulations performed in STO-3G basis for both pools, fermionic and qubit-excitation-based.

\subsection{~Fermionic pool}
\begin{figure}[H]
    \centering
    \includegraphics[width=10cm]{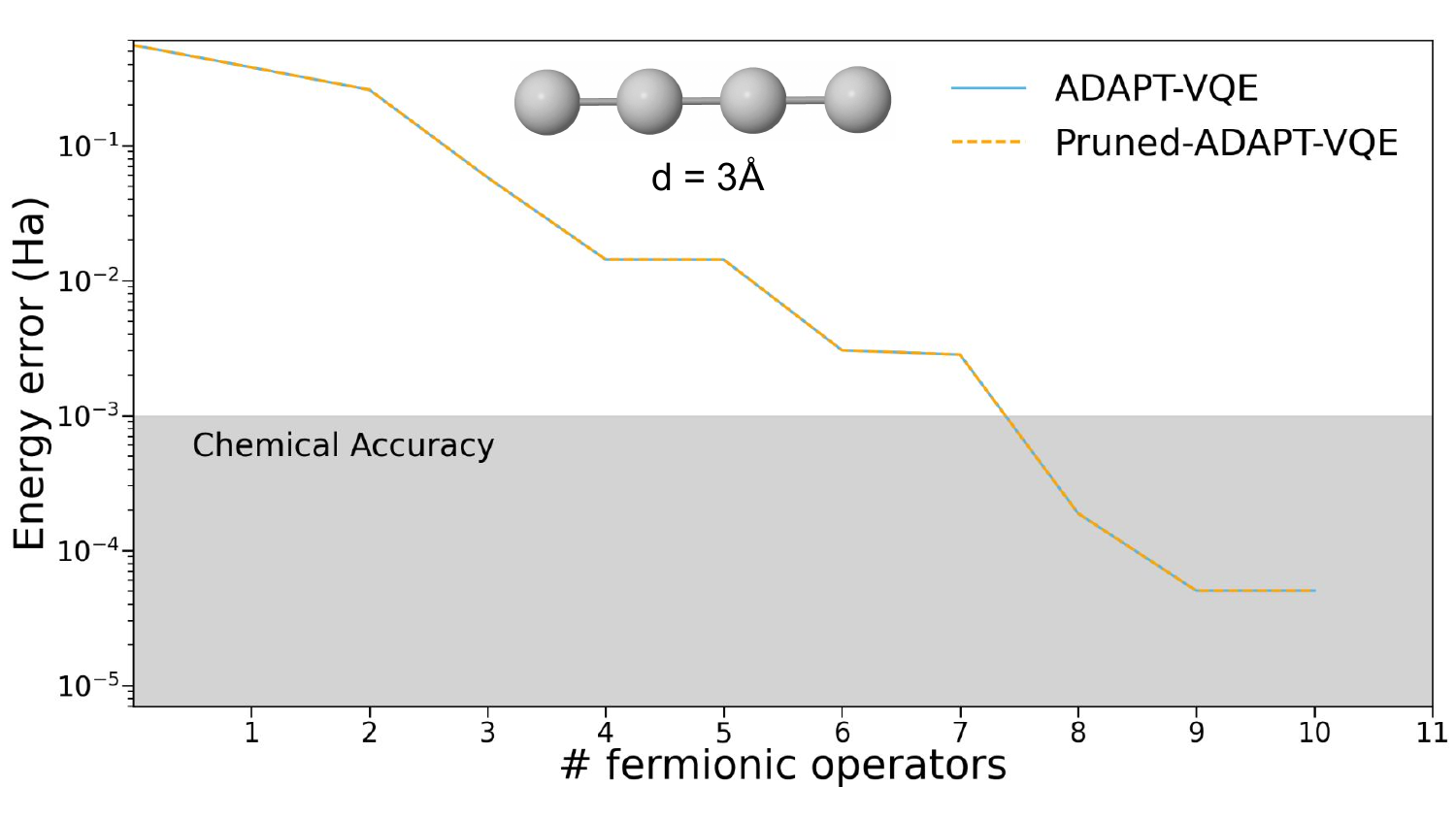}
    \caption{Energy errors (in Hartree) with respect to FCI obtained with ADAPT-VQE (solid blue) and Pruned-ADAPT-VQE (dashed orange) ans\"atze with fermionic pool for the linear {\ce{H4}} molecule with interatomic distance of 3.0~{\AA}, and computed with the STO-3G basis set and an active space of 8 orbitals.}
    \label{fig:h4_lin_sto3g}
\end{figure}

\begin{figure}[H]
    \centering
    \includegraphics[width=10cm]{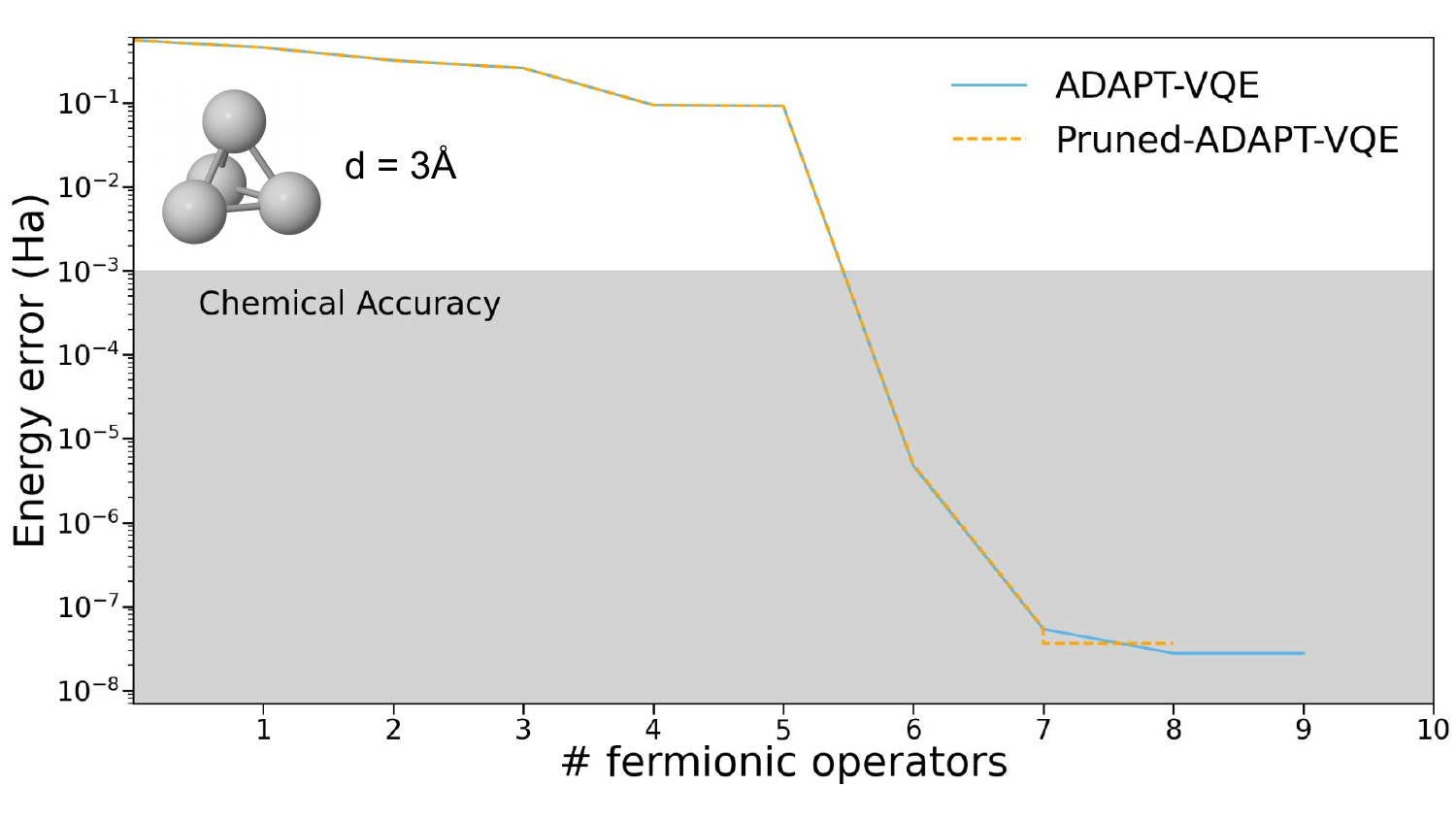}
    \caption{Energy errors (in Hartree) with respect to FCI obtained with ADAPT-VQE (solid blue) and Pruned-ADAPT-VQE (dashed orange) ans\"atze with fermionic pool for the tetrahedral {\ce{H4}} molecule with interatomic distance of 3.0~{\AA}, and computed with the STO-3G basis set and an active space of 8 orbitals.}
    \label{fig:h4_td_sto3g}
\end{figure}

\begin{figure}[H]
    \centering
    \includegraphics[width=10cm]{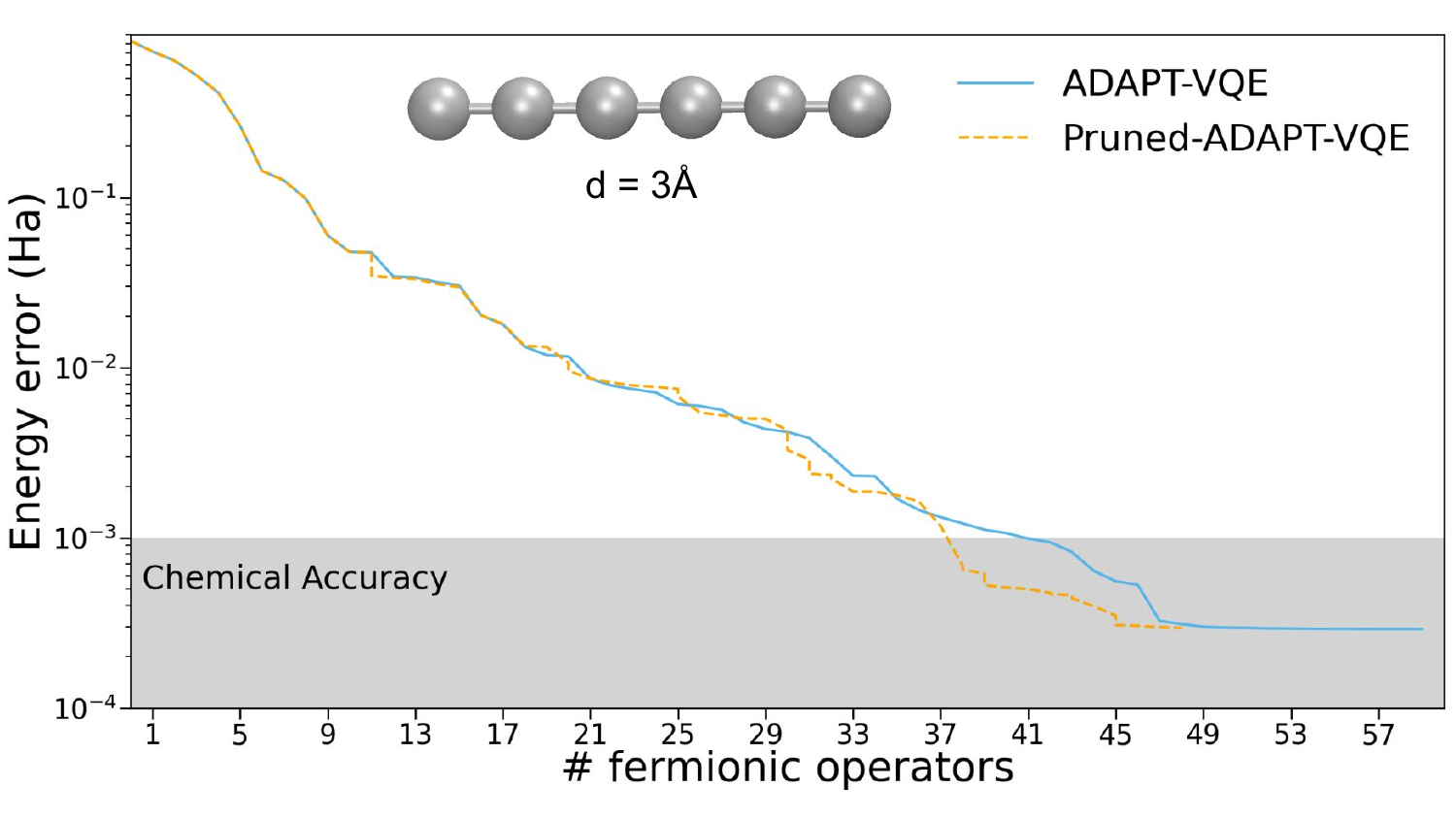}
    \caption{Energy errors (in Hartree) with respect to FCI obtained with ADAPT-VQE (solid blue) and Pruned-ADAPT-VQE (dashed orange) ans\"atze with fermionic pool for the {\ce{H6}} molecule with interatomic distance of 3.0~{\AA}, and computed with the STO-3G basis set and an active space of 6 orbitals.}
    \label{fig:h6_sto3g}
\end{figure}

\subsection{~Qubit-excitation-based pool}

\begin{figure}[H]
    \centering
    \includegraphics[width=10cm]{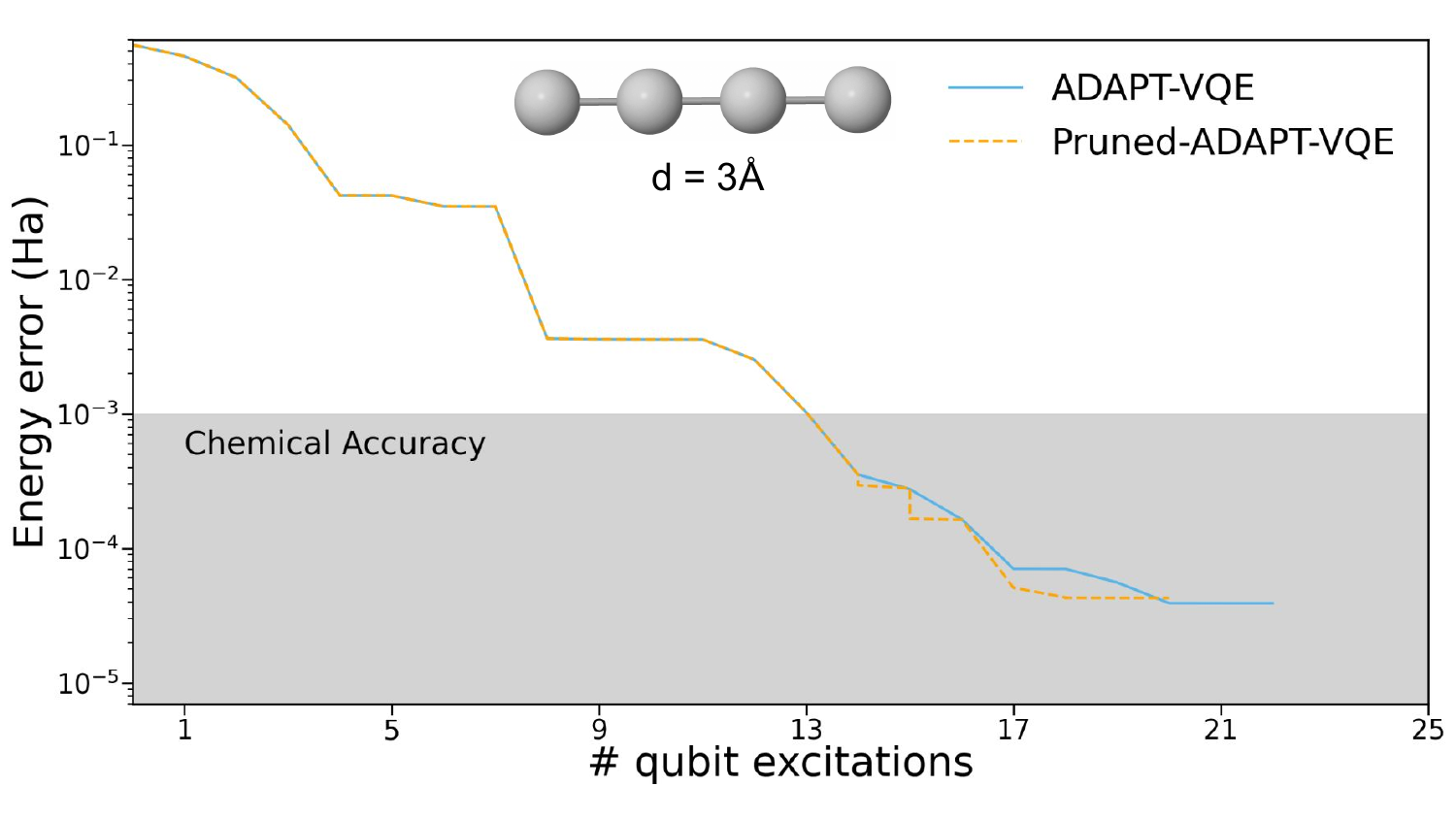}
    \caption{Energy errors (in Hartree) with respect to FCI obtained with ADAPT-VQE (solid blue) and Pruned-ADAPT-VQE (dashed orange) ans\"atze with qubit-excitation-based pool for the linear {\ce{H4}} molecule with interatomic distance of 3.0~{\AA}, and computed with the STO-3G basis set and an active space of 8 orbitals.}
    \label{fig:h4_lin_sto3g_qubit}
\end{figure}

\begin{figure}[H]
    \centering
    \includegraphics[width=0.5\linewidth]{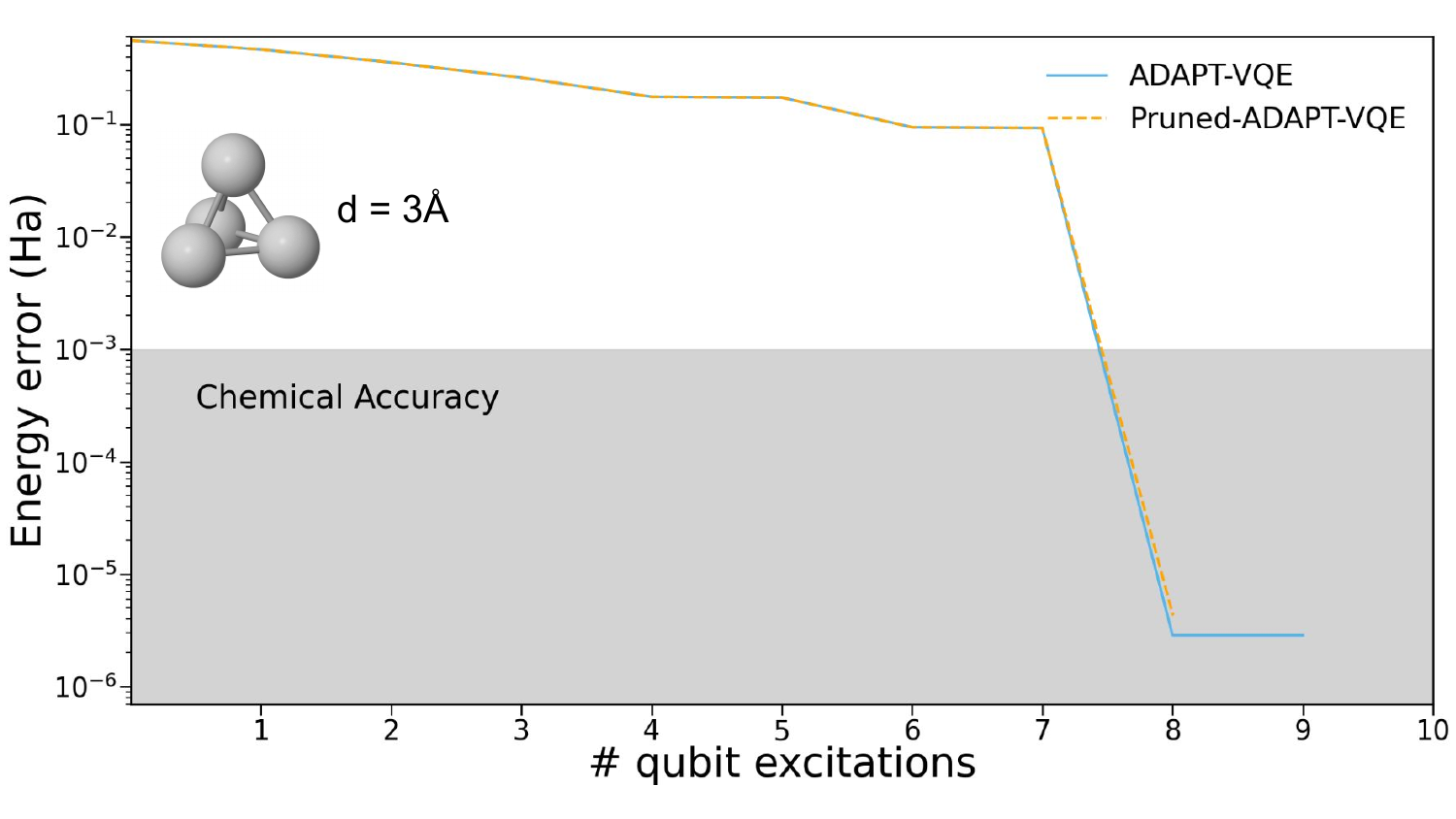}
    \caption{Energy errors (in Hartree) with respect to FCI obtained with ADAPT-VQE (solid blue) and Pruned-ADAPT-VQE (dashed orange) ans\"atze with qubit-excitation-based pool for the tetrahedral {\ce{H4}} molecule with interatomic distance of 3.0~{\AA}, and computed with the STO-3G basis set and an active space of 8 orbitals.}
    \label{fig:h4_td_sto3g_qubit}
\end{figure}

\begin{figure}
    \centering
    \includegraphics[width=0.5\linewidth]{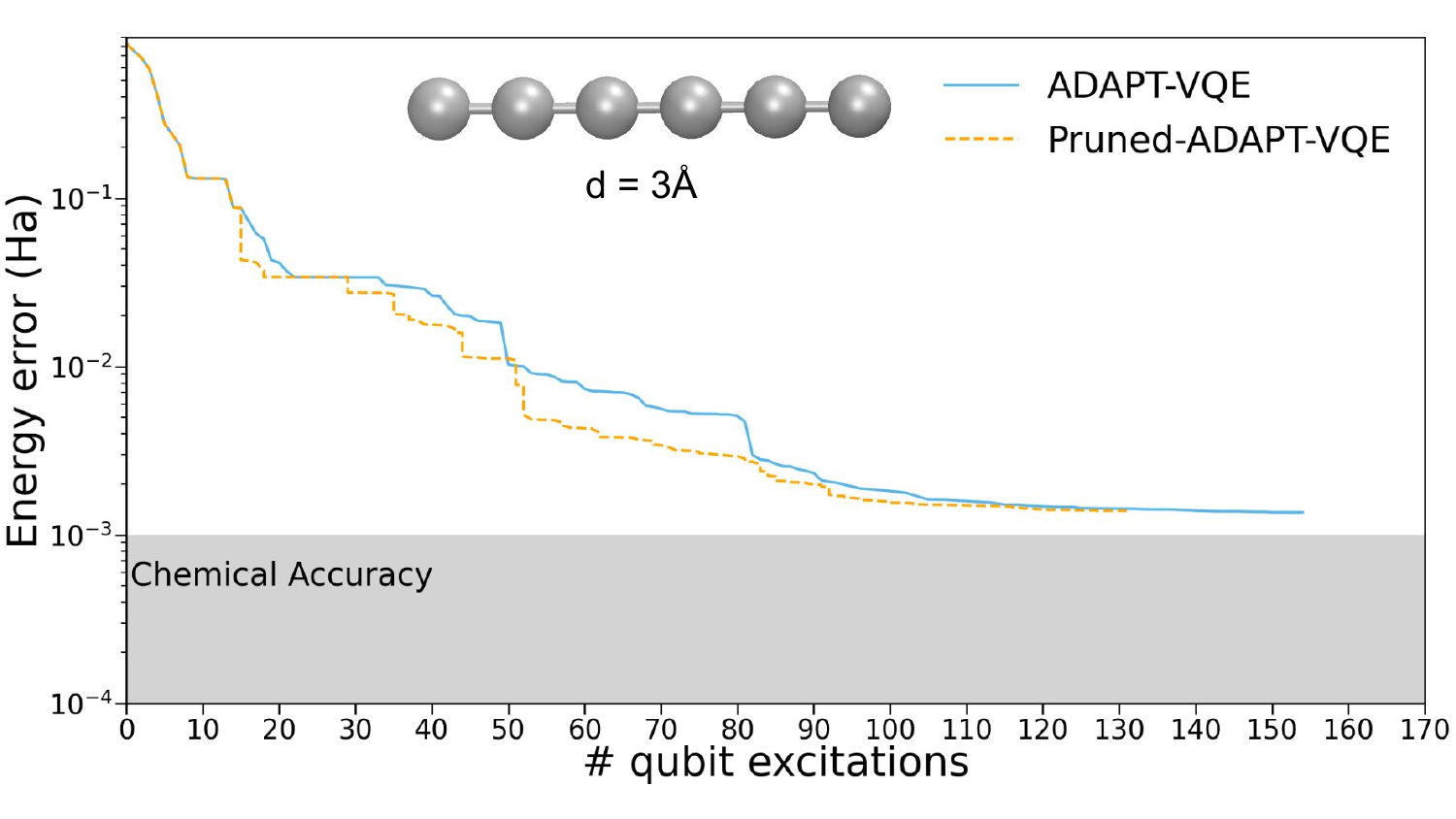}
    \caption{Energy errors (in Hartree) with respect to FCI obtained with ADAPT-VQE (solid blue) and Pruned-ADAPT-VQE (dashed orange) ans\"atze with qubit-excitation-based pool for the {\ce{H6}} molecule with interatomic distance of 3.0~{\AA}, and computed with the STO-3G basis set and an active space of 6 orbitals.}
    \label{fig:h6_sto3g_qubit}
\end{figure}

\section{~Energy difference vs parameter value} \label{sec:energy_diff_SI}

In this section we show the relationship between the parameter values and its impact on the VQE energy for the linear {\ce{H4}} molecule.

\begin{figure}[H]
    \centering
    \includegraphics[width=0.8\linewidth]{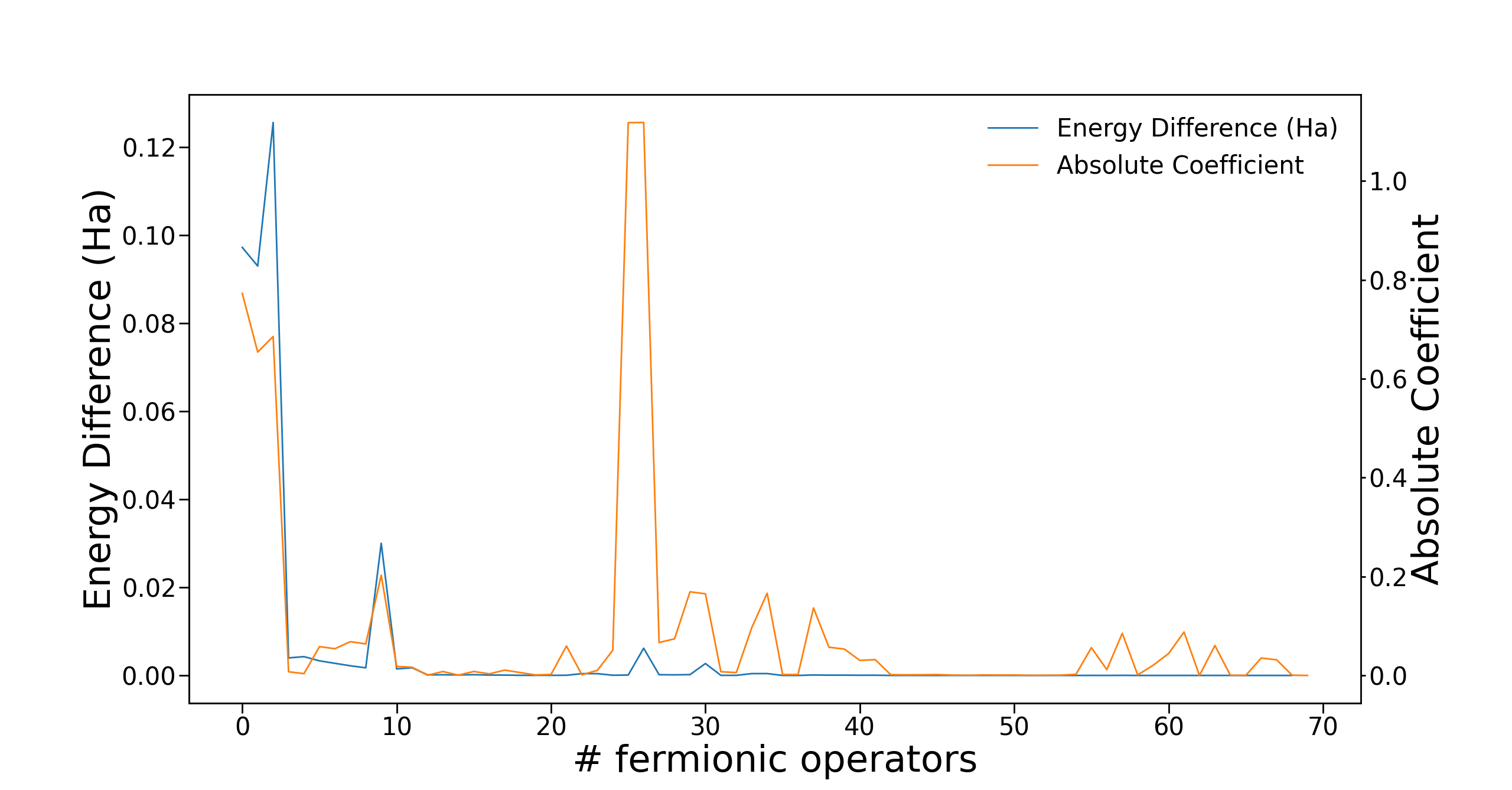}
    \caption{Energy differences (in a.u.) and parameter value for each added fermionic operator of ADAPT-VQE for the simulation of linear {\ce{H4}} with the 3-21G basis set.}
    \label{fig:energy_diff_coeffs}
\end{figure}

\begin{figure}[H]
    \centering
    \includegraphics[width=0.8\linewidth]{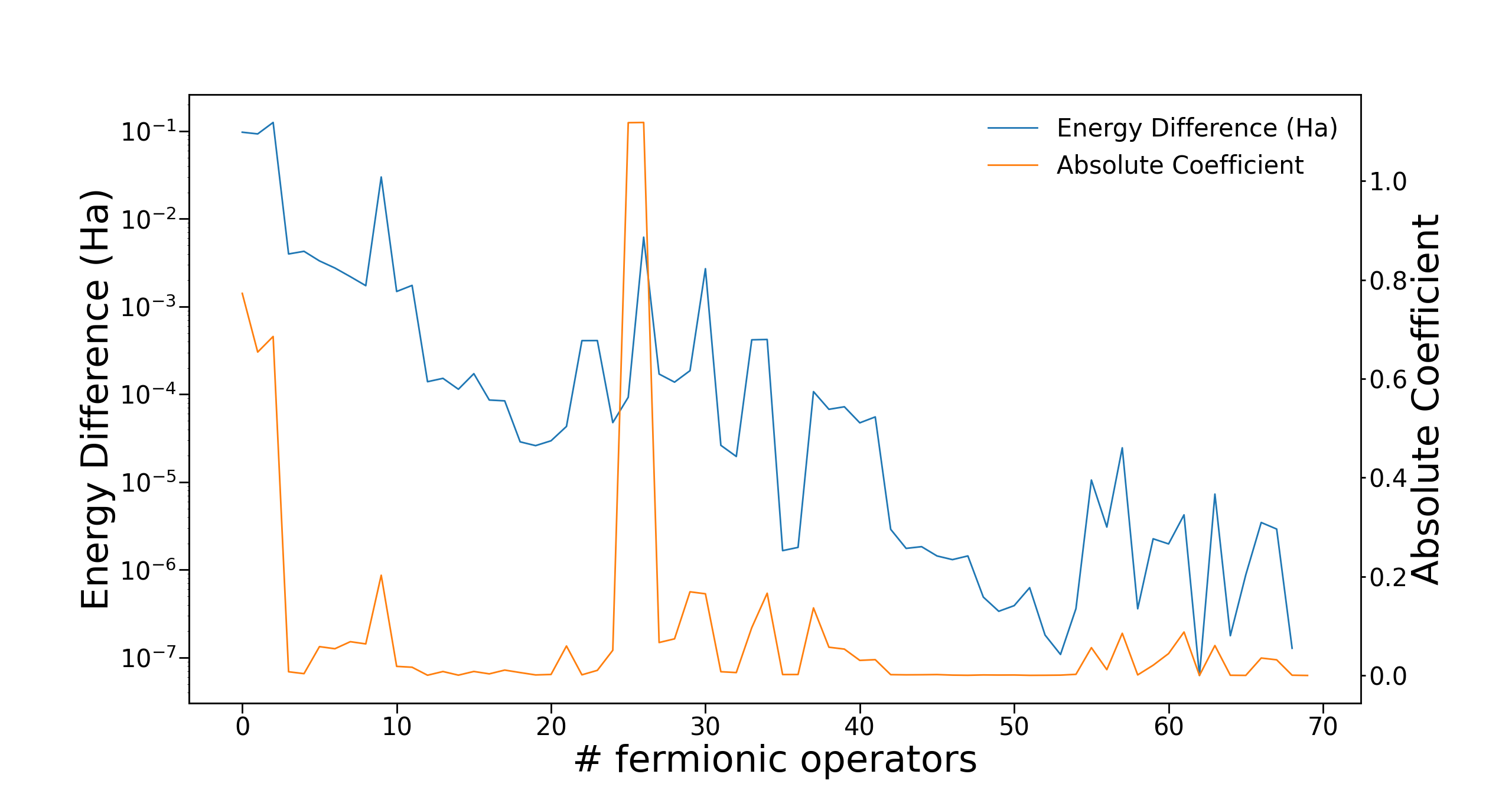}
    \caption{Energy differences (in a.u., logrithmic scale) and parameter value for each added fermionic operator of ADAPT-VQE for the simulation of linear {\ce{H4}} with the 3-21G basis set.}
    \label{fig:energy_diff_coeffs_log}
\end{figure}

\section{~Gate counts}

In this section, we compare Pruned-ADAPT-VQE and standard ADAPT-VQE by analyzing the number of CNOT gates in the ansatz at each iteration. The evaluation of the number of gates is performed by first mapping the fermionic operators to qubit operators using the Jordan–Wigner transformation. The resulting qubit operators are then encoded into quantum circuits using the staircase algorithm. Circuit optimization is subsequently applied using the transpilation routines available in Qiskit to reduce the final CNOT count. 
The analyzed examples include the linear {\ce{H4}} molecule using both the UCCSD ansatz and the qubit-excitation-based operator pool.

\begin{figure}
    \centering
    \includegraphics[width=10cm]{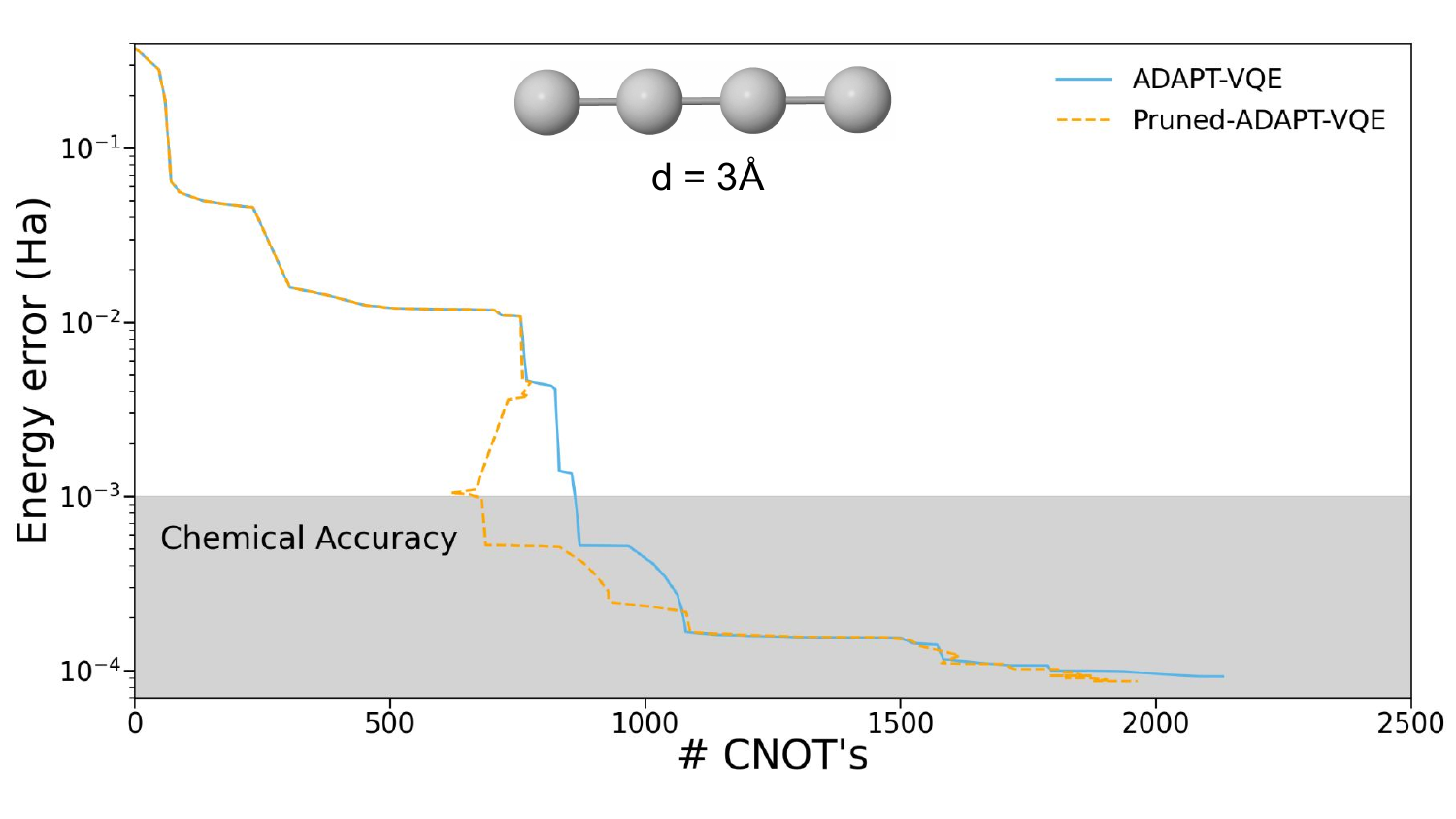}
    \caption{Energy errors (in Hartree) with respect to FCI as a function of the number of CNOT gates, obtained with ADAPT-VQE (solid blue) and Pruned-ADAPT-VQE (dashed orange) ans\"atze  for the linear {\ce{H4}} molecule with interatomic distance of 3.0~{\AA}.}
    \label{fig:h4_lin_cnot}
\end{figure}

\begin{figure}
    \centering
    \includegraphics[width=10cm]{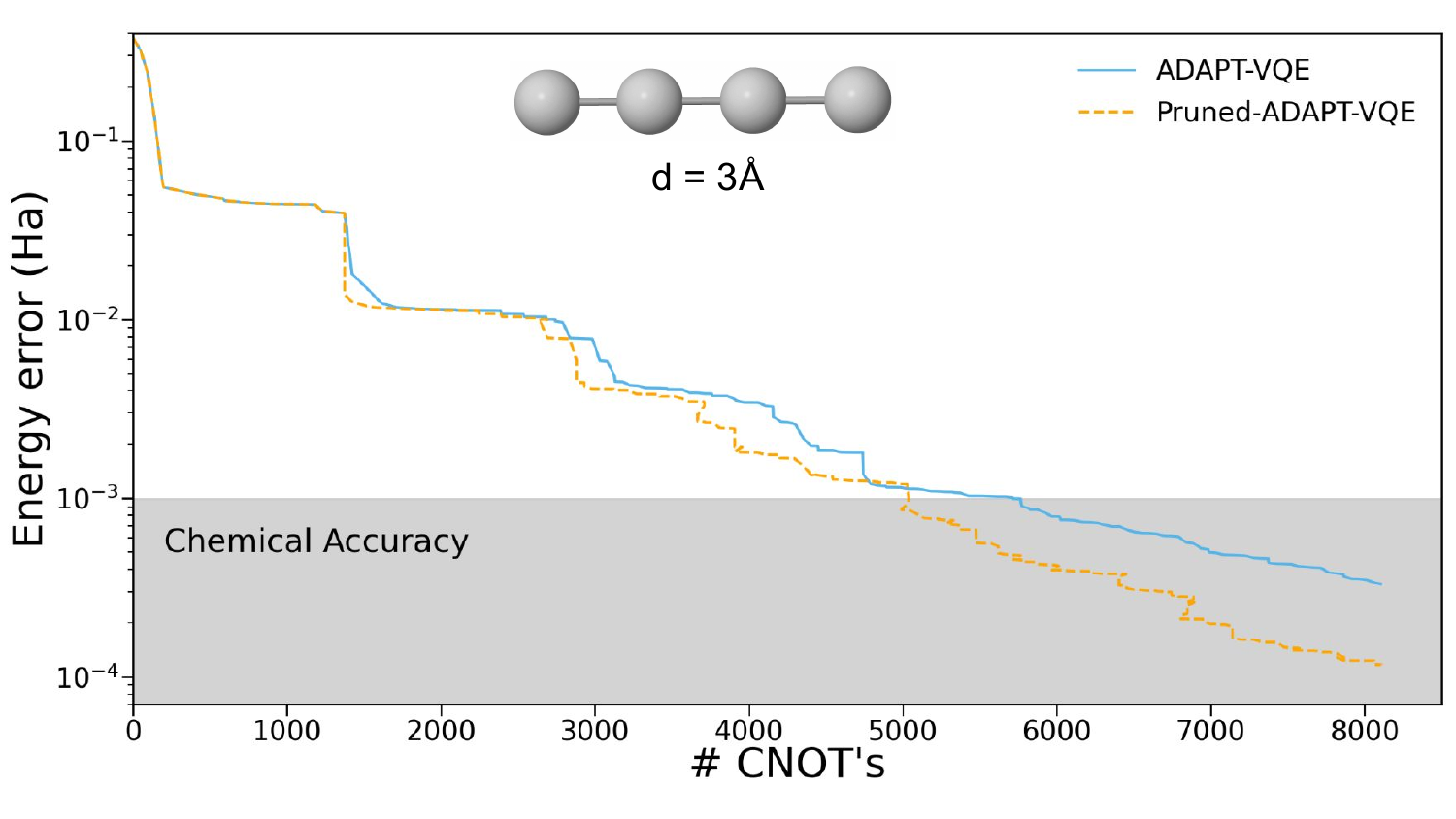}
    \caption{Energy errors (in Hartree) with respect to FCI as a function of the number of CNOT gates, obtained with ADAPT-VQE (solid blue) and Pruned-ADAPT-VQE (dashed orange) ans\"atze with the qubit-excitation-based pool operator for the linear {\ce{H4}} molecule with interatomic distance of 3.0~{\AA}.}
    \label{fig:h4_lin_cnot_qubit}
\end{figure}

\section{~Impact of Removal on Classical Optimization Cost}
In this section, we study whether the removal of an operator affects the next classical optimization loop, analyzing whether it leads to additional iterations.\\
The fairest comparison in this context is obtained by examining the optimization of the ansatz when the first operator is removed, since at that stage both ansätze are still identical. For the main simulation used as example in this work, \ce{H4} with the fermionic pool, the first removal occurs at iteration 27, where Pruned-ADAPT-VQE deletes the operator in position 4. Afterward, both methods select the same operator at iteration 27 and again at iteration 28. Thus, by iteration 28 both ansätze are structurally the same, except that Pruned-ADAPT-VQE lacks the operator at position 4. In this iteration, ADAPT-VQE converges in 13 classical steps, whereas Pruned-ADAPT-VQE requires 14.\\

\begin{figure}[H]
    \centering
    \includegraphics[width=0.8\linewidth]{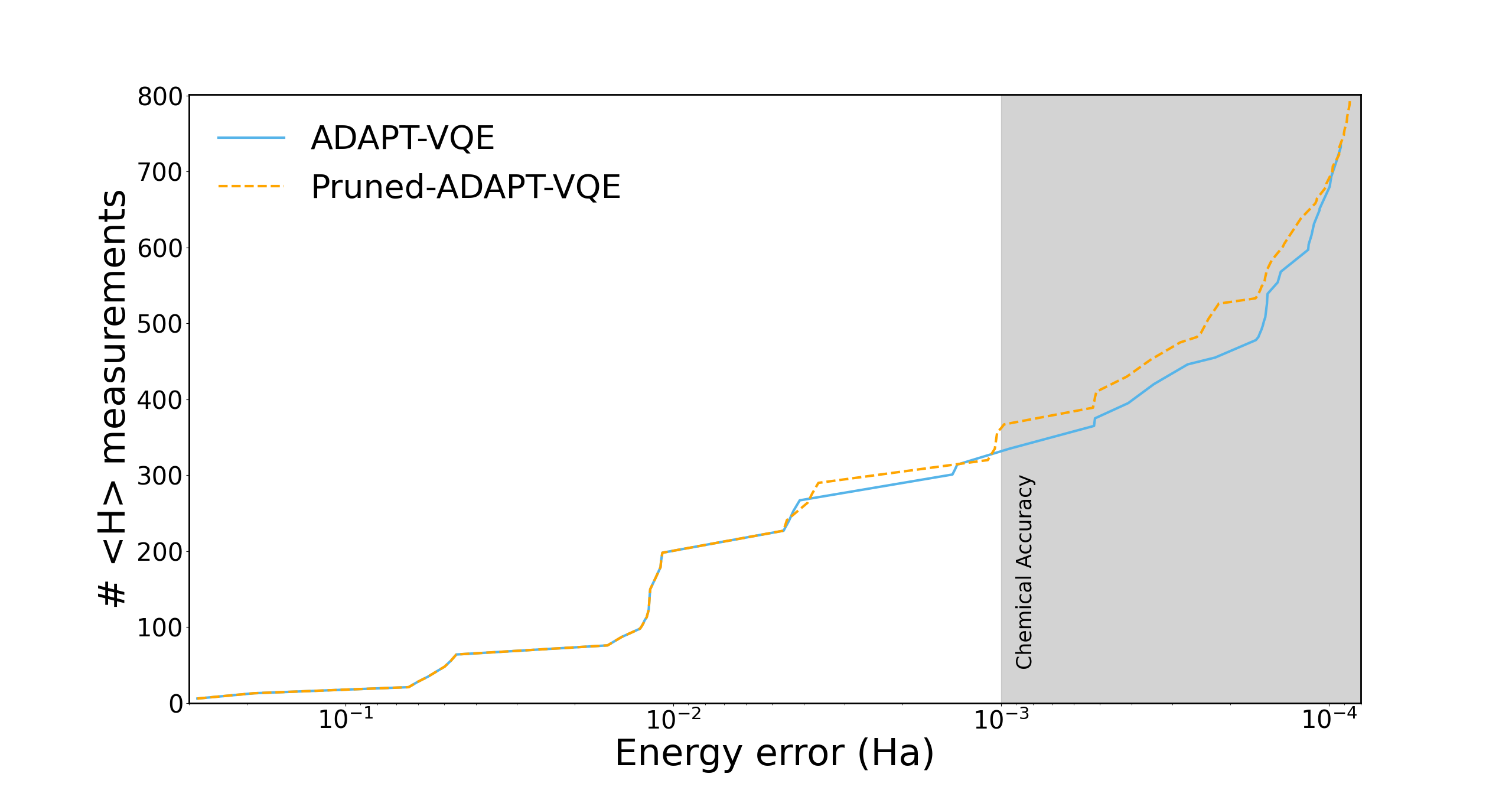}
    \caption{Number of energy measurements (or classical optimizer iterations) versus energy error (in Hartree) with respect to FCI for ADAPT-VQE (solid blue) and Pruned-ADAPT-VQE (dashed orange) ansätze with the fermionic pool for the linear \ce{H4} molecule with interatomic distance of 3.0~{\AA}.}
    \label{fig:si:classical_evs}
\end{figure}

To give a broader view of the cost associated with operator removal, Figure \ref{fig:si:classical_evs} shows the energy error as a function of the cumulative number of energy evaluations performed by the classical optimizer throughout the algorithm. This simulation corresponds to the \ce{H4} system at an interatomic distance of 3.0~{\AA} with a fermionic operator pool. The figure shows that Pruned-ADAPT-VQE introduces a small overhead in some segments, and the two methods converge to similar results by the end of the simulation.

\bibliography{abbr, references} % Produces the bibliography via BibTeX.